\documentclass[conference,onecolumn]{IEEEtran}
\IEEEoverridecommandlockouts

\makeatletter 
\newcommand{\linebreakand}{%
  \end{@IEEEauthorhalign}
  \hfill\mbox{}\par
  \mbox{}\hfill\begin{@IEEEauthorhalign}
}
\makeatother 

\usepackage{cite}
\usepackage{amsmath,amssymb,amsfonts}
\usepackage{algorithmic}
\usepackage{graphicx}
\usepackage{textcomp}
\usepackage{xcolor}
\usepackage{comment}

\usepackage{soul} 

\newcommand\mycom[2]{\genfrac{}{}{0pt}{}{#1}{#2}}

\newcommand {\done} {\quad\vrule height4pt WIDTH4PT}

\newtheorem{lemma}{Lemma}[section]
\newtheorem{proposition}{Proposition}[section]

\newtheorem{corollary}{Corollary}[section]
\newtheorem{remark}{Remark}[section]

\newcommand{\Prob}{\mathbb{P}}

\newcommand{\E}{\mathbb{E}}
\newcommand{\R}{\mathbb{R}}
\newcommand{\C}{\mathbb{C}}

\newcommand{\erf}{\mathrm{erf}}

\newcommand{\ar}[1]{{\color{black}{#1}}}
\newcommand{\pn}[1]{{\color{black}{#1}}}

\def\BibTeX{{\rm B\kern-.05em{\sc i\kern-.025em b}\kern-.08em
    T\kern-.1667em\lower.7ex\hbox{E}\kern-.125emX}}
\begin{document}

\title{Quickest Change Detection in Discrete-Time in Presence of a Covert Adversary\\
\thanks{A. Ramtin and D. Towsley were supported by the DEVCOM Army Research Laboratory under
Cooperative Agreement W911NF-17-2-0196 (IoBT CRA) and the National Science Foundation under Grant ECCS-2148159.}
}

\author{\IEEEauthorblockN{Amir Reza Ramtin}
\IEEEauthorblockA{\textit{College of Information $\&$ Computer Sciences} \\
\textit{University Massachusetts at Amherst}\\
Amherst, MA, USA\\
aramtin@umass.edu}
\and
\IEEEauthorblockN{Philippe Nain}
\IEEEauthorblockA{\textit{Inria centre at Universit\'e C\^ote d'Azur} \\
Sophia Antipolis, France  \\
philippe.nain@inria.fr}
\and
\linebreakand
\IEEEauthorblockN{Don Towsley}
\IEEEauthorblockA{\textit{College of Information $\&$ Computer Sciences} \\
\textit{University Massachusetts at Amherst}\\
Amherst, MA, USA\\
towsley@cs.umass.edu}
}

\maketitle

\begin{abstract}
We study the problem of covert quickest change detection in a discrete-time setting, where a sequence of observations undergoes a distributional change at an unknown time. Unlike classical formulations, we consider a covert adversary who has knowledge of the detector’s false alarm constraint parameter $\gamma$ and selects a stationary post-change distribution that depends on it, seeking to remain undetected for as long as possible. Building on the theoretical foundations of the CuSum procedure, we rigorously characterize the asymptotic behavior of the average detection delay (ADD) and the average time to false alarm (AT2FA) when the post-change distribution converges to the pre-change distribution as $\gamma \to \infty$. Our analysis establishes exact asymptotic expressions for these quantities, extending and refining classical results that no longer hold in this regime. We identify the critical scaling laws governing covert behavior and derive explicit conditions under which an adversary can maintain covertness, defined by ADD = $\Theta(\gamma)$, whereas in the classical setting, ADD grows only as $\mathcal{O}(\log \gamma)$. In particular, for Gaussian and Exponential models under adversarial perturbations of their respective parameters, we asymptotically characterize ADD as a function of the Kullback--Leibler divergence between the pre- and post-change distributions and $\gamma$.
\end{abstract}

\section{Introduction}
The problem  of {\it sequential change detection} or {\it quickest change detection} (QCD), arises in various engineering applications \cite{Poor2009,Tartakovsky2014,veer-bane-e-ref-2013}. In all of these applications, the decision-making agent receives a sequence of stochastic observations that undergo a change in distribution in response to a  change or disruption in the environment. As long as the behavior of the observations is consistent with the normal (pre-change) state, the decision-maker is content to allow the process to continue. If the distribution changes, then the decision-maker is interested in  detecting the change as soon as possible and taking any necessary action in response to the change. More precisely, the goal of the decision-maker is to make the average detection delay (ADD) as small as possible, while constraining the average time to false alarm (AT2FA) to be above some threshold. 

{Detecting adversarial behavior is one of the most critical applications of QCD \cite{Xie2021}. Traditional analyses, however, typically assume that potential adversaries are unaware of the underlying detection mechanism and therefore cannot adjust their actions to avoid being caught. In modern interconnected systems, this assumption is increasingly less realistic -- sophisticated adversaries can often infer or estimate aspects of the detection strategy and intentionally manipulate their behavior to remain undetected. Such covert adversarial behavior poses new challenges for QCD, especially in security-sensitive domains such as cyber defense, surveillance, and military operations, where even small delays in detection can lead to significant consequences.}

The notion of covertness, originally introduced in the context of covert communication \cite{bash2013limits}, refers to the ability of a transmitter to remain undetectable by an observing warden during communication. Most existing work on covert communication assumes that the detector knows the exact start time of any potential transmission. In contrast, we relax this assumption by extending the notion of covertness to the QCD setting, where adversaries deliberately act to remain hidden over prolonged time horizons.
This departure results in a fundamentally different adversarial strategy when considered within the framework of QCD. Unlike traditional adversaries whose actions result in detectable statistical deviations, covert adversaries aim to ensure that ADD scales with the same order as AT2FA, thereby rendering the QCD procedure ineffective \cite{huang_covert_2021}.

In this work, we specifically focus on the theoretical characterization of covert adversarial behavior against the Cumulative Sum (CuSum) procedure. The CuSum procedure, first proposed by Page \cite{Page1954}, is one of the most widely studied methods in quickest change detection. It operates by recursively accumulating log-likelihood ratios of the observed data and signaling an alarm when this cumulative statistic exceeds a prescribed threshold. The CuSum test is well known for its optimality under Lorden’s minimax criterion \cite{Lorden1971}, which minimizes the worst-case expected detection delay subject to a false-alarm constraint. Classical analyses, including Lorden’s asymptotic characterization of ADD, assume that the post-change distribution is fixed and independent of the false-alarm constraint $\gamma$. In more realistic adversarial settings, however, this assumption may not hold: a covert adversary can deliberately choose a post-change distribution that depends on $\gamma$, causing the post-change and pre-change distributions to become increasingly similar as $\gamma$ grows. This dependence invalidates classical asymptotic results and motivates the need for new analyses of scaling stationary post-change models, in which the post-change distribution converges to the pre-change distribution as $\gamma \to \infty$.

While prior work has examined covertness in both discrete-time \cite{huang_covert_2021, ramtin2024quickest} and continuous-time sequential detection settings \cite{ramtin2025continuous}, these studies differ substantially from our formulation. The work in \cite{huang_covert_2021} is primarily rooted in covert wireless communication, where the modeling assumptions are tailored to communication-theoretic contexts -- such as specific channel models and power constraints -- and the focus lies on communication-centric performance metrics rather than fundamental detection limits. In contrast, \cite{ramtin2024quickest} investigates an adversarial setting against the CuSum procedure under non-stationarity, without assuming that the adversary knows the false-alarm constraint parameter, whereas in our work the post-change distribution explicitly depends on this parameter. Consequently, the formulation in \cite{ramtin2024quickest} differs fundamentally from ours. The study most closely related to our framework is the continuous-time analysis in \cite{ramtin2025continuous}, which considers a Brownian-motion model where the post-change drift depends on the false-alarm constraint. However, that setting benefits from closed-form expressions for both ADD and AT2FA, allowing for exact characterizations. In contrast, the discrete-time case analyzed here is significantly more challenging, as no such exact equalities exist between ADD, AT2FA, the detection threshold, and the Kullback--Leibler divergence -- necessitating a refined asymptotic analysis to obtain precise scaling results.

We provide a rigorous asymptotic analysis under 
{a discrete-time stationary setting in which the post-change distribution depends on the false-alarm constraint parameter $\gamma$,} deriving a novel asymptotic characterization of ADD for the CuSum procedure when the difference between the pre-change and post-change distributions diminishes as $\gamma \to \infty$. In this regime, we demonstrate that the classical asymptotic result for ADD due to Lorden no longer applies. Our principal theoretical contribution -- motivated by the need to analyze the asymptotics of ADD in this challenging setting -- is the establishment of precise asymptotic expressions for both AT2FA and ADD, formulated in terms of the Kullback--Leibler divergence between the post-change and pre-change distributions and the detection threshold. These expressions, previously regarded as approximations in classical works such as \cite{Khan,Siegmund1985}, are rigorously proven to be asymptotically exact in our regime, thereby substantially strengthening the theoretical underpinnings of QCD.

Central to our proofs is demonstrating that, in our specific model, the overshoot -- the amount by which the detection statistic exceeds the threshold upon crossing -- vanishes asymptotically under the assumptions of our analysis. This behavior is not true in general; in fact, classical results (e.g., \cite{Siegmund1985}) characterize the asymptotic distribution of the overshoot rather than its disappearance. However, in our setting, the standard renewal-theoretic bounds on overshoots are too loose and fail to capture the fine-grained dependence on $\gamma$.
To {address} this, we develop a refined asymptotic analysis showing that under 
{two specific} conditions, the overshoots of both the upper and lower boundaries of the corresponding Sequential Probability Ratio Test (SPRT) -- which underlies the CuSum procedure -- indeed vanish in the asymptotic regime. 
This result is essential for rigorously justifying the validity of our asymptotic expressions for AT2FA and ADD. We note that these two conditions are not easy to check, since they involve the overshoot random variables of the cumulative sum at the upper and lower boundaries. For this reason, we introduce Lemma~\ref{lem:suffiicient-conditions}, which provides sufficient conditions under which these key conditions are guaranteed to hold. While these sufficient conditions are stronger than the original conditions, they offer a more structured route to verification. A substantial portion of our technical analysis is devoted to showing that these sufficient conditions hold for the Gaussian and exponential adversarial models considered in this paper. Accordingly, for any choice of pre- and post-change distributions for which these sufficient conditions hold, the overshoots vanish asymptotically under our analysis. 

Moreover, we investigate precise conditions under which an adversary can maintain covertness, which we define as the scenario in which the best achievable ADD, under the constraint that AT2FA is at least $\gamma$, grows asymptotically as $\Theta(\gamma)$. We formalize this definition in our analysis, noting that it aligns closely with the notion of covertness introduced in \cite{huang_covert_2021}. Through our theoretical framework, we characterize the required rate at which the post-change distribution must converge to the pre-change distribution in order to preserve covertness. In this formulation, the adversary's control over the post-change distribution models its ability to influence how the statistical properties of the observations evolve after the change. By selecting post-change distributions that are closer to the pre-change distribution, the adversary can delay detection while remaining covert, a perspective consistent with \cite{ramtin2025continuous}. We take this as motivation for studying adversarial choices of post-change distributions, since delayed detection increases the duration over which the adversary can influence the system before being detected. We also quantify the extent of this influence by building upon the notion of total damage introduced in \cite{ramtin2024quickest}, which measures the cumulative impact incurred prior to detection, and study its asymptotic scaling.

To illustrate our theory, we analyze Gaussian-distributed (resp. exponential-distributed) observations, explicitly studying adversaries that manipulate both the mean and variance parameters (resp. mean) as functions of $\gamma$. While adversaries that vary the variance have been previously considered in the covert communication literature, adversaries who adjust the mean are motivated by volume-based attacks that aim to increase activity levels without triggering detection \cite{ramtin2021fundamental}. We prove that our theoretical results apply to this model by conducting a detailed analysis showing that the conditions established in the preceding section -- which are sufficient to ensure that the overshoots vanish -- are indeed satisfied as $\gamma \to \infty$. This validation confirms that the asymptotic expressions for AT2FA and ADD remain accurate even in this parametric Gaussian/exponential setting, thereby demonstrating the broad applicability of our asymptotic framework. We then explore how the ADD scales based on the rate at which the post-change distribution converges to the pre-change distribution, and establish conditions under which covertness is maintained in the 
Gaussian (resp. exponential) case.

The remainder of the paper is structured as follows. Section \ref{sec:basic} presents the necessary preliminaries.
Section \ref{sec:2} develops our main asymptotic results under the setting where the post-change distribution approaches the pre-change distribution as $\gamma \to \infty$. We then specialize these results to Gaussian (resp. exponential) models in Section 
\ref{sec:Gaussian-exponential}, analyzing adversaries that manipulate the mean and variance (resp. mean).
Finally, we conclude with a summary of contributions in Section \ref{sec:conclusion}.

\section{Basic elements of change-point detection in discrete-time}
\label{sec:basic}

Let $X_1,X_2,\ldots$ be mutually independent random variables (rvs) taking values in a measurable space ${\cal X}$.
Denote by ${\cal F}_k=\sigma(X_1,\ldots,X_k)$ the  smallest $\sigma$-fied generated by $X_1,\ldots,X_k$.

Random variables $X_1,\ldots,X_{t-1}$ have common (pre-change) probability distribution $Q_0$ while rvs
$X_t,X_{t+1},\ldots$ have common (post-change) probability distribution $Q_1$, distinct from $Q_0$\footnote{Our notation follows that in \cite{Poor2009}. We note that some works in the quickest change detection literature instead use $\nu$ for the change-point, and $f_0$, $f_1$ for the pre- and post-change densities, respectively, along with $\Prob_k$, $\Prob_\infty$ and $\E_k$, $\E_\infty$ for the corresponding probability measures and expectations.}.  Denote by $\E_t$ the expectation operator associated with this model.
The change time $t$ is unknown but constant (non-Bayesian setting).
The case $t=\infty$ corresponds to no change occurring, i.e., only the distribution $Q_0$ is used, and $t=1$ corresponds to the situation where only $Q_1$ is used.
We assume that $Q_0$ and $Q_1$ have probability density functions (pdfs), denoted by $q_0$ and $q_1$, respectively, and that the 
log-likelihood $\log(q_1(X)/q_0(X))$ is finite a.s. under both $Q_0$ and $Q_1$.

Recall that the Kullback–Leibler (KL) divergence of a pdf $p_1$ from a pdf $p_2$, defined on the same measurable space,  is 
$D(p_1||p_2)=\E[\log(p_1(X)/p_2(X))$, with $X$ a rv with pdf $p_1$, whenever this expectation is finite; if so, $D(p_1||p_2)\geq 0$, with  $D(p_1||p_2)=0$ if and only if pdfs $p_1$ and $p_2$ are identical.

Denote by  ${\cal T}$ the set of all (possibly extended) stopping times with respect to  the filtration ${\cal F}:=\{{\cal F}_k, k\geq 1\}$.
Fix $\gamma\in (0,\infty)$. Lorden's classical problem \cite{Lorden1971}  consists in  minimizing  
\[
d(T):=\sup_{t\geq 1}\hbox{ess sup}\E_t[(T-t+1)^+ \,|\, {\cal F}_{t-1}],
\]
over all stopping times  $T\in {\cal T}$  such that  $\E_\infty[T]\geq \gamma$. 
In words, the goal is to minimize the worst-case average detection delay under a constraint on the false alarm rate (i.e., $1/\E_\infty[T]\leq 1/\gamma$).
Define  $n(\gamma)=\inf_{ \{T\in {\cal T}: \E_\infty[T]\geq \gamma\}} d(T)$, the optimal worst-case average detection delay.

Moustakides  \cite{Moustakides1986} (see also \cite[Section 6.2]{Poor2009}) showed that
Page's Cumulative Sum (CuSum) stopping rule\cite{Page1954}
\begin{equation}
	\label{cusum-st}
	\tau_h =\inf\left\{k\geq 0 : \max_{1\leq j\leq k+1}
    \sum_{i=j}^k \log\frac{q_1 (X_i)}{q_0(X_i)}
    \geq h\right\},\,\,h>0,
\end{equation}
solves Lorden's optimization problem  if $h$ is selected so that 
$\E_{q_0}[\tau_h]=\gamma$.

It is known that $d(\tau_h)=\E_{q_1}[\tau_h]$ for $h>0$ \cite[p. 143]{Poor2009}, \cite[p. 25]{Siegmund1985}, thereby implying that
\begin{equation}
	\label{n-gamma}
	n(\gamma)=\E_{q_1}[\tau_{h^\star}],
\end{equation}
with $h=h^\star$ solution of $\E_{q_0}[\tau_h]=\gamma$. Since $\tau_0=0$, notice that $h^\star>0$ for $\gamma>0$.

There is in general no closed-form expression for $n(\gamma)$. The behavior of $n(\gamma)$ as $\gamma\to\infty$ was obtained by Lorden \cite{Lorden1971},
\cite[Thm 6.17, p. 160]{Poor2009}.   It is given by 
\begin{equation}
	\label{Lorden}
	n(\gamma)\sim \frac{\log \gamma}{D(q_1||q_0)},\quad \gamma\to\infty.
\end{equation}
The aim of this work is to revisit Lorden's result when $q_1$ depends on $\gamma$ and $q_1$ converges pointwise to $q_0$ as $\gamma\to\infty$. 

Notation: For any event ${\cal A}$, ${\bf 1}_{\cal A}=1$ if ${\cal A}$ if true and zero otherwise. If a function  $f$ is differentiable at $x_0\in \R$, we denote by
$f^\prime(x_0)$ its derivative at this point. The shorthands $f(x)\sim_x g(x)$ and $\lim_x f(x)$ will stand for $f(x)=g(x)+o(1)$ ($x\to\infty$) and
$\lim_{x \to\infty} f(x)$, respectively.

\section{Limiting behavior of $n(\gamma)$ when $q_1$ depends on $\gamma$}
\label{sec:2}

We still consider the setting in Section \ref{sec:basic} but we now assume that $q_1$ depends on $\gamma$ with $\lim_{\gamma} q_1=q_0$. 
We further assume\footnote{Results in Section \ref{sec:2} hold if $\R$ is replaced by any measurable space ${\cal X}$ (e.g., ${\cal X}=\R^d$, for $d\geq 1$).} that $X_k\in \R=(-\infty,\infty)$ ($k\geq 1$).
To make explicit the dependence on $\gamma$, $q_1$ will now be denoted by $q_\gamma$. For simplicity, $q_0$ is denoted by $q$. We denote by
$\Prob_p$ and $\E_p$ the probability and expectation operators associated with the pdf $p\in \{q,q_\gamma\}$. In other words, under $\Prob_{q_\gamma}$ (resp. $\Prob_q$),
$X_1,X_2,\ldots$ are i.i.d. real-valued rvs with common pdf $q_\gamma$ (resp. $q$). We further assume that there exists $\gamma_0>0$ such that\footnote{Two pdfs are different if they differ on a set of positive measure.}
$q_\gamma\not=q$  for all $\gamma>\gamma_0$. Since we are only
interested in limiting results when $\gamma\to\infty$, we will actually assume, without loss of generality, that $q_\gamma\not=q$ for all $\gamma>0$. Last, we assume that the log-likelihood ratio $\log(q_\gamma(X_1)/q(X_1))$ is finite $\Prob_{q_\gamma}$-a.s. (resp. $\Prob_q$-a.s.) for all
$\gamma>0$. 

Our goal is to find conditions on $q_\gamma$ ensuring covertness of an attacker. We say that an attacker is covert if $n(\gamma)=\Theta(\gamma)$, namely, for large $\gamma$, the optimal worst-case average detection delay is of the same order of magnitude as the largest admissible lower bound on the expected time between false alarms. 

{In addition to analyzing $n(\gamma)$, we also introduce a measure of the impact that a change in distribution may have on the system prior to detection. We model this impact through a damage measure defined as a function of the KL divergence between the pre-change and post-change models \cite{ramtin2024quickest}. Specifically, let $g:\R_+\to\R_+$ be a strictly increasing function with $g(0)=0$
Then, the corresponding total expected damage accrued up to the detection time is defined as
\begin{equation}
\label{def:total-damage}
d(\gamma)= n(\gamma) g\big(D(q||q_\gamma)\big).
\end{equation}
}
Fix $\gamma>0$ unless otherwise mentioned. Introduce the Sequential Probability Ratio Test (SPRT) \cite[Chapter 4]{Poor2009}, \cite[Chapter 2]{Siegmund1985}
\begin{equation}
	\label{def-T}
	T_{a,b}=\inf\{k\geq 1 : S_k\not\in (a, b)\}, 
\end{equation}
with $S_k:=\sum_{j=1}^k Y_j$ and $Y_j:=\log \frac{q_\gamma(X_j)}{q(X_j)}$. We assume that $-\infty<a\leq 0<b<\infty$. 
Denote by $\alpha_{a,b}=\Prob_{q}(S_{T_{a,b}}\geq b)$ and $\beta_{a,b}=\Prob_{q_\gamma}(S_{T_{a,b}}\leq a)$ the error probabilities.

Notice that $T_{a,b}$ depends on $\gamma$ as $S_k$ does through $q_\gamma$, and so do $\alpha_{a,b}$ and $\beta_{a,b}$. To keep notation compact 
we do not make this dependence explicit.
In addition, when no confusion may occur, $T_{a,b}$, $\alpha_{a,b}$, and $\beta_{a,b}$ are simply denoted by $T$,  $\alpha$, and $\beta$, respectively.

Because\footnote{Let $A=\{x\in\R : q_\gamma(x)=q(x)\}$. Then, $\Prob_p(Y_1=0)=\int_A p(x)dx$. If the latter integral was equal to one then $q_\gamma=q$ p-a.e., which would contradict the assumption that $q_\gamma\not= q$. Hence, $\Prob_p(Y_1=0)<1$.}
$q_\gamma$ and $q$ are different pdfs,  $\Prob_p(Y_1=0)<1$ for $p\in \{q,q_\gamma\}$,  which ensures that $\E_p[T]<\infty$ for $p\in \{q,q_\gamma\}$
by Stein's lemma \cite[Proposition 2.18]{Siegmund1985}.  
Notice that  $\E_{q_\gamma}[Y_1]=D(q_\gamma|| q)$ and  $\E_{q}[Y_1]=-D(q||q_\gamma)$. We assume that both expectations are finite, so that
$\E_{q_\gamma}[Y_1]>0$ and $\E_{q}[Y_1]<0$ since $q_\gamma$ and $q$ are different pdfs.

Propositions \ref{prop:main}-\ref{prop:key-result} and Lemma \ref{lem:asympt-alpha-beta} will be proved under the 
conditions that the expected overshoots converge to zero as $\gamma\to\infty$ , namely,
	\begin{align}
		\label{result005}
		&\lim_\gamma\E_p[S_T-b \,|\, S_T\geq b]=0,\\ 
		&\lim_\gamma \E_p[S_T-a \,|\, S_T\leq a]=0,
		\label{result006}
	\end{align}
for $p\in \{q,q_\gamma\}$.
Conditions (\ref{result005})-(\ref{result006}) are not easy to check since they involve the (unknown) overshoot random variables $S_T-a$ and $S_T-b$.
In Lemma \ref{lem:suffiicient-conditions} we will provide sufficient conditions for (\ref{result005})-(\ref{result006}), 
easier to verify than conditions (\ref{result005})-(\ref{result006}).

\begin{proposition}
\label{prop:main}
Assume that conditions (\ref{result005})-(\ref{result006}) are satisfied. 
	Then,  for $a<0<b$,
	\begin{align}
		\E_{q_\gamma} [T]& \sim_\gamma\frac{  A (B-1) a+ B(1-A)b}{D(q_\gamma ||q) (B-A)},\label{prop:approx-T-q-gamma}\\
		\E_q[T]&\sim_\gamma -\frac{(B-1)a+(1-A)b}{D(q||q_\gamma)(B-A)},  \label{prop:approx-T-q}
	\end{align}
	with $A=e^a$ and $B=e^b$.
\end{proposition}

{\it Proof:} We have
\begin{align*}
	\E_{q_\gamma}[S_T {\bf 1}_{\{S_T\leq a\}}]
	&=\beta \,\E_{q_\gamma}[S_T -a \,|\,S_T\leq a]+ \beta a,\\
	\E_{q_\gamma}[S_T {\bf 1}_{\{S_T\geq b\}}]
	&=(1-\beta) \,\E_{q_\gamma}[S_T -b\,|\, S_T\geq b]+ (1-\beta) b.
\end{align*}
Summing both equations gives
\begin{equation}
	\label{result-gamma}
	\E_{q_\gamma}[S_T]= \beta a +(1-\beta)b  +c_1,
\end{equation}
with $c_1:=\beta \,\E_{q_\gamma}[S_T -a \,|\, S_T\leq a]+(1-\beta)\E_{q_\gamma}[S_T -b\,|\,S_T\geq b]$.
Similarly one finds
\begin{equation}
	\label{result-infinity}
	\E_q[S_T]=(1-\alpha) a + \alpha b+ c_2,
\end{equation}
with $c_2:=(1-\alpha) \,\E_q[S_T -a \,|\,S_T\leq a]+\alpha \E_q[S_T -b \,|\,S_T\geq b]$.

Since $\E_p[T]<\infty$ and $\E_p[Y_1]<\infty$ for $p\in\{q,q_\gamma\}$, Wald's identity applies to  $\E_p[S_T]$, to yield  $\E_p[S_T]=\E_p[T] \E_p[Y_1]$ for  $p\in\{q,q_\gamma\}$. 
Applying this identity to (\ref{result-gamma}) and (\ref{result-infinity}), we obtain
\begin{align}
	\E_{q_\gamma}[T]&=\frac{1}{D(q_\gamma || q)}\left(\beta a +(1-\beta)b  +c_1\right),
	\label{ET-q-gamma}\\
	\E_q[T]&=-\frac{1}{D(q||q_\gamma)}\left((1-\alpha) a + \alpha b + c_2\right).
	\label{ET-q}
\end{align}
Letting now $\gamma\to\infty$ in (\ref{ET-q-gamma})-(\ref{ET-q}) and using assumptions (\ref{result005})-(\ref{result006}), gives
\begin{align}
	\E_{q_\gamma}[T]&\sim_\gamma \frac{1}{D(q_\gamma || q)}(\beta a +(1-\beta)b),
	\label{approx-ET-q-gamma}\\
	\E_q[T]&\sim_\gamma -\frac{1}{D(q||q_\gamma)}((1-\alpha) a + \alpha b).
	\label{approx-ET-q}
\end{align}
The proof  is completed upon replacing $\alpha$ and $\beta$ in  (\ref{approx-ET-q-gamma})-(\ref{approx-ET-q}) by their
asymptotics given in Lemma \ref{lem:asympt-alpha-beta} below. \hfill\done

\begin{lemma}
	\label{lem:asympt-alpha-beta}
	If conditions (\ref{result005})-(\ref{result006}) hold, then
	\begin{equation}
		\label{asympt-alpha-beta}
		\alpha\sim_\gamma \frac{1-A}{B-A}\quad\hbox{and}\quad \beta\sim_\gamma A\frac{B-1}{B-A} ,
	\end{equation}
	with $A=e^a$ and $B=e^b$.
\end{lemma}
{\it Proof:} Fix $\epsilon>0$.  Let $A=e^a$ and $B=e^b$. Define $\Lambda_k=\prod_{j=1}^k \frac{q_k(X_k)}{q(X_k)}$. 
Conditioning on the value of $T$ (recall that $T<\infty$ $p$-a.s. since $\E_p[T]<\infty$ for
$p\in\{q,q_\gamma\}$), we obtain
\begin{align*}
	\Prob_q(B\leq \Lambda_T< B e^{\epsilon})
	&=\sum_{k\geq 1} \int_{{\cal S}_k}\prod_{j=1}^kq(x_j) dx_1\cdots dx_k,
\end{align*}
with  (here $\bar x_k=(x_1,\ldots,x_k)$)
\begin{align*}
	{\cal S}_k:=&\left\{\bar x_k\in \R^k,  \prod_{i=1}^j \frac{q_\gamma(x_i)}{q(x_i)} \in (A,B),  j=1,\ldots,k-1,
	B\leq \prod_{i=1}^k\frac{q_\gamma(x_i)}{q(x_i)} < B e^\epsilon\right\}. 
\end{align*}
Hence,
\begin{align*}
	&\frac{1}{B e^\epsilon} \sum_{k\geq 1} \int_{{\cal S}_k} \prod_{j=1}^k q_\gamma (x_j) dx_1\cdots dx_k\leq \Prob_q(B\leq \Lambda_T< B e^{\epsilon})
	\leq  \frac{1}{B} \sum_{k\geq 1} \int_{{\cal S}_k} \prod_{j=1}^k q_\gamma(x_j) dx_1\cdots dx_k.
\end{align*}
Since $ \sum_{k\geq 1} \int_{{\cal S}_k} \prod_{j=1}^k q_\gamma (x_j) dx_1\cdots dx_k= \Prob_{q_\gamma}(B\leq \Lambda_T< B e^{\epsilon})$, the above inequalities become
\begin{align}
	\frac{1}{Be^\epsilon} \Prob_{q_\gamma}(B\leq \Lambda_T< B e^{\epsilon})&\leq \Prob_q(B\leq \Lambda_T< B e^{\epsilon})
	\leq  \frac{1}{B} \Prob_{q_\gamma}(B\leq \Lambda_T< B e^{\epsilon}).
	\label{result024}
\end{align}
On the other hand,
\begin{subequations}
\begin{align}
\Prob_q(B\leq \Lambda_T<B e^\epsilon)&=\Prob_q(B\leq \Lambda_T)-\Prob_q(\Lambda_T\geq B e^\epsilon)=
	\alpha -\Prob_q(\Lambda_T\geq B e^\epsilon\,|\, \Lambda_T\geq B)\alpha,\nonumber\\
    &=\alpha \Prob_q(\Lambda_T<B e^\epsilon\,|\, \Lambda_T\geq B),
\label{ex-foot1}\\
\Prob_{q_\gamma}(B\leq \Lambda_T<B e^\epsilon)&=\Prob_{q_\gamma}(B\leq \Lambda_T)-\Prob_{q_\gamma}(\Lambda_T\geq B e^\epsilon)=
	1-\beta -\Prob_{q_\gamma}(\Lambda_T\geq B e^\epsilon\,|\, \Lambda_T\geq B)(1-\beta),\nonumber\\
    &=(1-\beta) \Prob_{q_\gamma}(\Lambda_T<B e^\epsilon\,|\, \Lambda_T\geq B).
\label{ex-foot2}
\end{align}
\end{subequations}
To establish (\ref{ex-foot2}) we used the identity $\Prob_{q_\gamma}(\Lambda_T\geq B)=\Prob_{q_\gamma}(\Lambda_T>A)=1-\beta$, which 
since $T$ is almost surely finite.
Combining (\ref{result024}), (\ref{ex-foot1}), and (\ref{ex-foot2}), we obtain
\begin{align}
	&\frac{1}{e^\epsilon}\left( \frac{1-\beta}{B}\right)\frac{\Prob_{q_\gamma}(\Lambda_T<B e^\epsilon\,|\, \Lambda_T\geq B)}{\Prob_q(\Lambda_T< Be^\epsilon\,|\, \Lambda _T\geq B)}
	\leq \alpha
	\leq  \left( \frac{1-\beta}{B}\right) \frac{\Prob_{q_\gamma}(\Lambda_T<B e^\epsilon\,|\, \Lambda_T\geq B)}{\Prob_q(\Lambda_T< Be^\epsilon\,|\, \Lambda _T\geq B)}.
	\label{result021}
\end{align}
For $p\in \{q,q_\gamma\}$,
\begin{align*}
	&\Prob_p(\Lambda_T \geq B e^\epsilon\,|\, \Lambda_T\geq B)= \Prob_p(S_T -b \geq \epsilon\,|\, S_T\geq b)
	\leq \frac{1}{\epsilon} \E_p[S_T -b \,|\, S_T\geq b]\to 0\,\,\hbox{as } \gamma\to\infty,
\end{align*}
where the inequality follows from conditional Markov's inequality\footnote{Let $X$ be a real-valued rv defined on a probability space $(\Omega, {\cal F}, \Prob)$ such that $\Prob(X\geq 0)>0$. For $x>0$, $\Prob(X\geq x\,|\, X\geq 0)=
\frac{\E[{\bf 1}_{\{X\geq x\}} {\bf 1}_{\{X\geq 0\}}]}{\Prob(X\geq 0)}\leq 
\frac{\E[\frac{X}{x} {\bf 1}_{\{X\geq 0\}}]}{\Prob(X\geq 0)}
= \frac{1}{x}\E[X\,|\, X\geq 0]$.} and the latter limit follows from  (\ref{result005}). This shows that 
$\lim_{\gamma}\Prob_p(\Lambda_T< B e^\epsilon\,|\, \Lambda_T\geq B)=1$.
Thus, upon letting $\gamma\to \infty$ and then $\epsilon\to 0$  in (\ref{result021}), we get 
\begin{equation}
	\label{result022}
	\alpha \sim_\gamma \frac{1-\beta}{B}, 
\end{equation}
Similarly, we find
\begin{equation} 
	\label{result023}
	\beta\sim_\gamma A (1-\alpha).
\end{equation}
From (\ref{result022})-(\ref{result023}) we derive (\ref{asympt-alpha-beta}). \hfill\done

The link between the CumSum test in (\ref{cusum-st}) and the SPRT in (\ref{def-T}) is made through the formula \cite[Theorem 6.5]{Poor2009}, \cite[Eq. (2.52)]{Siegmund1985}
\begin{equation}
	\label{Pages-result}
	\E_p[\tau_h]=\frac{\E_{p}[T_{0,h}]}{\Prob_p(S_{T_{0,h}}\geq h)}= \frac{\E_{p}[T_{0,h}]}{\E_p[{\bf 1}_{\{S_{T_{0,h}}\geq h\}}]},
\end{equation}
which holds for any pdf $p$ on $\R$. 
It allows us to establish the following proposition:
\begin{proposition}
	\label{prop:key-result}
	If conditions (\ref{result005})-(\ref{result006}) hold, then
	\begin{align}
		\E_{q_\gamma}[\tau_h]&\sim_\gamma \frac{1}{D(q_\gamma||q)}(e^{-h}+h -1),  \label{prop:asympt-tau-q-gamma}\\
		\E_q[\tau_h]&\sim_\gamma \frac{1}{D(q||q_\gamma)}(e^h- h -1). \label{prop:asympt-tau-q}
	\end{align}
\end{proposition}
{\it  Proof:}   From Proposition \ref{prop:main}, for $a<0<b$, 
\begin{align}
	\frac{\E_{q_\gamma}[T_{a,h}]}{\E_{q_\gamma}[{\bf 1}_{\{S_{T_{a,h}}\geq h\}}]}
	\sim_\gamma \frac{  e^a (e^h-1) a+ e^h(1-e^a)h}{D(q_\gamma ||q) e^h(1-e^a)},\label{prop:sim-1}\\
	\frac{\E_{q}[T_{a,h}]}{\E_q[{\bf 1}_{\{S_{T_{a,h}}\geq h\}}]}\sim_\gamma  \frac{  e^a (e^h-1) a+ e^h(1-e^a)h}{D(q||q_\gamma) (1-e^a)}.
	\label{prop:sim-2}
\end{align}
However,  letting $a=0$ in  the right-hand sides of (\ref{prop:sim-1}) and (\ref{prop:sim-2}) yieds $0/0$. We therefore need
to take a little detour. From (\ref{Pages-result})
\begin{align*}
	\E_p[\tau_h]&=\frac{\E_{p}[\lim_{m}T_{-1/m,h}]}{\E_p [\lim_{m}{\bf 1}_{\{S_{T_{-1/m,h}}\geq h\}}]}
	=\lim_{m} \frac{\E_{p}[T_{-1/m,h}]}{\E_p [{\bf 1}_{\{S_{T_{-1/m,h}}\geq h\}}]},
\end{align*}
where the interchange between the expectation operator and the limit in the numerator follows from the monotone convergence theorem applied to the non-increasing and non-negative sequence $\{T_{-1/m,h}\}_{m\geq 1}$, while  the interchange between the expectation operator and the limit in the denominator follows from the bounded convergence theorem. Hence, for $p=q_\gamma$, 
\begin{align}
	\lim_\gamma\frac{\E_{q_\gamma}[\tau_h]}{\frac{1}{D(q_\gamma||q)}(e^{-h}+h-1)}&
	=\frac{1}{e^{-h}+h-1}
	\lim_{\gamma} \lim_{m} \frac{\E_{q_\gamma}[T_{-1/m,h}]}{\frac{1}{D(q_\gamma||q)}(1-\beta_{-1/m,h})},\nonumber \\
	&=\frac{1}{e^{-h}+h-1}
	\lim_m \lim_\gamma\frac{\E_{q_\gamma}[T_{-1/m,h}]}{\frac{1}{D(q_\gamma||q)}(1-\beta_{-1/m,h})} ,\nonumber\\
	&=\frac{1}{e^{-h}+h-1}\times \lim_m 
	\frac{-\frac{1}{m}e^{-1/m} (e^h-1) +e^h(1-e^{-1/m})h}{e^h(1-e^{-1/m})},
	\label{result2000}\\
	&=\frac{1}{e^{-h}+h-1}\times \frac{-e^h +1 +e^h h}{e^h}=1, \nonumber
\end{align}
where (\ref{result2000}) follows from Proposition \ref{prop:main} and Lemma \ref{lem:asympt-alpha-beta} when conditions (\ref{result005})-(\ref{result006}) hold.
This proves (\ref{prop:asympt-tau-q-gamma}). The proof of (\ref{prop:asympt-tau-q}) is similar (Hint: $\E_q[{\bf 1}_{\{S_{T_{-1/m,h}}\geq h\}}]=\alpha_{-1/m,h}$)
and is therefore omitted.\hfill\done

The proofs of Propositions \ref{prop:main}-\ref{prop:key-result} and of Lemma \ref{lem:asympt-alpha-beta} are inspired by the proofs of several elementary results 
in sequential detection theory. These include the bounds (in our notation)
$\frac{1-\beta}{\alpha}\geq B$ and $\frac{\beta}{1-\alpha}\leq A$ \cite[Prop. 4.10]{Poor2009}, \cite[p. 45]{Wald1947},
$\E_{q}[T]\geq  D(q||q_\gamma)^{-1} [\alpha \log \frac{\alpha}{1-\beta}+(1-\alpha)\log \frac{1-\alpha}{\beta}]$ and 
$\E_{q_\gamma}[T]\geq D(q_\gamma ||q)^{-1} [(1-\beta) \log \frac{1-\beta}{\alpha}+\gamma \log \frac{\beta}{1-\alpha}]$ \cite[Prop. 4.11]{Poor2009}, \cite[Thm 2.39]{Siegmund1985}.
It is also worth noting that the r.h.s. of (\ref{prop:approx-T-q-gamma}) and (\ref{prop:approx-T-q}) provide known {\it approximations}
for $\E_{q_\gamma}[T]$ and $E_q[T]$, respectively,  where the approximation, due to Wald \cite{Wald1947},  originates from  neglecting the excess over the boundaries $a$ and $b$, i.e.,
$S_T\approx a$ or $S_T\approx b$ \cite[p. 77]{Poor2009}. Proposition \ref{prop:main} says that these approximations give the exact limiting behavior of $\E_{q_\gamma}[T]$ and $E_q[T]$ as $\gamma\to\infty$, when $\lim_\gamma q_\gamma =q$ and conditions (\ref{result005})-(\ref{result006}) are met.

The r.h.s. of (\ref{prop:asympt-tau-q-gamma})-(\ref{prop:asympt-tau-q}) are nothing but Khan's {\it approximations} for $\E_{q_\gamma}[\tau_h]$
and $\E_q[\tau_h]$ \cite[Corollary, p. 76]{Khan}, \cite[Eqns (2.54), p. 26]{Siegmund1985}.
Proposition \ref{prop:key-result} shows that these approximations are exact when $\lim_\gamma q_\gamma=q$ and  
conditions (\ref{result005})-(\ref{result006}) hold.

In Lemma \ref{lem:suffiicient-conditions} below we provide sufficient conditions for (\ref{result005})-(\ref{result006}) to hold. These new conditions only involve the rv $Y=\log\frac{q_\gamma(X)}{q(X)}$, making them easier to check. 
\begin{lemma}
	\label{lem:suffiicient-conditions} 
	Condition  (\ref{result005}) holds if 
	\begin{equation}
		\label{lem:sufficient1}
		\lim_\gamma \sup_{y\geq 0}\E_p[Y-y\,|\, Y\geq y]=0, 
	\end{equation}
	and condition (\ref{result006}) holds if 
	\begin{equation}
		\label{lem:sufficient2}
		\lim_\gamma \inf_{y\leq 0}\E_p[Y-y\,|\, Y\leq y]=0, 
	\end{equation}
	for $p\in\{q,q_\gamma\}$, with $Y=\log\frac{q_\gamma(X)}{q(X)}$ and  $X$ a real-valued random variable.
\end{lemma}
{\it Proof:} Fix $p\in\{q,q_\gamma\}$. Conditioning on $S_{T-1}=x$ and using $S_T=S_{T-1}+Y_T$, we get 
\begin{align}
	\E_p[S_T-b \,|\, S_T\geq b]
	&= \int_a^b \E_p[S_T-b\,|\, S_T\geq b, S_{T-1}=x] dF_{p,S_{T-1}\,|\, S_T\geq b}(x),\nonumber\\
	&=\int_a^b \E_p[Y_T -(b-x)\,|\, Y_T\geq b-x, S_{T-1}=x]  dF_{p,S_{T-1}\,|\, S_T\geq b}(x),
\label{lem:conditioning}
\end{align}
where $dF_{p,S_{T-1}\,|\, S_T\geq b}(x)$ is the pdf of $S_{T-1}$ under $p$ given that  $S_T\geq b$. 
By the strong Markov property for i.i.d. sequences, we have\footnote{For $C\in {\cal F}_{T-1}$, $\Prob_p(\{Y_T\geq y\} \cap C )=\sum_{k\geq 0}
\Prob_p(\{Y_{k+1}\geq y\}\cap C\cap\{ T-1=k\})=\sum_{k\geq 0}
\Prob_p(Y_{k+1}\geq y)\Prob_p(C\cap\{T-1=k\})$ since $Y_{k+1}$ is independent of
$C\cap\{T-1=k\}\in{\cal F}_{k}$. Hence, $\Prob_p(\{Y_T\geq y\} \cap C )=\Prob_p(Y_1\geq y)\sum_{k\geq 0}\Prob_p(C\cap\{T-1=k\})=\Prob_p(Y_1\geq y) \Prob_p(C)$. Therefore, for any $C\in {\cal F}_{T-1}$,
$\Prob_p(Y_T\geq y\,|\, C)=
\frac{\Prob_p(\{Y_T\geq y\}\cap C)}{\Prob_p(C)}=\Prob_p(Y_1\geq y)$, which proves (\ref{key-id1}). The proof of (\ref{key-id2}) is similar.}
\begin{align}
\Prob_p(Y_T\geq y\,|\, {\cal F}_{T-1}) &=\Prob_p(Y_1\geq y), \label{key-id1}\\ 
\E_p[Y_T {\bf 1}_{\{Y_T\geq y\}}\,|\,{\cal F}_{T-1}] &= \E_p[Y_1 {\bf 1}_{\{Y_1\geq y\}}].
\label{key-id2}
\end{align}
Thus, for any $y\in \R$,
\[
\E_p[Y_T\,|\, Y_T\geq y, {\cal F}_{T-1}]
=\frac{\E_p[Y_T {\bf 1}_{\{Y_T\geq y\}}\,|\,{\cal F}_{T-1}]}{\Prob_p(Y_T\geq y\,|\, {\cal F}_{T-1})}
=\frac{\E_p[Y_1{\bf 1}_{\{Y_1\geq y\}}]}{\Prob_p(Y_1\geq y)}=\E_p[Y_1\,|\, Y_1\geq y].
\]
Since $\{S_{T-1}=x\}\in {\cal F}_{T-1}$, the latter identity yields
\begin{equation}
\E_p[Y_T\,|\, Y_T\geq y,S_{T-1}=x]= \E_p[Y_1\,|\, Y_1\geq y].
\label{strong-Markov-iid}
\end{equation}
Subtracting $b-x$ from both sides of (\ref{strong-Markov-iid}) and  choosing $y=b-x$, gives
\[
\E_p[Y_T-(b-x)\,|\, Y_T\geq b-x, S_{T-1}=x]= \E_p[Y_1-(b-x)\,|\, Y_1\geq b-x].
\]
Introducing this result in (\ref{lem:conditioning}), we obtain
\begin{align*}
	0\leq \E_p[S_T-b \,|\, S_T\geq b]&=\int_a^b \E_p[Y_1 -(b-x)\,|\, Y_1\geq b-x]dF_{p,S_{T-1}\,|\, S_T\geq b}(x),\\
	&\leq \sup_{0\leq y\leq b-a}\E_p[Y_1-y\,|\, Y_1\geq y] \times \int_a^b dF_{p,S_{T-1}\,|\, S_T\geq b}(x),\\
	&\leq \sup_{y\geq 0}\E_p[Y_1-y\,|\, Y_1\geq y],
\end{align*}
where we have used that  $\int_a^b dF_{p,S_{T-1}\,|\, S_T\geq b}(x)=1$, which proves that (\ref{result005}) is true (\ref{lem:sufficient1}) is satisfied.
Similarly,
\begin{align*}
	&0\geq \E_p[S_T-a \,|\, S_T\leq a]=\int_a^b \E_p[Y_1 -(a-x)\,|\, Y_1\leq a-x] dF_{p,S_{T-1}\,|\, S_T\leq a}(x)
	\geq  \inf_{y\leq 0}\E_p[Y_1  - y\,|\, Y_1\leq y],
\end{align*}
which proves (\ref{result006}) is true if (\ref{lem:sufficient2}) is satisfied. \hfill\done

\begin{remark}[Sufficient condition for (\ref{result005})-(\ref{result006}) to hold]
Conditions (\ref{result005})-(\ref{result006}) will hold if (C) $\frac{q_\gamma(x)}{q(x)}\to 1$ uniformly on $\R$ as $\gamma\to\infty$.
This is so as $0\leq \E_p[S_T-b\,|\, S_T\geq b]\leq \sup_{x\in\R} \log \frac{q_\gamma(x)}{q(x)}$
and $\inf_{x\in\R} \log\frac{q_\gamma(x)}{q(x)}\leq \E_p[S_T-a\,|\, S_t\leq a]\leq 0$, and that
$\lim_\gamma\inf_{x\in\R} \log\frac{q_\gamma(x)}{q(x)} = \lim_\gamma \sup _{x\in\R} \log\frac{q_\gamma(x)}{q(x)}=0$ under (C).
Condition (C) is of course very restrictive; in particular, it does not hold for the Gaussian and exponential models discussed in 
Section \ref{sec:Gaussian-exponential}. It will hold, for instance, if rvs $\{X_k\}_k$ have a finite support in $\R$.
\end{remark}

Before stating the main result of this paper (Proposition \ref{prop:aympt-n} below), let us recall some properties of the Lambert function that we will use in its proof.
The Lambert function  is the solution of the equation $W(z)e^{W(z)}=z$, $z\in\C$  \cite{Corless}.  When $z$ is real, $W(z)e^{W(z)}=z$ has a solution if and only if $z\geq -e^{-1}$. When $z\geq 0$, this solution is
unique, given by the main branch  $W_0$ of $W$, and when $-e^{-1}\leq z<0$ there are two solutions given by the branches $W_0$ and $W_{-1}$ of $W$. In particular,
\begin{equation}
	\label{branches-Lambert}
	W_0(z)>-1, \,\, W_{-1}(z)< -1\,\hbox{ for }-e^{-1}<z<0,
\end{equation}
and $W_0(-e^{-1})=W_{-1}(-e^{-1})=-1$. Last \cite[p. 350]{Corless}
\begin{equation}
	\label{limit-Lambert}
	W_{-1}(x)\sim\log(-x)-\log(-\log(-x)) \,\hbox{as } x\to 0, x<0.
\end{equation}
%

\begin{proposition}[Limiting behavior of $n(\gamma)$]When conditions (\ref{result005})-(\ref{result006}) hold, \hfill
	\label{prop:aympt-n}
	\begin{equation}
		n(\gamma)\sim_\gamma \left\{ \begin{array}{ll}
			\frac{\log(\gamma D(q||q_\gamma))}{D(q_\gamma||q)}, &\mbox{if $\lim_\gamma \gamma D(q||q_\gamma)=\infty$},\\
            \frac{\gamma D(q||q_\gamma)}{D(q_\gamma||q)}\frac{G(y)}{y}, &\mbox{if  $\lim_\gamma \gamma D(q||q_\gamma)=y\in (0,\infty)$},\\
			\frac{\gamma D(q||q_\gamma)}{D(q_\gamma||q)}, &\mbox{if $\lim_\gamma \gamma D(q||q_\gamma)=0$},
		\end{array}
		\right.
		\label{asympt-n}
	\end{equation}
	where the mapping $G(y)$ given by
	\begin{equation}
		\label{def:G}
		G(y)=e^{1+y +W_{-1}(-e^{-1-y})}-W_{-1}(-e^{-1-y})-y-2,
	\end{equation}
satisfies $\tfrac{G(y)}{y}\in (0,1)$ when $0<y<\infty$, $\lim_{y\downarrow 0}\tfrac{G(y)}{y}=1$ and $\lim_{y\to\infty}\tfrac{G(y)}{y}=0$.
\end{proposition}
{\it Proof:}
Recall that for every $\gamma>0$, the optimal threshold $h^\star$ is strictly positive (cf. Section \ref{sec:basic}) and $D(q||q_\gamma)>0$ by assumption. 
The identity $\E_q[\tau_{h^\star}]=\gamma$ and (\ref{prop:asympt-tau-q}) imply that
$\frac{1}{D(q||q_\gamma)}(e^{h^\star}- h^\star -1)\sim_\gamma \gamma$. 
Hence, there exists a mapping $\zeta: (0,\infty)\to (-1,\infty)$ with  $\lim_\gamma \zeta(\gamma)=0$, such that 
\begin{equation}
\label{eq-h-star}
e^{h^\star}- h^\star -1 = \xi,
\end{equation}
with $\xi:= (1+\zeta) \gamma  D(q||q_\gamma)>0$. Eq.  (\ref{eq-h-star}) rewrites
\begin{equation}
	\label{eq-in-y}
	-(h^\star+r)e^{-(h^\star+r)}=-e^{-r},
\end{equation}
with $r:=1+\xi>1$. Since $-e^{-1}\leq -e^{-r}<0$, (\ref{eq-in-y}) has two solutions, $h_1+r=-W_0(-e^{-r})$ and $h_2+r=-W_{-1}(-e^{-r})$. 
From (\ref{branches-Lambert}), $h_1 <-\xi <0$, implying  that  $h^\star=h_2$, that is,
\[
h^\star(\gamma)=-1 -(1+\zeta) \gamma  D(q||q_\gamma)- W_{-1}(- e^{-1-(1+\zeta)\gamma  D(q||q_\gamma)}).
\]
Since $\zeta(\gamma)\sim_\gamma 0$, we deduce from the above that
\begin{equation}
	\label{approx-h}
	h^\star(\gamma)\sim_\gamma -1 - \gamma  D(q||q_\gamma) - W_{-1}(- e^{-1 -\gamma  D(q||q_\gamma)}).
\end{equation}
Combining  (\ref{n-gamma}), (\ref{prop:asympt-tau-q-gamma}), and (\ref{approx-h}) yields
\begin{align}
	n(\gamma)&\sim_\gamma \frac{G(\gamma  D(q||q_\gamma))} {D(q_\gamma||q)},
	\label{prop:asympt-n}
\end{align}
by using the definition of $G$ in (\ref{def:G}).
We are now in position to prove (\ref{asympt-n}).
Assume first that $\lim_\gamma \gamma D(q||q_\gamma)=\infty$. It is easily seen from (\ref{def:G}) and (\ref{limit-Lambert})
that $G(y)\sim_y \log y$, so that, by (\ref{prop:asympt-n}), 
\[
n(\gamma)\sim_\gamma  \frac{\log(\gamma D(q||q_\gamma))}{D(q_\gamma||q)}.
\]
When  $\lim_\gamma \gamma D(q||q_\gamma)=y\in (0,\infty)$ we see from (\ref{prop:asympt-n}) that 
\[
n(\gamma)\sim_{\gamma}  \frac{G(y)}{D(q_\gamma||q)}\sim_{\gamma}\frac{\gamma D(q||q_\gamma)}{D(q_\gamma||q_\gamma)} \cdot \frac{G(y)}{y}.
\]
Last, assume that $\lim_\gamma \gamma D(q||q_\gamma)=0$.  
Rewriting (\ref{prop:asympt-n}) as
\[
n(\gamma)\sim_\gamma \frac{G(\gamma D(q||q_\gamma)}{\gamma D(q||q_\gamma)}\times \frac{\gamma D(q||q_\gamma)}{D(q_\gamma||q)},
\]
we note that the last asymptotic result in (\ref{asympt-n}) holds if $\lim_{y\downarrow 0}\frac{G(y)}{y}=1$.
Let us prove the latter identity. Set $x=-W_{-1}(-e^{-1-y})$. Hence, $(-x)e^{-x}=-e^{-1-y}$ by definition of the Lambert function,
giving $y=x-1-\log x$. With this change of variable $G(y)/y$ becomes 
\begin{equation}
\label{ratio:G-y}
 \frac{G(y)}{y}=\frac{\log x -1 +\frac{1}{x}}{x-1-\log x},  
\end{equation}
yielding
\[
\lim_{\mycom{y\to 0}{y>0}}\frac{G(y)}{y}=\lim_{\mycom{x\to 1}{x<1}}\frac{\log x -1 +\frac{1}{x}}{x-1-\log x}=1,
\]
where the latter equality follows from L'H\^opital's rule. To conclude the proof, we readily obtain from (\ref{ratio:G-y}) that
$G(y)/y\in (0,1)$ for $y\in (0,\infty)$ and $\lim_{y\to\infty}G(y)/y=0$. \hfill\done

\begin{corollary}
The largest asymptotic order of $n(\gamma)$ is reached when $\gamma D(q||q_0)=y+o(1)$ ($\gamma\to \infty)$, and it corresponds to 
covertness, i.e., $n(\gamma)=\Theta(\gamma)$.
\end{corollary}

\begin{remark}[Asymptotic behavior of $h^\star(\gamma)$]
When conditions (\ref{result005})-(\ref{result006}) hold, 
\begin{equation}
		h^\star(\gamma)\sim_\gamma \left\{ \begin{array}{ll}
			\log(\gamma D(q||q_\gamma)), &\mbox{if $\lim_\gamma \gamma D(q||q_\gamma)=\infty$},\\
			-1-y-W_{-1}(-e^{-1-y}), &\mbox{if  $\lim_\gamma \gamma D(q||q_\gamma)=y\in (0,\infty)$},\\
			\sqrt{2\gamma D(q||q_\gamma)}, &\mbox{if $\lim_\gamma \gamma D(q||q_\gamma)=0$}.
		\end{array}
		\right.
		\label{asympt-h-star}
        \end{equation}
The first asymptotic result in (\ref{asympt-h-star}) follows from (\ref{approx-h}) and (\ref{limit-Lambert}). The second asymptotic result 
directly follows from (\ref{approx-h}). The third asymptotic result follows from (\ref{approx-h}) together with
$-1-x-W_{-1}(-e^{-1-x})\sim \sqrt{2x}$ when $x\to 0$ with $x>0$ \cite[pp. 350-351, Eq. (4.22)]{Corless} (take $p=-\sqrt{2(ez+1)}$ for $W_{-1}$ as mentioned at the top of p. 351).
It is tempting to conclude 
that (\ref{asympt-h-star}) generalizes the corresponding result, given by $h^\star(\gamma)\sim_\gamma \log\gamma$,
obtained in the classical setting when the post-distribution does not depend on the false alarm rate. Indeed, if $q_\gamma$ does not
depend on $\gamma$, then $\gamma D(q||q_\gamma)\approx \gamma$, and $h^\star(\gamma)\sim_\gamma \log\gamma$ from the 
first asymptotic result in (\ref{asympt-h-star}). This conclusion is however incorrect since (\ref{asympt-h-star}) only holds if
conditions (\ref{result005})-(\ref{result006}) are met, which will clearly be not true if $q_\gamma$ does not depend on $\gamma$ or,
more specifically, if $q_\gamma$ does not converge to $q$ as $\gamma\to\infty$.
\end{remark}

We now analyze the asymptotic behavior of the adversary's total damage as defined in \eqref{def:total-damage}.
\begin{proposition}
\label{prop:damage}
Assume $g(x)=x^\rho$ with $\rho\in(0,1)$ and that
$\lim_{\gamma} D(q||q_\gamma)/D(q_\gamma||q)=c$ for some $c>0$.
Then, if conditions \eqref{result005}--\eqref{result006} hold, the total damage
$d(\gamma)=n(\gamma)g(D(q||q_\gamma))$ has its largest asymptotic order when
$\gamma D(q||q_\gamma)=\Theta(1)$, with $d(\gamma)=\Theta(\gamma^{1-\rho})$.
\end{proposition}

{\it Proof:} Since conditions \eqref{result005}--\eqref{result006} hold, Proposition~\ref{prop:aympt-n} applies. 
With $g(x)=x^\rho$, we have $d(\gamma)=n(\gamma)\big(D(q||q_\gamma)\big)^\rho$. Moreover, since $\lim_{\gamma} D(q||q_\gamma)/D(q_\gamma||q)=c$, it follows that $1/D(q_\gamma||q)\sim_\gamma c/D(q||q_\gamma)$.

First, if $\lim_\gamma \gamma D(q||q_\gamma)=\infty$, with $t_\gamma=\gamma D(q||q_\gamma)\to\infty$,
\[
d(\gamma)\sim_\gamma c\big(D(q||q_\gamma)\big)^{\rho-1}\log(t_\gamma)
= c\gamma^{1-\rho} t_\gamma^{\rho-1}\log(t_\gamma),
\]
which, since $\lim_{t\to\infty} t^{\rho-1}\log t = 0$ for $\rho\in(0,1)$, implies
\[
d(\gamma) = o(\gamma^{1-\rho}).
\]

Next, if $\lim_\gamma \gamma D(q||q_\gamma)=y\in(0,\infty)$, then $D(q||q_\gamma)\sim_\gamma y/\gamma$ and
\[
d(\gamma)\sim_\gamma c\gamma\frac{G(y)}{y}\big(D(q||q_\gamma)\big)^\rho
\sim_\gamma c\frac{G(y)}{y}y^\rho \gamma^{1-\rho}
=\Theta(\gamma^{1-\rho}).
\]

Finally, if $\lim_\gamma \gamma D(q||q_\gamma)=0$, then $D(q||q_\gamma)=o(1/\gamma)$ and
\[
d(\gamma)\sim_\gamma c\gamma\big(D(q||q_\gamma)\big)^\rho=o(\gamma^{1-\rho}).
\]

Therefore, $d(\gamma)$ has its largest asymptotic order when $\gamma D(q||q_\gamma)=\Theta(1)$, and in this case
$d(\gamma)=\Theta(\gamma^{1-\rho})$.
\hfill\done

Note that, in the setting of the above proposition and following Proposition~\ref{prop:aympt-n}, the regime $\gamma D(q||q_\gamma)=\Theta(1)$ coincides with the transition between detectability and covertness.

\section{Application to Gaussian and exponential models}
\label{sec:Gaussian-exponential}

In this section, we first address the situation when both the pre-change and post-change distributions are Gaussian distributions, then
we investigate the case when both distributions are exponential.

\subsection{Gaussian probability density functions}
\label{ssec:Gaussian-pdfs}
Let 
\begin{equation}
q(x)=\frac{1}{\sqrt{2\pi}}e^{-\frac{1}{2}x^2},\quad q_\gamma(x)=\frac{1}{\sqrt{2\pi(1+\sigma^2)}}e^{-\frac{(x-\mu)^2}{2(1+\sigma^2)}},
\label{Gaussian-pdfs}
\end{equation}
for $x\in\R$,  where  $\mu$ and $\sigma$ are both functions of $\gamma$. We assume that  $(\mu,\sigma) \not= (0,0)$
	for all $\gamma>0$ (this assumption ensures that $q\not\equiv q_\gamma$), $\lim_{\gamma}\mu=0$ when $\mu\not\equiv 0$, and $\lim_{\gamma}\sigma=0$
	when $\sigma\not\equiv 0$.

\begin{proposition}[Gaussian pdfs]
	\label{prop:asymptotic-gaussian}
Conditions (\ref{lem:sufficient1})-(\ref{lem:sufficient2}) are satisfied for Gaussian pdfs $q$ and $q_\gamma$ defined in (\ref{Gaussian-pdfs}).
\end{proposition}

The proof can be found in Appendix \ref{sec:conditions-Gaussian}.

\begin{proposition}
	\label{prop:gaussian}
	For Gaussian pdfs $q$ and $q_\gamma$ given in (\ref{Gaussian-pdfs}),
	\begin{equation}
		\label{n-gaussian}
		n(\gamma)\sim_\gamma \left\{\begin{array}{ll}
			\frac{\log \bigl(\gamma  \bigl(\frac{\mu^2}{2} +\frac{\sigma^4}{4}\bigr)\bigr)}{\frac{1}{2}\mu^2  +\frac{1}{4}\sigma^4},
			&\mbox{if $|\mu| \sqrt{\gamma}\sim_\gamma \infty$ or if $\sigma^2 \sqrt{\gamma}\sim_\gamma\infty$,}\\
			\gamma \frac{G(y)}{y},  &\mbox{if $\frac{1}{2}\gamma(\mu^2 +\frac{1}{2}\sigma^4)\sim_\gamma y\in (0,\infty)$,}\\
			\gamma, &\mbox{if $|\mu|\sqrt{\gamma}\sim_\gamma 0 \, \& \,\sigma^2\sqrt{\gamma} \sim_\gamma 0$,}
		\end{array}
		\right.
	\end{equation}
where $G(y)$ is defined in (\ref{def:G}).
\end{proposition} 
{\it Proof:} We have (see e.g., \cite[Example 1, p. 76]{Khan}) 
\begin{align*}
	D(q_\gamma||q)&=\frac{1}{2}\mu^2 +\frac{1}{2}\sigma^2 -\frac{1}{2}\log (1+\sigma^2),\\
	D(q||q_\gamma) &=\frac{\mu^2}{2(1+\sigma^2)} -\frac{\sigma^2}{2(1+\sigma^2)} +\frac{1}{2}\log (1+\sigma^2),
\end{align*}
yielding $D(q_\gamma ||q)\sim_\gamma \tfrac{1}{2}\mu^2 +\tfrac{1}{4}\sigma^4$ and $D(q||q_\gamma)\sim_\gamma \tfrac{1}{2}\mu^2 +\tfrac{1}{4}\sigma^4$.
Then, (\ref{n-gaussian}) follows from Proposition \ref{prop:aympt-n}. \hfill \done

Proposition \ref{prop:gaussian} shows that an attacker is covert if either (i) $\frac{1}{2}\gamma (\mu^2 +\frac{1}{2}\sigma^4)$ 
converges to a positive constant as $\gamma \to\infty$, or if (ii)
$|\mu|\sqrt{\gamma}\sim_\gamma 0$ and $\sigma^2\sqrt{\gamma} \sim_\gamma 0$.
Hence, covertness will occur, in particular, when $(\mu,\sigma^2) = (\gamma^{-\delta_1},\gamma^{-\delta_2})$ with 
$\delta_1\geq \frac{1}{2}$ and $\delta_2\geq\frac{1}{2}$.

We now focus on the total damage through two representative examples. We take
$g(x)=\sqrt{x}$ and let $\sigma=0$. In this case, $g(D(q||q_\gamma))\sim |\mu|/\sqrt{2}$. Consequently, the total damage $d(\gamma)=n(\gamma)g(D(q||q_\gamma))$ is proportional to $n(\gamma)|\mu|$,
which corresponds to the total mean shift introduced by the adversary prior to detection. Then, following Proposition~\ref{prop:damage}, when $\gamma D(q||q_\gamma)=\Theta(1)$, which here implies $\mu=\Theta(1/\sqrt{\gamma})$, the total damage is maximized and satisfies $d(\gamma)=\Theta(\sqrt{\gamma})$.
Similarly, if we instead set $\mu=0$ and allow $\sigma\neq 0$, then $d(\gamma)$ is proportional to $n(\gamma)\sigma^2$, which corresponds to the total power injected by the adversary prior to detection in a wireless communication setting. Then, under $\sigma^2=\Theta(1/\sqrt{\gamma})$, the total damage is maximized with $d(\gamma)=\Theta(\sqrt{\gamma})$. The same scaling law was also reported in \cite{huang_covert_2021} for this communication setting.


\subsection{Exponential probability density functions}
\label{ssec:exponential-pdfs}
For $x\geq 0$, let
\begin{equation}
q(x)=\lambda e^{-\lambda x},\quad q_\gamma(x)=\lambda_\gamma e^{-\lambda_\gamma x}, 
\label{Exponentials-pdfs}
\end{equation}
with $\lambda>0$ and $\lambda_\gamma>0$ for all $\gamma>0$.
We assume that $\lambda_\gamma\not=\lambda$ for all $\gamma>0$ and  $\lim_\gamma \lambda_\gamma =\lambda$.

\begin{proposition}[Exponential pdfs]
	\label{prop:asymptotic-expo}
Conditions (\ref{lem:sufficient1})-(\ref{lem:sufficient2}) are satisfied for exponential pdfs $q$ and $q_\gamma$ defined in 
(\ref{Exponentials-pdfs}).
\end{proposition}
The proof of Proposition \ref{prop:asymptotic-expo} can be found in Appendix \ref{sec:proof-prop:asymptotic-expo}.

\begin{proposition}
	\label{prop:exponential}
	For exponential pdfs $q$ and $q_\gamma$ given in (\ref{Exponentials-pdfs}),
	\begin{equation}
		\label{n-expo}
		n(\gamma)\sim_\gamma \left\{\begin{array}{ll}
			\frac{2\log\left(\frac{1}{2}\gamma \left(1-\lambda_\gamma/\lambda\right)^2\right)}{\left(1-\lambda/\lambda_\gamma\right)^2},
			&\mbox{if $\sqrt{\gamma} \left|1-\lambda_\gamma/\lambda\right|\sim_\gamma \infty$},\\		
            \gamma\frac{G(y)}{y},
			&\mbox{if $\sqrt{\gamma} \left|1-\lambda_\gamma/\lambda\right|\sim_\gamma y \in (0,\infty)$,}\\	
  		\gamma,		&\mbox{if $\sqrt{\gamma} \left|1-\lambda_\gamma/\lambda\right|\sim_\gamma 0$,}      
        \end{array}
		\right.
	\end{equation}
    where $G(y)$ is defined in (\ref{def:G}).
\end{proposition} 
{\it Proof:} Proposition \ref{prop:exponential} follows from Proposition \ref{prop:aympt-n} upon noticing that
\[
D(q||q_\gamma) = \log \frac{\lambda}{\lambda_\gamma}+\frac{\lambda_\gamma-\lambda}{\lambda}\sim_\gamma \frac{1}{2}
\left(1-\frac{\lambda_\gamma}{\lambda}\right)^2 \,\, \hbox{and}\,\,
D(q_\gamma||q) = \log \frac{\lambda_\gamma}{\lambda}+\frac{\lambda-\lambda_\gamma}{\lambda_\gamma}\sim_\gamma \frac{1}{2}
\left(1-\frac{\lambda}{\lambda_\gamma}\right)^2.
\]
In particular, the result that $D(q||q_\gamma) \sim_\gamma D(q_\gamma||q)$ gives the third asymptotic in (\ref{n-expo}).
\hfill\done

We conclude from Proposition \ref{prop:exponential} that an attacker is covert if $\sqrt{\gamma} \left|1-\lambda_\gamma/\lambda\right|\sim_\gamma y\in [0,\infty)$.
This will occur, for instance, if $\lambda_\gamma=\lambda(1\pm\gamma^{-\delta})$ with $\delta\geq \frac{1}{2}$.

Taking $g(x)=\sqrt{x}$, we have
$d(\gamma)\sim_\gamma n(\gamma)\frac{1}{\sqrt{2}}
\left|1-\frac{\lambda_\gamma}{\lambda}\right|$.
The difference between the post-change and pre-change means is
$|\lambda_\gamma^{-1}-\lambda^{-1}|=\lambda_\gamma^{-1}|1-\lambda_\gamma/\lambda|$. Since $\lambda$ is fixed and $\lim_\gamma \lambda_\gamma = \lambda$, it follows that $d(\gamma)$ is proportional to the total mean shift introduced by the adversary prior to detection. Then, following Proposition~\ref{prop:damage}, when $\gamma D(q||q_\gamma)=\Theta(1)$, i.e., when the mean difference is of
order $\Theta(1/\sqrt{\gamma})$, the total damage is maximized and satisfies $d(\gamma)=\Theta(\sqrt{\gamma})$.


\section{Conclusion}
\label{sec:conclusion}
\ar{In this work, we developed an asymptotic framework for covert quickest change detection in discrete time, analyzing the CuSum procedure in a stationary setting where the post-change distribution $q_\gamma$ depends on the false-alarm constraint $\gamma$ and approaches the pre-change distribution as $\gamma \to \infty$. Classical QCD analyses do not allow such $\gamma$-dependent post-change models, and therefore cannot capture this covert regime.}

\ar{We showed that standard asymptotic characterizations of detection delay fail under this dependence, and we established conditions under which the overshoot of the CuSum statistic vanishes asymptotically, enabling sharp and rigorous characterizations of both AT2FA and ADD. We further verified, in Gaussian and exponential models, that these conditions hold under convergence of $q_\gamma$ to $q$ in parameter space as $\gamma \to \infty$. We also derived the scaling laws under which an adversary can preserve covertness.}


A natural direction for future research is to identify structural properties of the pdfs $q_\gamma$ 
and $q$ that would guarantee that sufficient conditions \eqref{lem:sufficient1} and \eqref{lem:sufficient2}, for the vanishing of the overshoots, are automatically satisfied.

{\it Acknowledgments:} The authors thank Prof. Venu Veeravalli for helpful comments and suggestions.

\appendices 

\section{Proof of Proposition \ref{prop:asymptotic-gaussian}}
\label{sec:conditions-Gaussian}
The error function $\erf(x)=\frac{2}{\sqrt{\pi}}\int_0^x e^{-t^2}dt\in [-1,1]$, $x\in\R$,  will be used throughout. Recall that $\erf(-x)=-\erf(x)$, $\lim_{x\to-\infty}
\erf(x)=-1$, and $\lim_{x\to\infty}\erf(x)=1$.
The asymptotic expansion \cite[p. 298, 7.1.23]{AS65}
\begin{equation}
	\label{erf-asymp}
	\erf(x)=1-\frac{e^{-x^2}}{\sqrt{\pi}}\left(\frac{1}{x}-\frac{1}{2x^3}\right),\quad x\to\infty, 
\end{equation}
will be used repeatedly.
We will use the well-known results that for any real-valued random variable $U$ with pdf $\varphi$ and $u\in\R$
\[
\E[U\,|\, U\geq u] = \frac{\int_u^\infty t \varphi(t)dt}{\int_u^\infty \varphi(t)dt},\,\,
\E[U\,|\, U\leq u] = \frac{\int_{-\infty}^u t\varphi(t)dt}{\int_{-\infty}^u \varphi(t)dt},
\]
so that
\begin{align}
	\E[U-u\,|\, U\geq u] &= \frac{\int_u^\infty (t-u) \varphi(t)dt}{\int_u^\infty \varphi(t)dt},\label{lifetime-sup} \\
	\E[U-u\,|\, U\leq u] &= \frac{\int_{-\infty}^u (t-u) \varphi(t)dt}{\int_{-\infty}^u \varphi(t)dt}.
	\label{lifetime-inf}
\end{align}

Denote by  $f_p$ the pdf of $Y=\log \frac{q_\gamma(X)}{q(X)}$, with $X$ a real-valued rv with Gaussian pdf $p\in\{q,q_\gamma\}$,
$q$ and $q_\gamma$ given in (\ref{Gaussian-pdfs}).  The pdf $f_p$ is given in (\ref{cdf-Y-p}) in Appendix \ref{ssec:pdf-Y}.

We are now in position to prove the  validity of conditions  (\ref{lem:sufficient1}) and (\ref{lem:sufficient2}) in Proposition \ref{prop:asymptotic-gaussian}.
In each case, we will treat separately the cases $\sigma\equiv 0$ and $\sigma \not\equiv 0$. Throughout $p\in \{q,q_\gamma\}$.

\subsection{Proof that  (\ref{lem:sufficient1}) holds for Gaussian pdfs}
\label{app-proof-sufficient1} 

\subsubsection{Case when $\sigma\equiv 0$}
\label{app-proof-sufficient1-1}
The latter implies that  $\mu\not\equiv 0$ as, by assumption, we cannot have simultaneously $\sigma\equiv 0$ and $\mu\equiv 0$.
By (\ref{lifetime-sup}) and  (\ref{cdf-Y-p}), for $y\geq 0$,
\begin{align}
	\E_{p}[Y-y\,|\, Y\geq y]&=
	\frac{\int_y^\infty (z-y) e^{-\frac{1}{2}(\frac{z}{\mu}-\frac{\mu \xi_p}{2})^2} dz}{\int_y^\infty e^{-\frac{1}{2}(\frac{z}{\mu}-\frac{\mu \xi_p}{2})^2} dz}\nonumber\\
	&=|\mu|\Delta_1\left(\frac{y}{|\mu|}-\frac{|\mu| \xi_p}{2}\right), 
	\label{lifetime-Y-1}
\end{align}
where 
\begin{align}
	\Delta_1(x)&:=\frac{\int_{x}^\infty (t-x) e^{-\frac{1}{2}t^2} dt}{\int_x^\infty e^{-\frac{1}{2}t^2} dt},\nonumber\\
	&=\frac{\sqrt{2}e^{-\frac{1}{2}x^2}-x\sqrt{\pi}(1- \erf(\frac{x}{\sqrt{2}}))}{\sqrt{\pi}(1-\erf(\frac{x}{\sqrt{2}}))}.\label{value-Delta1}
\end{align}

\begin{lemma}
	\label{lem-monotonicity}
	The mapping $x\to \Delta_1(x)$  is non-increasing in $\R$.
\end{lemma}
{\it Proof:}
We have
\begin{align}
	\Delta_1^\prime (x)&=\frac{\Gamma(x)} {\pi(1-\erf(\frac{x}{\sqrt{2}}))^2},
	\label{result530}
\end{align}
with $\Gamma(x):=-\pi+2e^{-x^2} -\sqrt{2\pi}x (1-\erf(\frac{x}{\sqrt{2}}))e^{-\tfrac{1}{2}x^2}-\pi\erf(\frac{x}{\sqrt{2}})(\erf(\frac{x}{\sqrt{2}})-2)$, and
$\Gamma^\prime(x)=\Phi(x)$
with $\Phi(x):=\sqrt{2\pi}(x^2+1)(1-\erf(\frac{x}{\sqrt{2}}))-2xe^{-\tfrac{1}{2}x^2}$.
Since $\erf(x)\in [-1,1]$ for all $x\in\R$,  obviously $\Phi(x)\geq 0$ for $x\leq 0$.  Assume that now that $x>0$. By using the bound 
$1-\erf(x)\geq \frac{2e^{-x^2}}{\sqrt{\pi}(x+\sqrt{x^2+2})}$ ($x>0$) \cite[7.1.13]{AS65} we get
$\Phi(x)\geq 2 e^{-\frac{1}{2}x^2}\left(\frac{x^2+2 -x\sqrt{x^2+4}}{x+\sqrt{x^2+4}} \right)\geq 0$
since it is easily seen that $x^2+2 -x\sqrt{x^2+4}\geq 0$ for all $x\in\R$. This shows that  $\Gamma(x)$ is non-decreasing on $\R$.
But since  $\lim_{x\to\infty} \Gamma(x)=0$ (Hint: $\lim_{x\to\infty} \erf(x)=1$), we conclude that $\Gamma(x)\leq 0$ for all $x\in\R$, which in turn proves
from  (\ref{result530}) that $\Delta_1(x)$ in non-increasing in $\R$.

\hfill\done

By (\ref{lifetime-Y-1}) and Lemma \ref{lem-monotonicity}, $y\to \E_{p}[Y-y\,|\, Y\geq y]$ is non-increasing in $[0,\infty)$,
Therefore,
\begin{align*}
	0&\leq \lim_{\gamma\to \infty } \sup_{y\geq 0} \E_{p}[Y-y\,|\, Y\geq y]\leq \lim_{\gamma}\E_{p}[Y\,|\, Y\geq 0]\\
	&= \lim_{\mu\to 0} |\mu| \Delta_1(-\frac{|\mu| \xi_p}{2})=0,
\end{align*}
since $\lim_{x\to 0}\Delta_1(x)=\sqrt{\tfrac{2}{\pi}}$ (Hint: $\erf(0)=0$). This proves (\ref{lem:sufficient1})  when $\mu\not=0$ and $\sigma=0$.


\subsubsection{Case when $\sigma\not\equiv 0$}
\label{app-proof-sufficient1-2}
By (\ref{cdf-Y-p}) and (\ref{lifetime-sup}), for $y\geq 0$ (Hint:  $\delta<0$ so that $\max\{y,\delta\}=y$),
\begin{align}
	\label{value-Egamma}
	\E_{p}[Y-y\,|\, Y\geq y]&=\frac{\int_y^\infty  \frac{t-y}{\sqrt{t-\delta}}\,\rho_1(t)dt +\int_y^\infty  \frac{t-y}{\sqrt{t-\delta}}\,\rho_2(t)dt}
	{\int_y^\infty \frac{1}{\sqrt{t-\delta}}\rho_1(t) dt + \int_y^\infty \frac{1}{\sqrt{t-\delta}}\rho_2(t) dt },
\end{align}
where $\rho_1(t)$ and $\rho_2(t)$ are defined in (\ref{definitions-constants-2}).
With  the change of variable  $v=\frac{\tau\chi_p+\sqrt{\nu(t-\delta)}}{\sqrt{\chi_p}}$, 
we find
\begin{align*}
	\int_y^\infty  \frac{t-y}{\sqrt{t-\delta}}\,\rho_1(t)dt &=\frac{2\sqrt{\chi_p}}{\nu\sqrt{\nu}}
	\int_{ \frac{\tau\chi_p+\sqrt{\nu(y-\delta)}}{\sqrt{\chi_p}}}^\infty
	((v\sqrt{\chi_p} -\tau\chi_p)^2 -\nu(y-\delta))\,e^{-\tfrac{1}{2}v^2} dv, \\
	\int_y^\infty \frac{\rho_1(t)}{\sqrt{t-\delta}}dt&=
	\frac{2\sqrt{\chi_p}}{\sqrt{\nu}}\int_{ \frac{\tau\chi_p+\sqrt{\nu(y-\delta)}}{\sqrt{\chi_p}}}^\infty e^{-\tfrac{1}{2}v^2} dv.
\end{align*}
Similarly, with the change of variable  $v=\frac{\tau\chi_p-\sqrt{\nu(t-\delta)}}{\sqrt{\chi_p}}$, we find
\begin{align*}
	\int_y^\infty  \frac{t-y}{\sqrt{t-\delta}}\,\rho_2(t)dt &=\frac{2\sqrt{\chi_p}}{\nu\sqrt{\nu}}
	\int_{-\infty}^{\frac{\tau\chi_p-\sqrt{\nu(y-\delta)}}{\sqrt{\chi_p}}}
	((v\sqrt{\chi_p}-\tau\chi_p)^2 -\nu(y-\delta))\,e^{-\tfrac{1}{2}v^2} dv,\\
	\int_y^\infty \frac{\rho_2(t)}{\sqrt{t-\delta}}dt&=
	\frac{2\sqrt{\chi_p}}{\sqrt{\nu}}\int_{-\infty}^{\frac{\tau\chi_p-\sqrt{\nu(y-\delta)}}{\sqrt{\chi_p}}}e^{-\tfrac{1}{2}v^2} dv.
\end{align*}
Putting the pieces together, we obtain for $y\geq 0$
\begin{align}
	\E_{p}[Y-y\,|\, Y\geq y]&=\frac{\chi_p}{\nu}
	\left(\frac{ \int_{-\infty}^{\frac{\tau\chi_p-\sqrt{\nu(y-\delta)}}{\sqrt{\chi_p}}} H(v,y)e^{-\tfrac{1}{2}v^2} dv}
	{\int_{-\infty}^{\frac{\tau\chi_p-\sqrt{\nu(y-\delta)}}{\sqrt{\chi_p}}}e^{-\tfrac{1}{2}v^2} dv
		+\int_{\frac{\tau\chi_p-\sqrt{\nu(y-\delta)}}{\sqrt{\chi_p}}}^\infty e^{-\tfrac{1}{2}v^2} dv}\right.
	\nonumber\\
	&\left. +\frac{\int_{ \frac{\tau\chi_p+\sqrt{\nu(y-\delta)}}{\sqrt{\chi_p}}}^\infty H(v,y) e^{-\tfrac{1}{2}v^2} dv}
	{\int_{-\infty}^{\frac{\tau\chi_p-\sqrt{\nu(y-\delta)}}{\sqrt{\chi_p}}}e^{-\tfrac{1}{2}v^2} dv
		+\int_{\frac{\tau\chi_p-\sqrt{\nu(y-\delta)}}{\sqrt{\chi_p}}}^\infty e^{-\tfrac{1}{2}v^2} dv}\right),
	\label{result201}
\end{align}
with $H(v,y):=(v-\tau\sqrt{\chi_p})^2 -\frac{\nu(y-\delta)}{\chi_p}$. With  the mapping 
\begin{align}
	J(x,\theta)&:=\frac{\int_{-\infty}^{\theta-x}\, ((v-\theta)^2-x^2)  e^{-\tfrac{1}{2}v^2} dv}
	{\int_{-\infty}^{\theta-x} e^{-\tfrac{1}{2}v^2} dv+\int_{\theta+x}^\infty e^{-\tfrac{1}{2}v^2} dv}
	+\frac{\int_{\theta+x}^\infty((v-\theta)^2-x^2) e^{-\tfrac{1}{2}v^2} dv}
	{\int_{-\infty}^{\theta-x} e^{-\tfrac{1}{2}v^2} dv+\int_{\theta+x}^\infty e^{-\tfrac{1}{2}v^2} dv},
	\label{def-J}
\end{align}
formula (\ref{result201}) rewrites, for $y\geq 0$,
\begin{equation}
	\label{result961}
	\E_{p}[Y-y \,|\, Y\geq y]=
	\frac{\sigma^2 \chi_p}{2(1+\sigma^2)}J\left(\sqrt{\frac{\nu (y-\delta)}{\chi_p}},\tau \sqrt{\chi_p}\right).
\end{equation}
Hence,
\begin{align}
	\sup_{y\geq 0} \E_p[Y-y\,|\, Y\geq y]&=\frac{\sigma^2 \chi_p}{2(1+\sigma^2)}
    \sup_{x\geq \sqrt{-\frac{\nu\delta}{1+\sigma^2}}}
	J\left(x,\frac{\tau\chi_p}{\sqrt{1+\sigma^2}}  \right).
	\label{result645}
\end{align}
Routine algebra shows that
\begin{align}
	J(x,\theta)&=\sqrt{\frac{2}{\pi}}\cdot\frac{(x+\theta)e^{-\frac{1}{2}(x-\theta)^2}+(x-\theta)e^{-\frac{1}{2}(x+\theta)^2}}{2-\erf\left(\frac{x-\theta}{\sqrt{2}}\right) - \erf\left(\frac{x+\theta}{\sqrt{2}}\right)}+\theta^2 -x^2 +1.
	\label{value-J}
\end{align}
Note that 
\begin{equation}
	\label{theta-sym}
	J(x,\theta)=J(x,-\theta),\quad \forall x\in \R, \,\forall \theta\in \R,
\end{equation}
so that,  by (\ref{result961}) and the definition of $\nu$, $\delta$, and $\chi_p$ in  
(\ref{definitions-constants}) and (\ref{def-chi-xi}), we see that $\E_{p}[Y-y\,|\, Y\geq 0]$ does not depend on the sign of $\mu$, or equivalently, on the sign of $\tau$
($=\mu/\sigma^2$).

We now show that $\lim_\gamma \sup_{y\geq 0} \E_{p}[Y-y\,|\, Y\geq y]=0$.
To achieve this, we distinguish between the cases  (recall that $\tau=\frac{\mu}{\sigma^2}$) a) $-\infty < \liminf_\gamma \tau \leq \limsup_\gamma
\tau <\infty$  and b) $\lim_\gamma\tau=\pm\infty$.


\paragraph{{Case when $-\infty < \liminf_\gamma \tau \leq \limsup_\gamma\tau <\infty$}} \hfill

In  this case $|\tau|$ is bounded, say by a constant $\eta_2>0$, so that $|\tau  \sqrt{\chi_p}|<2\eta_2$ 
for $\gamma$ large enough by definition of $\chi_p$ in (\ref{def-chi-xi}).  Therefore, there exists $\gamma_0>0$ such that for all $\gamma>\gamma_0$ 
 the arguments of the mapping $J\left(\sqrt{\frac{\nu (y-\delta)}{\chi_p}},\tau\sqrt{\chi_p}\right)$ are contained in the set $[0,\infty)\times [-2\eta_2,2\eta_2]$
when $y\geq 0$. 
Since $\lim_{x\to 0} J(x,\theta)=0$ (Hint: use L'H\^opital's rule) and $\lim_{x\to\infty} J(x,\theta)=2$ (cf. Lemma \ref{lem:limit-2}) below),  we deduce that, for all $\gamma>\gamma_0$,
$|J(x,\theta)|$ is bounded, say by a constant $C_1>0$,  for all $(x,\theta)\in [0,\infty)\times [-2\eta_2,2\eta_2]$. Hence, 
\[
0\leq \lim_\gamma \sup_{y\geq 0}\E_p[Y-y\,|\, Y\geq y]\leq C_1 \lim_\gamma \frac{ \sigma^2\chi_p}{2(1+\sigma^2)}=0,
\]
which proves (\ref{lem:sufficient1}).
\begin{lemma}
	\label{lem:limit-2}
	For any $\theta\in\R$,  $\lim_x J(x,\theta)=2$.
\end{lemma}
{\it Proof:} Applying  (\ref{erf-asymp}) to (\ref{value-J}) yields
\begin{align}
	J(x,\theta)&\sim_x
	\frac{e^{-\frac{1}{2}(x-\theta)^2}\left(x+\theta +(\theta^2 -x^2 +1)\left(\frac{1}{x-\theta}-\frac{1}{(x-\theta)^3}\right)\right)}
	{e^{-\frac{1}{2}(x-\theta)^2}\left(\frac{1}{x-\theta}-\frac{1}{(x-\theta)^3}\right)+e^{-\frac{1}{2}(x+\theta)^2}\left(\frac{1}{x+\theta}-\frac{1}{(x+\theta)^3}\right)}\nonumber\\
	&+\frac{e^{-\frac{1}{2}(x+\theta)^2}\left(x-\theta +(\theta^2 -x^2 +1)\left(\frac{1}{x+\theta}-\frac{1}{(x+\theta)^3}\right)\right)}
	{e^{-\frac{1}{2}(x-\theta)^2}\left(\frac{1}{x-\theta}-\frac{1}{(x-\theta)^3}\right)+e^{-\frac{1}{2}(x+\theta)^2}\left(\frac{1}{x+\theta}-\frac{1}{(x+\theta)^3}\right)}.
	\label{result980}
\end{align}
Assume that $\theta\geq 0$. Multiplying both the numerator and the denominator in (\ref{result980}) by $e^{\frac{1}{2}(x-\theta)^2}$ yields
\begin{align*}
	J(x,\theta)&\sim_x \frac{x+\theta +(\theta^2 -x^2 +1)\left(\frac{1}{x-\theta}-\frac{1}{(x-\theta)^3}\right)}
	{\frac{1}{x-\theta}-\frac{1}{(x-\theta)^3}+e^{-2x\theta}\left(\frac{1}{x+\theta}-\frac{1}{(x+\theta)^3}\right)}
	+\frac{e^{-2x\theta}\left(x-\theta +(\theta^2 -x^2 +1)\left(\frac{1}{x+\theta}-\frac{1}{(x+\theta)^3}\right)\right)}
	{\frac{1}{x-\theta}-\frac{1}{(x-\theta)^3}+e^{-2x\theta}\left(\frac{1}{x+\theta}-\frac{1}{(x+\theta)^3}\right)},\\
	&\sim_x \frac{x+\theta +(\theta^2 -x^2 +1)\left(\frac{1}{x-\theta}-\frac{1}{(x-\theta)^3}\right)}{\frac{1}{x-\theta}-\frac{1}{(x-\theta)^3}},\\
	&=\frac{2 x^2 -2\theta x-1}{x^2 -2\theta x +\theta^2 -1}\sim_x 2.
\end{align*}
Thanks to (\ref{theta-sym}) the lemma also holds when $\theta < 0$, which concludes the proof. \hfill\done

\paragraph{{Case when $\lim_{\gamma} \tau=\pm \infty$}}\hfill

As observed earlier (see comment below (\ref{theta-sym})),  $\E_{p}[Y-y\,|\, Y\geq y]$ does not depend on the sign of $\mu$ (or $\tau$).  We will therefore assume, without loss of generality, that $\lim_\gamma \tau=\infty$. 
\begin{proposition}
\label{prop:cond-27}
For any $\theta\geq \frac{3}{\sqrt{2}}$, the mapping $x\to J(x,\theta)$ is non-increasing in $[1/\sqrt{2},\infty)$.
\end{proposition}

The proof of Proposition \ref{prop:cond-27} can be found in Section \ref{app:proof-prop-A1} below.

Since $\lim_\gamma \sqrt{-\frac{\nu\delta}{1+\sigma^2}}=\infty$ and 
$\lim_\gamma \frac{\tau \chi_p }{\sqrt{1+\sigma^2}}=\infty$,
there exists $\gamma_0>0$ such that $\sqrt{-\frac{\nu\delta}{1+\sigma^2}}\geq \frac{1}{\sqrt{2}}$ and
$\frac{\tau \chi_p }{\sqrt{1+\sigma^2}}\geq \frac{3}{\sqrt{2}}$ for $\gamma>\gamma_0$.
Hence, (\ref{result645}) and the fact that the mapping $x\to J(x,\theta)$ is non-increasing in  $[1/\sqrt{2},\infty)$ 
for all $\theta\geq 3/\sqrt{2}$, which follows from (\ref{result222}) and Proposition \ref{prop:cond-27},
yield
\[
\sup_{y\geq 0} \E_{p}[Y-y\,|\, Y\geq y]\leq  \frac{\sigma^2\chi_p}{2(1+\sigma^2)} J\left(\sqrt{-\frac{\nu \delta}{\chi_p}},\tau \sqrt{\chi_p}\right).
\]
for $\gamma>\gamma_0$.
Since $\lim_\gamma \frac{\chi_p}{2(1+\sigma^2)}=\frac{1}{2}$, the proof is concluded by applying  Lemma \ref{lem:limit-J-x-theta} below.
\begin{lemma}
	\label{lem:limit-J-x-theta}
	If $\lim_{\gamma}\tau=\infty$, 
	\[
	\lim_{\gamma}\sigma^2
	J\left(\sqrt{-\frac{\nu \delta}{\chi_p}},\tau \sqrt{\chi_p}\right)=0.
	\]
\end{lemma}
{\bf Proof.}
Define $x_0:=\sqrt{-\frac{\nu\delta}{\chi_p}}=\sqrt{\frac{1+\sigma^2}{\chi_p}(\frac{\mu^2}{\sigma^4}+\frac{\log(1+\sigma^2)}{\sigma^2})}$ 
and $\theta_0:=\tau \sqrt{\chi_p}=\frac{\mu}{\sigma^2}\sqrt{\chi_p}$. Since
\[
\sigma^2(\theta_0^2-x_0^2+1)= \left\{\begin{array}{ll}
	\mu^2 - \log(1+\sigma^2)+\sigma^2 ,&\mbox{if $p=q_\gamma$},\\
	-\frac{\mu^2}{1+\sigma^2}- \log(1+\sigma^2)+\sigma^2  ,&\mbox{if $p=q$},
\end{array}
\right. 
\]
we see that $\sigma^2(\theta_0^2-x_0^2+1)\sim_\gamma 0$  for $p\in \{q,q_\gamma\}$. Therefore, from (\ref{value-J}), 
\begin{align}
	\sigma^2 J(x_0,\theta_0)\sim_\gamma  
	\sqrt{\frac{2}{\pi}}\cdot\frac{\sigma^2(x_0+\theta_0)e^{-\frac{1}{2}(x_0-\theta_0)^2}+\sigma^2(x_0-\theta_0)e^{-\frac{1}{2}(x_0+\theta)^2}}{
		2-\erf\left(\frac{x_0-\theta_0}{\sqrt{2}}\right) - \erf\left(\frac{x_0+\theta_0}{\sqrt{2}}\right)}.
	\label{result981}
\end{align}
Until this end of this proof we assume that $\gamma$ is large enough so that  $\mu>0$, which is justified by the assumption that $\lim_\gamma \tau=\infty$.
We have, using again the definition of $\chi_p$ in (\ref{def-chi-xi}),
\begin{align}
x_0-\theta_0&=\tau\left(\sqrt{\frac{1+\sigma^2}{\chi_p}\left(1+ \frac{1}{\tau^2}\frac{\log(1+\sigma^2)}{\sigma^2}\right)} -\sqrt{ \chi_p}\right)
\sim_\gamma
   \left\{\begin{array}{ll}
\frac{\sigma^2}{2\mu}-\frac{\mu}{2}  \sim_\gamma 0, &\mbox{if $p=q_\gamma$},\\
\frac{\sigma^2}{2\mu}\sim_\gamma 0, &\mbox{if $p=q$},
              \end{array}
    \right.
\label{result913}
\end{align}
where we have used  that  $\frac{\log(1+\sigma^2)}{\sigma^2} \sim_\gamma 1$, $\sqrt{1+x}=1+\frac{1}{2}x +o(x)$, and  $\lim_\gamma \frac{\sigma^2}{\mu}=\lim_\gamma \frac{1}{\tau}=0$ by assumption. Similarly, we find
\begin{equation}
\label{result914}
x_0+\theta_0\sim_\gamma \frac{2\mu}{\sigma^2}  \sim_\gamma \infty,
\end{equation}
yielding
\begin{equation}
\label{result915}
\sigma^2 (x_0+\theta_0)\sim_\gamma 2\mu \sim_\gamma 0.
\end{equation}
By (\ref{result913})-(\ref{result915}) as $\gamma\to\infty$, the numerator  in the r.h.s of (\ref{result981}) goes to zero while
 the denominator goes to $\sqrt{\pi}$. As a result, the r.h.s of (\ref{result981}) converges to zero as $\gamma\to\infty$, which completes the
proof of  the lemma. \hfill\done


\subsection{Proof that (\ref{lem:sufficient2}) holds for Gaussian pdfs}
\label{app-proof-sufficient2}

\subsubsection{Case when $\sigma\equiv 0$}
\label{app-proof-sufficient2-1}
In this section $\sigma\equiv 0$, so that $\mu\not\equiv 0$.  By (\ref{lifetime-inf}) and (\ref{cdf-Y-p}), for $y\leq 0$,
\begin{eqnarray}
	\E_{p}[Y-y\,|\, Y\leq y]&=&
	\frac{\int_{-\infty}^y (z-y) e^{-\frac{1}{2}(\frac{z}{\mu}-\frac{\mu \xi_p}{2})^2} dz}{\int_{-\infty}^y e^{-\frac{1}{2}(\frac{z}{\mu}-\frac{\mu \xi_p}{2})^2} dz},\nonumber\\
	&=&|\mu|\Delta_2\left(\frac{y}{|\mu|}-\frac{|\mu| \xi_p}{2}\right),
	\label{lifetime-Y}
\end{eqnarray}
where
\begin{align}
	\Delta_2(x)&:=
	\frac{\int_{-\infty}^{x} (t-x) e^{-\frac{1}{2}t^2} dt}{\int_{-\infty}^x e^{-\frac{1}{2}t^2} dt},\nonumber\\
	&=\frac{\sqrt{2}e^{-\frac{1}{2}x^2}+x\sqrt{\pi}\left(1+ \erf\left(\frac{x}{\sqrt{2}}\right)\right)}
	{\sqrt{\pi} \left(1+\erf \left(\frac{x}{\sqrt{2}}\right)\right)}.\label{value-Delta2}
\end{align}

We observe that $\Delta_2(x)=\Delta_1(-x)$, where $\Delta_1(x)$ is given in (\ref{value-Delta1}). Therefore, by Lemma \ref{lem-monotonicity}, 
we see that $x\to\Delta_2(x)$ is non-decreasing in $\R$. 

Consequently,
\begin{align*}
	\inf_{y\leq 0}\E_p[Y-y\,|\, Y\leq y] &\geq \lim_{y\to -\infty} \E_p[Y-y\,|\, Y\leq y],\\
	&=|\mu|\lim_{y\to-\infty}\Delta_2\left(\frac{y}{|\mu|}-\frac{|\mu| \xi_p}{2}\right)=0,
\end{align*}
as\footnote{Write $\Delta_2(x)$ as $\Delta_2(x)=\frac{\sqrt{2}e^{-\frac{1}{2}x^2}+x\sqrt{\pi}(1- \erf(\frac{-x}{\sqrt{2}}))}{\sqrt{\pi}(1-\erf(\frac{-x}{\sqrt{2}}))}$ and
	use (\ref{erf-asymp})  to get
	$\Delta_2(x)\sim \frac{1}{\frac{1}{x}-x}\sim 0$ as $x\to -\infty$.} $\lim_{x\downarrow -\infty} \Delta_2(x)=0$. Consequently, $\lim_{\gamma} \inf_{y\leq 0}\E_p[Y-y\,|\, Y\leq y] =0$, which 
proves (\ref{lem:sufficient2}) when $\mu\not\equiv 0$ and $\sigma \equiv 0$.


\subsubsection{Case when $\sigma\not\equiv 0$}
\label{app-proof-sufficient2-2}
Since $\E_{p}[Y-y\,|\, Y\leq y]\pn{=\E_{p}[(Y-y){\bf 1}_{\{Y\leq y\}}\,|\, Y\leq y]}$,  by  (\ref{lifetime-inf}) and (\ref{cdf-Y-p}) we see that
$\E_{p}[Y-y\,|\, Y\leq y]=0$ for $y\in (-\infty,\delta]$, so that  (\ref{lem:sufficient2}) reduces  to showing that 
\begin{equation}
\label{new-sufficient2}
\lim_{\gamma\to \infty}\inf_{\delta<y\leq 0} \E_p[Y-y\,|\, Y\leq y]=0.
\end{equation}
For $\delta<y\leq 0$, we have from   (\ref{lifetime-inf}) and (\ref{cdf-Y-p})
\begin{equation}
\label{value-Egamma-2}
\E_{p}[Y-y\,|\, Y\leq y]=\frac{\int_\delta^y  \frac{t-y}{\sqrt{t-\delta}}\,\rho_1(t)dt +\int_\delta^y  \frac{t-y}{\sqrt{t-\delta}}\,\rho_2(t)dt}
 {\int_\delta^y \frac{1}{\sqrt{t-\delta}}\rho_1(t) dt + \int_\delta^y \frac{1}{\sqrt{t-\delta}}\rho_2(t) dt },
 \end{equation}
where $\rho_1(x)$ and  $\rho_2(x)$ are defined in (\ref{definitions-constants-2}) and $\delta$ is defined in (\ref{definitions-constants}).

By duplicating the derivation of (\ref{result201}), we easily find that, for $\delta< y\leq 0$, 
\begin{align}
&\E_{p}[Y-y\,|\, Y\leq y]=
\frac{\left(\frac{\chi_p}{\nu}\right) \int_{\frac{\tau\chi_p-\sqrt{\nu(y-\delta)}}{\sqrt{\chi_p}}}^{\tau\sqrt{\chi_p}}
H(v,y)e^{-\tfrac{1}{2}v^2} dv}
{\int_{\frac{\tau\chi_p-\sqrt{\nu(y-\delta)}}{\sqrt{\chi_p}}}^{\tau\sqrt{\chi_p}}e^{-\tfrac{1}{2}v^2} dv+
\int_{\tau\sqrt{\chi_p}}^{\frac{\tau\chi_p+\sqrt{\nu(y-\delta)}}{\sqrt{\chi_p}}} e^{-\tfrac{1}{2}v^2} dv}
+\frac{\left(\frac{\chi_p}{\nu}\right) \int_{\tau\sqrt{\chi_p}}^{\frac{\tau\chi_p+\sqrt{\nu(y-\delta)}}{\sqrt{\chi_p}}}H(v,y) e^{-\tfrac{1}{2}v^2} dv}
{\int_{\frac{\tau\chi_p-\sqrt{\nu(y-\delta)}}{\sqrt{\chi_p}}}^{\tau\sqrt{\chi_p}} e^{-\tfrac{1}{2}v^2} dv+
\int_{\tau\sqrt{\chi_p}}^{\frac{\tau\chi_p+\sqrt{\nu(y-\delta)}}{\sqrt{\chi_p}}} e^{-\tfrac{1}{2}v^2} dv},
\label{result2001}
\end{align}
where $H(v,y):=(v-\tau\sqrt{\chi_p})^2 -\frac{\nu(y-\delta)}{\chi_p}$. With  the mapping 
\begin{align}
G(x,\theta)&:=
\frac{\int_{\theta-x}^\theta\, ((v-\theta)^2-x^2)  e^{-\tfrac{1}{2}v^2} dv}
{\int_{\theta-x}^\theta e^{-\tfrac{1}{2}v^2} dv+\int_\theta^{\theta+x}e^{-\tfrac{1}{2}v^2} dv}+\frac{\int_\theta^{\theta+x}((v-\theta)^2-x^2) e^{-\tfrac{1}{2}v^2} dv}
{\int_{\theta-x}^\theta e^{-\tfrac{1}{2}v^2} dv+\int_\theta^{\theta+x}e^{-\tfrac{1}{2}v^2} dv},
\label{def-G}
\end{align}
the conditional expectation $\E_{p}[Y-y \,|\, Y\leq y]$ in (\ref{value-Egamma-2}) rewrites
\begin{equation}
\label{result969}
\E_{p}[Y-y \,|\, Y\leq y]=
\frac{\sigma^2\chi_p}{2(1+\sigma^2)} G\left(\sqrt{\frac{\nu(y-\delta)}{\chi_p}},\tau\sqrt{\chi_p} \right),
\end{equation}
for $\delta<y\leq 0$.
Hence, 
\begin{align}
0&\geq \lim_\gamma\inf_{\delta<y\leq 0} \E_p[Y-y\,|\, Y\leq y]
=\frac{1}{2}\lim_\gamma\inf_{0< t\leq 1}\sigma^2  G\left(\sqrt{-\frac{\nu \delta t}{\chi_p}},\tau\sqrt{\chi_p} \right),
\label{result455}
\end{align}
since $\lim_\gamma \frac{ \chi_p}{1+\sigma^2}=1$ from the definition of $\chi_p$.
Standard algebra gives 
\begin{align}
G(x,\theta)
&=\sqrt{\frac{2}{\pi}}\cdot\frac{(\theta-x)e^{-\frac{1}{2}(\theta+x)^2}-(\theta+x)e^{-\frac{1}{2}(\theta-x)^2}}
{\erf\left(\frac{\theta+x}{\sqrt{2}}\right)-\erf\left(\frac{\theta-x}{\sqrt{2}}\right)}+\theta^2 -x^2+1.
\label{value-G}
\end{align}
Notice that $G(x,\theta)=G(x,-\theta)$, which implies from (\ref{result969}), that 
 $\E_{p}[Y-y\,|\, Y\leq y]$ does not depend on the sign of $\tau $ or, equivalently, on the sign of $\mu$ since $\tau=\frac{\mu}{\sigma^2}$.\

Like in Section \ref{app-proof-sufficient1} we will distinguish the case (a) $-\infty < \liminf_{\gamma}\tau \leq \limsup_{\gamma}\tau<\infty$ from the case (b) 
$\lim_\gamma\tau=\pm \infty$.

\paragraph{{Case when $-\infty < \liminf_{\gamma}\tau \leq \limsup_{\gamma}\tau<\infty$}}\hfill

 In this case $|\tau|$ is bounded, say by a constant $\eta_2>0$, so that $|\tau  \sqrt{\chi_p}|<2\eta_2$ 
for $\gamma$ large enough by definition of $\chi_p$ in (\ref{def-chi-xi}). On the other hand, 
for  $\delta\leq y\leq 0$, $\sqrt{\nu(y-\delta)/\chi_p}$ lies in the interval $[0,\sqrt{-\nu\delta/\chi_p}]$.  For  $\gamma$ large enough, this interval is bounded as  
$\sqrt{-\frac{\nu\delta}{\chi_p}}=\sqrt{\frac{1+\sigma^2}{\chi_p}(\tau^2 +\frac{\log(1+\sigma^2)}{\sigma^2})}
\leq \sqrt{2(\eta_2^2 +1)}:=\eta_1$  by using that $\frac{\log(1+x)}{x}\leq 1$ for all $x\geq 0$ and $\frac{1+\sigma^2}{\chi_p}\leq 2$ for $\gamma$ large enough.
Therefore, for $\gamma$ large enough, the arguments of $G(\sqrt{\nu(y-\delta)/\chi_p},\tau \sqrt{\chi_p})$ take values in the compact set $[0,\eta_1]\times [-2\eta_2,2\eta_2]$ for
all $y\in [\delta,0]$. Since the mapping  $(x,\theta)\to G(x,\theta)$ is continuous on  $\R\times \R$,  by the extreme value theorem, $|G(x,\theta)|$ is bounded 
on any   compact set  of  $\R\times \R$, which implies that, for $\gamma$ large enough,  there exists a constant $C_1>0$ such that
$|G(\sqrt{\nu(y-\delta)/\chi_p},\tau \sqrt{\chi_p})|< C_1$ for any $y\in [\delta,0]$.  Hence, by (\ref{result455}),
\[
0\geq \lim_{\gamma}\inf_{\delta\leq y\leq 0}\E_p[Y-y\,|\, Y\leq y]\geq \frac{C_1}{2}\lim_{\gamma} \sigma^2=0.
\]


\paragraph{Case when $\lim_{\gamma} \tau =\pm \infty$}\hfill

As noted earlier, $\E_{p}[Y-y\,|\, Y\leq y]$ does not depend on the sign of  $\tau$. We may therefore assume,  without loss of generality, 
that $\lim_{\gamma} \tau=\infty$.  The latter implies that  $\tau>0$ for large enough, which we will assume from now on.

\begin{proposition}
\label{prop:cond-28}
For any $\theta\geq 0$, the mapping $x\to G(x,\theta)$ is non-increasing in $[0,\infty)$.
\end{proposition}
The proof of Proposition \ref{prop:cond-28} can be found in Section \ref{app:proof-prop-A2} below.

Combining (\ref{result455}) and Proposition \ref{prop:cond-28} yields
\begin{align}
\lim_\gamma\inf_{\delta<y\leq 0} \E_p[Y-y\,|\, Y\leq y]
&=\frac{1}{2}\lim_\gamma \sigma^2  G\left(\sqrt{-\frac{\nu \delta}{\chi_p}},\tau\sqrt{\chi_p} \right).
\label{result1255}
\end{align}
The proof that condition (\ref{lem:sufficient2}) holds for Gaussian pdfs when $\lim_\gamma \tau=\infty$ then follows from the lemma below.
\begin{lemma}
\label{lem-limit-28}
If $\lim_\gamma \tau=\infty$ then
\[
\lim_\gamma \sigma^2 G\left(\sqrt{-\frac{\nu \delta}{\chi_p}},\tau\sqrt{\chi_p} \right)=0.
\]
\end{lemma}
{\bf Proof.}
Define $x_0:=\sqrt{-\frac{\nu\delta}{\chi_p}}$ and $\theta_0:=\tau \sqrt{\chi_p}$.  
As established in the proof of Lemma \ref{lem:limit-J-x-theta},
$\sigma^2(\theta_0^2 -x_0^2 +1)\sim_\gamma 0$, $x_0-\theta_0\sim_\gamma 0$, $x_0+\theta_0\sim_\gamma \infty$, and 
$\sigma^2(x_0+\theta_0)\sim_\gamma 0$, so that  (cf. (\ref{value-G})),
\begin{align*}
&\sigma_2 G(x_0,\theta_0)\sim_\gamma \sqrt{\frac{2}{\pi}}\frac{\sigma^2(\theta_0-x_0)e^{-\frac{1}{2}(\theta_0+x_0)^2}-
\sigma^2(\theta_0+x_0)e^{-\frac{1}{2}(\theta_0-x_0)^2}}
{\erf\left(\frac{\theta_0+x_0}{\sqrt{2}}\right)-\erf\left(\frac{\theta_0-x_0}{\sqrt{2}}\right)} \sim_\gamma 0,
\end{align*}
by using that  $\lim_{x\to \infty}\erf(x)=1$ and $\lim_{x\to 0}\erf(x)=0$, so that the denominator goes to one
while the numerator goes to zero as $\gamma \to \infty$.\hfill\done


\section{Probability density function of $Y=\log \frac{q_\gamma(X)}{q(X)}$}
\label{ssec:pdf-Y}
Let $q(x)=\frac{1}{\sqrt{2\pi}} e^{-\frac{1}{2}x^2}$, $q_\gamma(x)=\frac{1}{\sqrt{2\pi (1+\sigma^2)}}e^{-\frac{(x-\mu)^2}{2(1+\sigma^2)}}$, and $Y=\log \frac{q_\gamma(X)}{q(X)}$.
We denote by $f_p$ the pdf of $Y$ under $p\in\{q,q_\gamma\}$.

Assume first that $\sigma\not\equiv 0$.  Set $\widetilde \sigma^2=1+\sigma^2$. We first determine $f_{q_\gamma}$. 
We have
\begin{align}
	\Prob_{q_\gamma}(Y<x)
	= \frac{1}{\sqrt{2\pi \widetilde\sigma^2}} \int_\R e^{-\frac{(z-\mu)^2}{2\widetilde \sigma^2}}  {\bf 1}_{\{-\frac{1}{2}\log\widetilde\sigma^2 
		- (\frac{(z-\mu)^2}{2\widetilde\sigma^2}-\frac{1}{2}z^2)<x\}} dz.\label{law-LZ-ub}
\end{align}
We observe from (\ref{law-LZ-ub}) that
\[
\Prob_{q_\gamma}(Y<x) =0 \quad \hbox{if}\quad \frac{\mu^2 \widetilde \sigma^2}{\sigma^4}%
+\frac{2\widetilde \sigma^2}{\sigma^2}\left( x+\frac{1}{2}\log \widetilde \sigma^2\right)\leq 0,
\]
or, equivalently,
\begin{equation}
	\label{constraint-x-ub}
	\Prob_{q_\gamma}(Y<x) =0 \quad \hbox{if}\quad x\leq \delta,
\end{equation}
where $\delta:= -\frac{\mu^2}{2\sigma^2}-\frac{1}{2}\log \widetilde\sigma^2\leq 0$.
When $x> \delta$, with
\[
\nu:=\frac{2\widetilde\sigma^2}{\sigma^2},\quad
\tau:=\frac{\mu}{\sigma^2}, \quad \rho(x):= \sqrt{\nu (x-\delta)}, 
\]
we obtain from (\ref{law-LZ-ub}) 
\begin{align}
	\Prob_{q_\gamma}(Y<x) &=\frac{1}{\sqrt{2\pi  \widetilde\sigma^2}}\int_\R {\bf 1}_{\{|z+\tau|<\rho(x)\}} e^{-\frac{(z-\mu)^2}{2\widetilde \sigma^2}}dz,\nonumber\\
	&=\frac{1}{\sqrt{2\pi \widetilde\sigma^2}}\int_{-\tau-\rho(x)}^{-\tau +\rho(x)} e^{-\frac{(z-\mu)^2}{2\widetilde \sigma^2}}dz.
	\label{int888-ub}
\end{align}
In summary (cf. (\ref{constraint-x-ub}), (\ref{int888-ub})),
\begin{equation}
	\Prob_{q_\gamma}(Y<x) =\left\{\begin{array}{ll}
		0,&\mbox{for $x\leq\delta$},\\
		\frac{1}{\sqrt{2\pi  \widetilde\sigma^2}}\displaystyle \int_{-\tau-\rho(x)}^{-\tau +\rho(x)} e^{-\frac{(z-\mu)^2}{2\widetilde \sigma^2}}dz, &\mbox{for $x>\delta$}.
	\end{array}
	\right.
	\label{cdf-LZ1-sigma-positive-ub}
\end{equation}
From (\ref{cdf-LZ1-sigma-positive-ub}) we get 
\begin{align}
	f_{q_\gamma}(x)=\frac{d}{dx} \Prob_{q_\gamma}(Y<x)
	=\left\{\begin{array}{ll}
		0,&\mbox{for $x\leq\delta$},\\
		\frac{\nu(\widetilde \rho_1(x)+\widetilde\rho_2(x))}{2\sqrt{\nu(x-\delta)}\sqrt{2\pi  \widetilde\sigma^2}}, &\mbox{for $x>\delta$},
	\end{array}
	\right.
	\label{pdf-Y-q-gamma-1} 
\end{align}
where $\widetilde\rho_1(x):=e^{-\frac{(\tau+\mu-\rho(x))^2}{2\widetilde \sigma^2}}$ and $\widetilde\rho_2(x):=e^{-\frac{(\tau+\mu+\rho(x))^2}{2\widetilde \sigma^2}}$.
When $\sigma\equiv 0$  and $\mu\not\equiv 0$, 
\begin{align*}
	\Prob_{q_\gamma}(Y<x)&=\frac{1}{\sqrt{2\pi}}\int_\R {\bf 1}_{\{\mu z-\frac{\mu^2}{2}<x\}} e^{-\frac{1}{2}(z-\mu)^2} dz
	=\left\{\begin{array}{ll}
		\frac{1}{\sqrt{2\pi}}\int_{-\infty}^{\frac{x}{\mu}+\frac{\mu}{2}}e^{-\frac{1}{2}(z-\mu)^2} dz,&\mbox{if $\mu>0$,}\\
		\frac{1}{\sqrt{2\pi}}\int_{\frac{x}{\mu}+\frac{\mu}{2}}^\infty e^{-\frac{1}{2}(z-\mu)^2} dz,&\mbox{if $\mu<0$,}
	\end{array}
	\right.
	\label{cdf-LZ1-sigma-zero-ub}
\end{align*}
for all $x\in\R$, yielding
\begin{equation}
	\label{pdf-Y-q-gamma-2}
	f_{q_\gamma}(x)=\frac{d}{dx}\Prob_{q_\gamma}(Y<x)=\frac{1}{|\mu| \sqrt{2\pi}} \,e^{- \frac{1}{2}(\frac{x}{\mu}-\frac{\mu}{2})^2}\,dx , \quad \forall x\in\R.
\end{equation}
We now calculate $f_q$. When  $\sigma\not\equiv 0$, mimicking the derivation of (\ref{cdf-LZ1-sigma-positive-ub}), we find 
\begin{align*}
	\Prob_{q}(Y<x)&=\left\{\begin{array}{ll}
		0,&\mbox{for $x\leq\delta$},\\
		\frac{1}{\sqrt{2\pi}}\int_{-\tau-\rho(x)}^{-\tau +\rho(x)}  e^{-\frac{1}{2}z^2}  dz,&\mbox{for $x>\delta$},
	\end{array}
	\right.
	\label{law-LZ-ub}
\end{align*}
yielding
\begin{equation}
	f_q(x)=\frac{\nu \left(e^{-\frac{1}{2}(\tau+\rho(x))^2} + e^{-\frac{1}{2}(\tau-\rho(x))^2}\right)}{2\sqrt{\nu(x-\delta)}\sqrt{2\pi}}\,
	{\bf 1}_{\{x>\delta\}}.
	\label{pdf-Y-q-1}
\end{equation}
Finally, for $\mu\not\equiv 0$ and $\sigma\equiv 0$, we  readily get
\begin{equation}
	f_q(x)=\frac{1}{|\mu| \sqrt{2\pi}} \,e^{- \frac{1}{2}(\frac{x}{\mu}+\frac{\mu}{2})^2}\,dx , \quad \forall x\in\R.
	\label{pdf-Y-q-2}
\end{equation}
Pdfs in  (\ref{pdf-Y-q-gamma-1})-(\ref{pdf-Y-q-2}) can be represented by the unified formula
\begin{align}
	\label{cdf-Y-p}
	f_p(x)=\left\{\begin{array}{ll}
		{\bf 1}_{\{x>\delta\}}\ \frac{\alpha (\rho_1(x)+\rho_2(x))}{2\sqrt{\alpha(x-\delta)} \sqrt{2\pi(1+\sigma^2)}}, &\mbox{if $\sigma \not=0$,}\\
		\frac{1}{|\mu| \sqrt{2\pi} } e^{-\tfrac{1}{2}(\tfrac{x}{\mu}-\tfrac{\mu \xi_p}{2})^2}, &\mbox{if $\mu\not=0$, $\sigma=0$},
	\end{array}
	\right.
\end{align}
for $x\in\R$, where
\begin{align}
	&\rho_1(x):=e^{-\frac{(\tau \chi_p+\sqrt{\nu(x-\delta)})^2}{2\chi_p}},
	\rho_2(x):=e^{-\frac{(\tau \chi_p-\sqrt{\nu(x-\delta)})^2}{2\chi_p}},
	\label{definitions-constants-2}\\
	&\nu:=\frac{2(1+\sigma^2)}{\sigma^2}, \,\, \tau:=\frac{\mu}{\sigma^2}, \,\,\delta:=-\frac{\mu^2}{2\sigma^2}-\frac{1}{2}\log(1+\sigma^2),
	\label{definitions-constants}\\
	&\chi_p:=\left\{
	\begin{array}{ll}
		1+\sigma^2, &\mbox{if $p=q_\gamma$},\\
		1, &\mbox{if  $p=q$},
	\end{array}
	\right.
	\,\,
	\xi_p:=\left\{
	\begin{array}{ll}
		1, &\mbox{if $p=q_\gamma$},\\
		-1, &\mbox{if  $p=q$}.
	\end{array}
	\right.
	\label{def-chi-xi}
\end{align}

                                                                                                                           
\section{Proof of Proposition \ref{prop:cond-27} in Section \ref{app-proof-sufficient1-2}}
\label{app:proof-prop-A1}

We want to show (cf. Remark \ref{rem:prop})  that $x\to J(x,\theta)$ in non-increasing in $[1/\sqrt{2},\infty)$ when $\theta \geq \frac{3}{\sqrt{2}}$, 
where $J(x,\theta)$ is given in  (\ref{value-J}). 
We have
\begin{equation}
\label{result222}
\frac{\partial}{\partial x} J(x,\theta)= \frac{Q(x,\theta)}{\pi \left(2-\left(\erf\left(\frac{\theta+x}{\sqrt{2}}\right)-\erf\left(\frac{\theta-x}{\sqrt{2}}\right)\right)\right)^2},
\end{equation}
with
\begin{align}
Q(x,\theta)&:=-8 \pi x + 4xe^{-\theta^2-x^2}+2 (\theta+x) e^{-(\theta-x)^2}- 2(\theta-x) e^{-(\theta+x)^2}\nonumber\\
&+\sqrt{2\pi}(\theta^2 -x^2 +1)\left( e^{-\frac{(\theta+x)^2}{2}}+e^{-\frac{(\theta-x)^2}{2}}\right)
\left(2-\left(\erf\left(\frac{\theta+x}{\sqrt{2}}\right) - \erf\left(\frac{\theta-x}{\sqrt{2}}\right) \right)\right) \nonumber \\
&+8\pi x \left(\erf\left(\frac{\theta+x}{\sqrt{2}}\right)-\erf\left(\frac{\theta-x}{\sqrt{2}}\right)\right)
-2\pi x \left(\erf\left(\frac{\theta+x}{\sqrt{2}}\right)-\erf\left(\frac{\theta-x}{\sqrt{2}}\right)\right)^2.
\label{def-Q}
\end{align}
Substituting $x$ by $\frac{u-v}{\sqrt{2}}$ and $\theta$ by $\frac{u+v}{\sqrt{2}}$ in (\ref{def-Q}) (notice that necessarily $u\geq v$ since $x\geq 0$), we obtain
\begin{align}
\label{relation-Q-Vhat}
Q\left(\frac{u-v}{\sqrt{2}}, \frac{u+v}{\sqrt{2}}\right)&=
 -4\sqrt{2}\pi (u-v)+ 2\sqrt{2}(u-v) e^{-u^2 -v^2} + 2\sqrt{2} u e^{-2v^2} -2\sqrt{2} v e^{-2u^2}\nonumber\\
 &+\sqrt{2\pi}(2uv+1)\left(e^{-u^2}+e^{-v^2}\right)(2-(\erf(u)-\erf(v))) \nonumber\\
&+4\sqrt{2}\pi (u-v) (\erf(u)-\erf(v))-\sqrt{2}\pi (u-v) (\erf(u)-\erf(v))^2,\nonumber\\
&\leq \sqrt{2}V(u,v),
\end{align}
for $u\geq v$, where
\begin{align}
V(u,v):=& -4\pi (u-v)+ 2(u-v) e^{-u^2 -v^2} + 2 u e^{-2v^2} -2 v e^{-2u^2}\nonumber\\
&+\sqrt{\pi}(2uv+1)\left(e^{-v^2}+e^{-u^2}\right)(2-(\erf(u)-\erf(v)) +4\pi (u-v) (\erf(u)-\erf(v)).
\label{def-V}
\end{align}
\begin{remark}
 \label{rem:prop}
The reader may legitimately wonder  why we do not show that $x\to J(x,\theta)$ is non-increasing in $[0,\infty)$ 
when $\theta \geq 0$. Actually, this result is not correct, as  $Q(x,\theta)$ in (\ref{result222})  can be positive for $x\geq 0$ and $\theta\geq 0$. 
For instance, $Q(0,1)\approx 12.16$. However, numerical results clearly indicate, for instance, 
that  $Q(x,\theta)$ is negative  for $x\geq 0$ and $\theta\geq 4$ as well as for $x\geq 1/\sqrt{2}$ and $\theta\geq 3/\sqrt{2}$. 
It will be more convenient to show the latter, thereby explaining the statement in Proposition  \ref{prop:cond-27}.
\end{remark}
\begin{lemma}
For any $x\geq \frac{1}{\sqrt{2}}$ and $\theta\geq \frac{3}{\sqrt{2}}$ one can always find $u$ and $v$ such that $x=\frac{u-v}{\sqrt{2}}$, $\theta=\frac{u+v}{\sqrt{2}}$, $v\geq 1$ and $u\geq v+1$.
\end{lemma}
{\it Proof:} $x=\frac{u-v}{\sqrt{2}}$ and $\theta=\frac{u+v}{\sqrt{2}}$ give $u=\frac{\theta+x}{\sqrt{2}}$ and $v=\frac{\theta-x}{\sqrt{2}}$. We have
\[
x=\frac{u-v}{\sqrt{2}}\geq \frac{1}{\sqrt{2}},
\]
yielding $u\geq v+1$. It remains to check that $v\geq 1$. We have  $u=\frac{\theta+x}{\sqrt{2}}\geq \frac{\frac{3}{\sqrt{2}}+\frac{1}{\sqrt{2}}}{\sqrt{2}}=2$, 
which coupled with the inequality $u\geq v+1$ obtained above, necessarily implies that $v\geq 1$. This concludes the proof of the  claim. \hfill\done

As a result, Proposition  \ref{prop:cond-27}  will hold if the following proposition is true:
\begin{proposition}
\label{prop:V}
For any $v\geq 1$,  $V(u,v)\leq 0$ for $u\geq v+1$.
\end{proposition}

The rest is this section devoted to the proof of Proposition \ref{prop:V}.

We have
\begin{align*}
\frac{d}{du} V(u,v)
&=-4 u^2e^{-u^2 - v^2} + 2\sqrt{\pi}(\erf(u)-\erf(v))(2v u^2 + u - v)e^{-u^2}-4\sqrt{\pi}( 2 v u^2 - u +v)e^{-u^2}\\
& - 2 v \sqrt{\pi} (\erf(u) - \erf(v) - 2)e^{-v^2} + 2(2u v - 1)e^{-2u^2} + 4\pi(\erf(u) - \erf(v)) - 4\pi + 2e^{-2v^2},
\end{align*}
\begin{align*}
\frac{d}{dv}\left(\frac{d}{du} V(u,v)\right)=&-4(u -v) e^{-u^2 - v^2} + 4\sqrt{\pi}\left(\left(u^2 - \frac{1}{2}\right)(\erf(u)-\erf(v)) - 2u^2 - 1\right)e^{-u^2}\\
& + 4\sqrt{\pi}\left(\left(v^2 - \frac{1}{2}\right)(\erf(u) -\erf(v)) - 2v^2 - 1\right)e^{-v^2} + 4 u e^{-2 u^2} - 4 v e^{-2v^2},
\end{align*}
\begin{align}
\frac{d^2}{dv^2}\left(\frac{d}{du} V(u,v)\right)=e^{-v^2} K(u),
\label{dvdvduV}
\end{align}
with
\begin{align*}
K(u):=&8(-u^2 + uv - v^2 + 1)e^{-u^2} - 8v\left[\sqrt{\pi}\left(\left(v^2 - \frac{3}{2}\right)(\erf(u)-\erf(v)) - 2v^2 + 1\right) - ve^{-v^2}\right].
\end{align*}
Differentiating $K(u)$ w.r.t. $u$ gives
\[
K^\prime(u):=\frac{d}{du}K(u)=16 e^{-u^2} (u-v)(u^2 + v^2 - 2).
\]
We observe that $K^\prime(u)$ vanishes at $u=v$, and at $u=-\sqrt{2-v^2}$ and $u=\sqrt{2-v^2}$ when $v^2 \leq 2$.

Assume that $v\geq \sqrt{2}$. Then, $K^\prime(u)\geq 0$ for all $u\geq v$. Therefore, the mapping
$u\to K(u)$ is increasing, in particular, in $[v+1,\infty)$, yielding $K(u)\geq K(v+1)$ for all $u\geq v+1$.

Assume now that $1\leq v< \sqrt{2}$.  Notice that $v\geq \sqrt{2-v^2}$ in this case, so that $K^\prime(u) \geq  0$ for $u\geq v$.
Hence, $u\to K(u)$ is increasing, in particular, in $[v+1,\infty)$, and $K(u)\geq K(v+1)$.

In summary,  when $v\geq 1$, 
\begin{equation}
\label{bound-H2-u}
K(u)\geq K(v+1) , \quad \forall u\geq v+1,
\end{equation}
where 
\begin{align}
K(v+1)
&=8v\left( -(v +1)e^{-(v + 1)^2} + v e^{-v^2} + \sqrt{\pi} \left( \frac{3}{2}-v^2\right)(\erf(v + 1)-\erf(v)) +\sqrt{\pi}(2v^2- 1)\right).
\label{H2-v-plus-1}
\end{align}
It is shown in Section \ref{ssec:proof0} below that 
\begin{equation}
K(v+1)>0, \quad \forall v\geq 1.
\label{Maple-1}
\end{equation}
We have therefore shown (cf. (\ref{dvdvduV}), (\ref{bound-H2-u}), and (\ref{Maple-1})) that $\frac{d^2}{dv^2}\left(\frac{d}{du} V(u,v)\right)>0$ 
for all $u\geq v+1$ and $v\geq 1$. This shows that the mapping  $v\to \frac{d}{dv}\left(\frac{d}{du} V(u,v)\right)$ is increasing in $[1,u-1]$ for $u\geq 2$.

It is shown in Section \ref{ssec:proof-1} below that $\frac{d}{dv}\left(\frac{d}{du} V(u,v)\right)|_{v=u-1}$, given by
\begin{align}
&\frac{d}{dv}\left(\frac{d}{du} V(u,v)\right)|_{v=u-1}\nonumber\\
&= 4e^{-(u-1)^2} \Biggl[ - \sqrt{\pi} -2\sqrt{\pi}(u - 1)^2 -e^{-u^2}  + u e^{-u^2-2u+1}- (u - 1)e^{-(u - 1)^2} - \sqrt{\pi} (2u^2 + 1)e^{-2u+1}\nonumber\\
&+ \sqrt{\pi}\left( \left(u^2 - \frac{1}{2}\right)e^{-2u+1}+\left( (u - 1)^2 - \frac{1}{2}\right)\right) (\erf(u) -\erf(u-1))\Biggr],
 \label{dvduV-u-minus-1}
 \end{align}
is negative for $u\geq 2$. Therefore, the mapping $v\to \frac{d}{du} V(u,v)$ is decreasing  in $[1,u-1]$ for $u\geq 2$.
 
 It is shown in Section \ref{ssec:proof-2} below that  $\frac{d}{du} V(u,v)|_{v=1}$, given by 
 \begin{align}
 \frac{d}{du} V(u,v)|_{v=1}&=
 -4\pi + 2e^{-2}  + 4\sqrt{\pi}e^{-1}- 4u^2 e^{-u^2 -1}+ 2(2u - 1)e^{-2u^2} - 4\sqrt{\pi} (2u^2 - u + 1)e^{-u^2} \nonumber\\
 &+ 2\sqrt{\pi} (2u^2 + u - 1)(\erf(u) - \erf(1))e^{-u^2} + 2\sqrt{\pi} (2\sqrt{\pi}-e^{-1} )(\erf(u) - \erf(1)),
 \label{duV-v=1}
 \end{align}
 is negative for all $u\geq 2$. Hence, $\frac{d}{du} V(u,v)<0$ for all $u\geq v+1$ and $v\geq 1$,  which  in turn implies that the mapping $u\to V(u,v)$ is decreasing
  in $[v+1,\infty)$ for $v\geq 1$.
 But $V(v+1,v)$ given by
 \begin{align}
V(v+1,v)&=-4\pi +2e^{-2v^2 - 2v - 1} - 2ve^{-2(v + 1)^2} + 2(v+1)e^{-2v^2}   + 4\pi(\erf(v+1) -\erf(v)) \nonumber\\
&+\sqrt{\pi}(2v^2 +2v+1)(\erf(v)-\erf(v+1)+2)(e^{-(v + 1)^2} +e^{-v^2}),
\label{V-v-plus-1}
\end{align}
 is negative  for all $v\geq 1$ as shown in Section \ref{ssec:proof-4} below, so that $V(u,v)<0$ for all $u\geq v+1$ and $v\geq 1$.
 This completes the proof of Proposition \ref{prop:cond-27} and, consequently, the proof of  Proposition \ref{prop:V}


\subsection{Proof that $K(v+1)>0$ for $v\geq 1$}
\label{ssec:proof0}

We have (cf. (\ref{H2-v-plus-1}))
\begin{align*}
K(v+1)= 8 vg(v),
\end{align*}
with $g(v):= -(v +1)e^{-(v + 1)^2} + v e^{-v^2} + \sqrt{\pi} \left( \frac{3}{2}-v^2\right)(\erf(v + 1)-\erf(v)) +\sqrt{\pi}(2v^2- 1)$. 

Since $\erf(v+1)-\erf(v)=\frac{2}{\sqrt{\pi}}\int_v^{v+1} e^{-t^2} dt $, we can rewrite $g(v)$ as 
\[
g(v)= -(v +1)e^{-(v + 1)^2} + v e^{-v^2} + (3-2v^2) \int_v^{v+1} e^{-t^2} dt +\sqrt{\pi}(2v^2- 1).
\]
Differentiating $g(v)$ gives
\begin{align*}
g^\prime(v)&=4\left(v\sqrt{\pi}+(v+1)e^{-(v+1)^2}-v \int_v^{v+1} e^{-t^2} dt-\frac{1}{2}e^{-v^2}\right).
\end{align*}
Using the inequalities  $(v+1)e^{-(v+1)^2}>0$ and $\int_v^{v+1} e^{-t^2} dt\leq e^{-v^2}$,  we obtain that
\[
g^\prime(v)\geq 4\left(\sqrt{\pi}-\left(v +\frac{1}{2}\right) e^{-v^2}\right),
\]
for $v\geq 1$.

The mapping $v\to \left(v+\frac{1}{2}\right)e^{-v^2}$ is decreasing in $[1,\infty)$ (its derivative is $-(2v^2+v-1) e^{-v^2}$, which is negative when $v\geq 1$).
Hence,  $\left(v+\frac{1}{2}\right)e^{-v^2}\leq \frac{3}{2}e^{-1}$, yielding
\begin{align*}
g^\prime(v)&\geq 4\left(\sqrt{\pi}-\frac{3}{2}e^{-1}\right)\approx 4.88, \quad \forall v\geq 1.
\end{align*}
Therefore, the mapping $v\to g(v)$ is increasing in $[1,\infty)$, and since $g(1)\approx 2.23$, we conclude that $g(v)\geq 0$ for $v\geq 1$.


\subsection{Proof that $\frac{d}{dv} \frac{d}{du}V(u,v)|_{v=u-1}<0$ for $u\geq 2$.}
\label{ssec:proof-1}

Denote by $g(u)$ the term between square brackets in the r.h.s. of  (\ref{dvduV-u-minus-1}), that is
\begin{align*}
g(u)&=  - \sqrt{\pi} -2\sqrt{\pi}(u - 1)^2 -e^{-u^2}  + u e^{-u^2-2u+1}- (u - 1)e^{-(u - 1)^2} - \sqrt{\pi} (2u^2 + 1)e^{-2u+1}\nonumber\\
&+ \sqrt{\pi}\left( \left(u^2 - \frac{1}{2}\right)e^{-2u+1}+\left( (u - 1)^2 - \frac{1}{2}\right)\right) (\erf(u) -\erf(u-1)).
\end{align*}
By using the inequality $\erf(u)-\erf(u-1) \leq \frac{2}{\sqrt{\pi}}e^{-(u-1)^2}$ we obtain for $u\geq 2$ (Hint: the coefficient of $\erf(u)-\erf(u-1)$ is positive when $u\geq 2$)
\begin{align}
g(u)&
\leq - \sqrt{\pi} -2\sqrt{\pi}(u - 1)^2 -e^{-u^2}  + u e^{-u^2-2u+1}- (u - 1)e^{-(u - 1)^2} - \sqrt{\pi} (2u^2 + 1)e^{-2u+1}\nonumber\\
&+2\left(u^2 - \frac{1}{2}\right)e^{-u^2}+2\left((u - 1)^2 - \frac{1}{2}\right)e^{-(u-1)^2},\nonumber\\
&= - \sqrt{\pi} -2\sqrt{\pi}(u - 1)^2 +2(u^2-1)e^{-u^2}  + u e^{-u^2-2u+1}+(2u^2-5u+2)e^{-(u - 1)^2}\nonumber\\
& - \sqrt{\pi} (2u^2 + 1)e^{-2u+1}.
\label{bound-g-prime}\end{align}
The mapping $u\to (u^2-1)e^{-u^2}$ is decreasing in $[2,\infty)$ (its derivative is $2u(2-u^2)e^{-u^2}$ which is negative for $u\geq 2$), yielding
$2(u^2-1)e^{-u^2}\leq 6e^{-4}$ for $u\geq 2$.

The mapping $u\to u e^{-u^2-2u+1}$  is decreasing in $[2,\infty)$ (its derivative is $(-2u^2 -2u +1) e^{-u^2-2u+1}$ which is negative for $u\geq 2$), yielding
$u e^{-u^2-2u+1}\leq 2 e^{-7}$ for $u\geq 2$.

Finally, $(2u^2-5u+2)e^{-(u - 1)^2}\leq 2u^2e^{-(u - 1)^2}$ for $u\geq 2$. Since the mapping $v\to  2u^2e^{-(u - 1)^2}$ is decreasing in $[2,\infty)$
(its derivative is $-4u(u^2-u-1)e^{-(u - 1)^2}$ which is negative for $u\geq 2$), yielding  $2u^2e^{-(u - 1)^2}\leq 4 e^{-1}$ for $u\geq 2$, and
therefore $(2u^2-5u+2)e^{-(u - 1)^2}\leq 4e^{-1}$ for $u\geq 2.$

Introducing these bounds into (\ref{bound-g-prime}) gives
\begin{align}
g(u)&\leq - \sqrt{\pi}  +6 e^{-4}+ 2e^{-7}+ 4e^{-1} -2\sqrt{\pi}(u - 1)^2- \sqrt{\pi} (2u^2 + 1)e^{-2u+1}.
\label{final-bound-g}
\end{align}
for $v\geq 1$. Since $- \sqrt{\pi}  +6 e^{-4}+ 2e^{-7}+ 4e^{-1} \approx -0.189$, we conclude from (\ref{final-bound-g}) that $g(u)<0$ for all $u\geq 2$.


\subsection{Proof that $\frac{d}{du} V(u,v)|_{v=1}<0$ for $u\geq 2$.}
\label{ssec:proof-2}

Denote by $g(u)$ the rh.s. of (\ref{duV-v=1}), that is,
\begin{align}
g(u) &= -4\pi + 2e^{-2}  + 4\sqrt{\pi}e^{-1}- 4u^2 e^{-u^2 -1}+ 2(2u - 1)e^{-2u^2} - 4\sqrt{\pi} (2u^2 - u + 1)e^{-u^2} \nonumber\\
 &+ 2\sqrt{\pi} (2u^2 + u - 1)(\erf(u) - \erf(1))e^{-u^2} + 2\sqrt{\pi} (2\sqrt{\pi}-e^{-1} )(\erf(u) - \erf(1)).
\label{def-g-proof-2}
 \end{align}
By using the two-sided inequalities (Hint: $\erf(x)-\erf(y)=\frac{2}{\sqrt{\pi}}\int_y^x e^{-t^2} dt$ for $x>y$)
\begin{equation}
\label{two-sided-inq-Erf}
\frac{2}{\sqrt{\pi}}e^{-y^2}\leq \erf(x)-\erf(y) \leq \frac{2}{\sqrt{\pi}}e^{-x^2},\,\, x> y,
\end{equation}
 in (\ref{def-g-proof-2}),  we obtain  (Hint: $(2u^2 + u - 1>0$ for $u\geq 2$ and $2\sqrt{\pi}-e^{-1} >0$)
 \begin{align*}
 g(u)&\leq  -4\pi + 2e^{-2}  + 4\sqrt{\pi}e^{-1}- 4u^2 e^{-u^2 -1}+ 2(2u - 1)e^{-2u^2} - 4\sqrt{\pi} (2u^2 - u + 1)e^{-u^2} \nonumber\\
&+ 4(2u^2 + u - 1)e^{-1-u^2}+4(2\sqrt{\pi}-e^{-1} )e^{-1},\\
&\leq 
 -4\pi + 2e^{-2}  + 4\sqrt{\pi}e^{-1}+ 4(2\sqrt{\pi}-e^{-1} )e^{-1}+2(2u - 1)e^{-2u^2}+ 4e^{-1}(2u^2 + u - 1)e^{-u^2},
 \end{align*}
 for $u\geq 2$.
 
 The mapping $u\to (2u - 1)e^{-2u^2}$ is decreasing in $[2,\infty)$ (its derivative is ($-2(2u^2-u-1)e^{-2u^2}$ which is negative for $u\geq 2$), yielding
 $2(2u - 1)e^{-2u^2}\leq 6e^{-8}$ for $u\geq 2$.
 
 The mapping $u\to (2u^2 + u - 1)e^{-u^2}$  is decreasing in  $[2,\infty)$ (its derivative is $(-2u^2(u-1)-2u(u^2-1)-1)e^{-u^2}$ which is negative for $u\geq 2$), 
 yielding $4e^{-1}(2u^2 + u - 1)e^{-u^2}\leq 36 e^{-5}$  for $u\geq 2$. Hence, for $u\geq 2$,
 \[
 g(u)\leq -4\pi + 2e^{-2}  + 4\sqrt{\pi}e^{-1}+ 4(2\sqrt{\pi}-e^{-1} )e^{-1}+6e^{-8}+36 e^{-5} \approx -4.76.
 \]
 

\subsection{Proof that $V(v+1,v)<0$ for $v\geq 1$}
\label{ssec:proof-4}
Applying the  two-sided inequalities (\ref{two-sided-inq-Erf}) to (\ref{V-v-plus-1}) gives
\begin{align*}
V(v+1,v)
\leq&-4\pi + 2e^{-2v^2 - 2v - 1} - 2ve^{-2(v + 1)^2} + 2(v+1)e^{-2v^2}   + 8\sqrt{\pi} e^{-v^2}\nonumber\\
&-2(2v^2 + 2v+1)e^{-(v+1)^2} (e^{-(v + 1)^2} +e^{-v^2})+2\sqrt{\pi}(2v^2 + 2v+1)(e^{-(v + 1)^2} +e^{-v^2}),\\
=&-4\pi + 2e^{-2v^2 - 2v - 1} - 2ve^{-2(v + 1)^2} + 2(v+1)e^{-2v^2}   + 8\sqrt{\pi} e^{-v^2}\nonumber\\
&-2(2v^2 + 2v+1)(e^{-2(v + 1)^2} +e^{-2v^2-2v-1})+2\sqrt{\pi}(2v^2 + 2v+1)(e^{-(v + 1)^2} +e^{-v^2}),\\
=&-4\pi + 2(v+1)e^{-2v^2}   + 8\sqrt{\pi} e^{-v^2}-2(2v^2 + 3v+1)e^{-2(v + 1)^2} \\
& -4v(v + 1)e^{-2v^2-2v-1}+2\sqrt{\pi}(2v^2 + 2v+1)(e^{-(v + 1)^2} +e^{-v^2}).
\end{align*}
The mapping $v\to (v+1)e^{-2v^2} $ is decreasing in $[1,\infty)$ (its derivative is  $(1-2v(v+1))e^{-2v^2}$ which is negative for $v\geq 1$),
so that $2(v+1)e^{-2v^2}\leq 4e^{-2}$. Therefore,
\begin{align*}
V(v+1,v) \leq&
-4(\pi -e^{-2}) + 8\sqrt{\pi} e^{-v^2}-2(2v^2 + 3v+1)e^{-2(v + 1)^2} \\
& -4v(v + 1)e^{-2v^2-2v-1}+2\sqrt{\pi}(2v^2 + 2v+1)(e^{-(v + 1)^2} +e^{-v^2}).
\end{align*}
The mapping $v\to (2v^2 + 2v+1)(e^{-(v + 1)^2} +e^{-v^2})$ is decreasing in $[1,\infty)$. Indeed,
\begin{align*}
\frac{d}{dv}(2v^2 + 2v+1)(e^{-(v + 1)^2} +e^{-v^2}) &=-2v(2v^2 +4v +1)e^{-(v+1)^2} -2 (2v^3 +2v^2 -v -1)e^{-v^2}.
\end{align*}
When $v\geq 1$, $2v^3 +2v^2 -v -1\geq 0$, which shows that $\frac{d}{dv}(2v^2 + 2v+1)(e^{-(v + 1)^2} +e^{-v^2})<0$ for all $v\geq 1$, which
in turn proves that  $v\to (2v^2 + 2v+1)(e^{-(v + 1)^2} +e^{-v^2})$ is decreasing in $[1,\infty)$. Hence,
$2\sqrt{\pi}(2v^2 + 2v+1)(e^{-(v + 1)^2} +e^{-v^2})\leq 10\sqrt{\pi}(e^{-4}+e^{-1})$ for all $v\geq 1$, and
\begin{align*}
V(v+1,v) \leq&-4(\pi -e^{-2}) + 8\sqrt{\pi} e^{-v^2}-2(2v^2 + 3v+1)e^{-2(v + 1)^2}  -4v(v + 1)e^{-2v^2-2v-1}+10\sqrt{\pi}(e^{-4}+e^{-1}),
\end{align*}
for $v\geq 1$.

Define $g(v)=8\sqrt{\pi} e^{-v^2}-2(2v^2 + 3v+1)e^{-2(v + 1)^2}  -4v(v + 1)e^{-2v^2-2v-1}$ so that
\[
V(v+1,v)\leq -4(\pi -e^{-2}) +10\sqrt{\pi}(e^{-4}+e^{-1})+g(v).
\]
Assume that $v\to g(v)$ is decreasing in $[1,\infty)$. If so, for $v\geq 1$,
\[
V(v+1,v)\leq -4(\pi -e^{-2}) +10\sqrt{\pi}(e^{-4}+e^{-1})+g(1)\approx -0.021,
\]
which would complete the proof that  $V(v+1,v)<0$ for all $v\geq 1$.

It remains to prove that $v\to g(v)$ is decreasing in $[1,\infty)$. We have
\[
g^\prime(v)=-16\sqrt{\pi}v e^{-v^2}+2(8v^3+20v^2+12v+1)e^{-2(v+1)^2}+4(4v^3 +6v^2-1)e^{-2v^2-2v-1}.
\]
$g^\prime(v)<0$ is equivalent to 
\begin{align}
16\sqrt{\pi}v &\geq 2(8v^3+20v^2+12v+1)e^{-2(v+1)^2+v^2}+4(4v^3 +6v^2-1)e^{-2v^2-2v-1+v^2},
\label{result333}
\end{align}
Since
\[
\frac{d}{dv}\left((8v^3+20v^2+12v+1)e^{-2(v+1)^2+v^2}\right)= -2(8v^4 +38v^3 + 32v^2+13v +1) )e^{-2(v+1)^2+v^2}<0,
\]
for $v\geq 1$, the mapping $v\to 2(8v^3+20v^2+12v+1)e^{-2(v+1)^2+v^2}$ is decreasing in $[1,\infty)$, yielding
$2(8v^3+20v^2+12v+1)e^{-2(v+1)^2+v^2}\leq 2(8+38+32+13+1)e^{-5}=184e^{-5}$ for all $v\geq 1$.
On the other hand,
\[
\frac{d}{dv}\left((4v^3 +6v^2-1)e^{-v^2-2v-1}\right)=2(-4v^4 -10v^3+6v^2+v+1)e^{-v^2-2v-1}.
\]
But 
\begin{align*}
-4v^4 -10v^3+6v^2+v+1&=-v^4 -3v^4 -10v^3+6 v^2+v +1,\\
&\leq -1 - 3 v - 10 v^2 +6v^2 +v+1, \,\,\hbox{as } v^4\geq 1, v^4 \geq v, v^3\geq v^2 \hbox{ when } v\geq 1,\\
&= -4 v^2 -2v<0,
\end{align*}
when $v\geq 1$, which shows that the mapping $v\to (4v^3 +6v^2-1)e^{-v^2-2v-1}$ is decreasing in $[1,\infty)$, yielding
$4(4v^3 +6v^2-1)e^{-v^2-2v-1}\leq 4(4 +6-1)e^{-4}=36 e^{-4}$ for all $v\geq 1$.

We have shown that
\[
 2(8v^3+20v^2+12v+1)e^{-v^2-2v-2}+4(4v^3 -6v^2-1)e^{-v^2-2v-1} \leq 184e^{-5} +36 e^{-4} \approx  1.899.
\]
for all $v\geq 1$.

On the other hand, $16\sqrt{\pi}v \geq 16\sqrt{\pi}=28.359$ for $v\geq 1$. This shows from (\ref{result333}) that $g^\prime(v)<0$ for all $v\geq 1$.


\section{Proof of Proposition \ref{prop:cond-28}  in Section \ref{app-proof-sufficient2-2}}
\label{app:proof-prop-A2}
We want to show  $x\to G(x,\theta)$ is non-increasing in $[0,\infty)$ for all $\theta\geq 0$, 
where $G(x,\theta)$ is given in  (\ref{value-G}).  We have 
\begin{equation}
\label{value-dG}
\frac{\partial}{\partial x}G(x,\theta)=\frac{R(x,\theta)}{\pi\left(\erf\left(\frac{x+\theta}{\sqrt{2}}\right) +\erf\left(\frac{x-\theta}{\sqrt{2}}\right)\right)^2},
\end{equation}
with
\begin{align*}
R(x,\theta)&:= 4 x e^{-x^2 -\theta^2} 
-\sqrt{2\pi}(\theta^2 -x^2 +1) \left(e^{-\frac{(x-\theta)^2}{2}}+e^{-\frac{(x+\theta)^2}{2}}\right)\left(\erf\left(\frac{x+\theta}{\sqrt{2}}\right)+\erf\left(\frac{x-\theta}{\sqrt{2}}\right)\right)\\
&+2(x+\theta) e^{-(x-\theta)^2} +2(x-\theta) e^{-(x+\theta)^2} -2\pi x \left(\erf\left(\frac{x+\theta}{\sqrt{2}}\right)+\erf\left(\frac{x-\theta}{\sqrt{2}}\right)\right)^2.
\end{align*}
Since it is hard to determine the sign of $R(x,\theta)$, let us look at its  derivative with respect to $x$, which is given by 
\begin{align}
&\frac{\partial}{\partial x}R(x,\theta)
= -\Biggl[4(x^2+\theta^2) e^{-x^2-\theta^2 } + 2\pi\left(\erf\left(\frac{x+\theta}{\sqrt{2}}\right)+\erf\left(\frac{x-\theta}{\sqrt{2}}\right)\right)^2
+2(x^2-\theta^2)\left(e^{-(x-\theta)^2}+e^{-(x+\theta)^2}\right)\nonumber\\
&+\sqrt{2\pi}\left(\erf\left(\frac{x+\theta}{\sqrt{2}}\right)+\erf\left(\frac{x-\theta}{\sqrt{2}}\right)\right) 
\left( (x+\theta)((x-\theta)^2+1) e^{-\frac{(x-\theta)^2}{2}} +(x-\theta)((x+\theta)^2+1) e^{-\frac{(x+\theta)^2}{2}} \right) \Biggr].
\label{result1010}
\end{align}
A quick  inspection of the terms inside the square brackets reveals that they are all non-negative when $x\geq \theta\geq 0$ (Hint: $\erf(z)\geq 0$ if $z\geq 0$),
so that $\frac{\partial}{\partial x}R(x,\theta)\leq 0$ for $x\geq \theta\geq 0$.

Assume that  $\frac{\partial}{\partial x}R(x,\theta)\leq 0$ for $0\leq x\leq \theta$. Then, $\frac{\partial}{\partial x}R(x,\theta)\leq 0$ 
for all $x\geq 0$ and $\theta\geq 0$. This implies that, for every $\theta\geq 0$,
\begin{equation}
 \label{R:decreasing}
 x\to R(x,\theta)\,\, \hbox{ is decreasing in } [0,\infty).
 \end{equation}
Since $R(0,\theta)=0$, we conclude from (\ref{R:decreasing}) that $R(x,\theta)\leq 0$ for $x\geq 0$ and $\theta\geq 0$. This in turn implies from 
(\ref{value-dG}) that, for every  $\theta\geq 0$,  the mapping $x\to G(x,\theta)$ is non-increasing in $[0,\infty)$, which proves Proposition \ref{prop:cond-28}.

The rest of Section \ref{app:proof-prop-A2} is devoted to the proof that $\frac{\partial}{\partial x}R(x,\theta)\leq 0$ for $0\leq x\leq \theta$.


\label{sec:proof-lemma}
Introduce the mapping $F(u,v)$ defined by
\begin{align}
F(u,v):=&2(u^2 + v^2)e^{-u^2- v^2}- 2uv(e^{-2u^2} + e^{-2v^2})\nonumber\\
&+\sqrt{\pi}(\erf(u) - \erf(v))(u(2v^2 + 1)e^{-v^2} - v(2u^2 + 1)e^{-u^2}).
\label{def-S-u-v}
\end{align}
We have, from (\ref{result1010}),
\begin{equation}
\label{bound-partial-R}
\frac{\partial}{\partial x}R(x,\theta)=-2\pi\left(\erf\left(\frac{\theta+x}{\sqrt{2}}\right)+\erf\left(\frac{\theta-x}{\sqrt{2}}\right)\right)^2
 -2 F\left(\frac{\theta+x}{\sqrt{2}},\frac{\theta-x}{\sqrt{2}}\right).
\end{equation}
Therefore,  $\frac{\partial}{\partial x}R(x,\theta)\leq 0$ for  $0\leq x\leq \theta$  if $F(u,v)\geq 0$ for all $0\leq v\leq u$.

The remainder of Section \ref{sec:proof-lemma} is now devoted to the proof of the following proposition: 
\begin{proposition}
\label{prop:F-positive}
 $F(u,v)\geq 0$ for all $0\leq v\leq u$. 
 \end{proposition}

The derivative of $F(u,v)$ with respect to $u$ is
\begin{align}
\frac{\partial}{\partial u} F(u,v)=& 2u(3-2u^2)e^{-u^2 - v^2} +  4 (u^2-1) v e^{-2u^2} - 2 v  e^{-2v^2}\nonumber\\
& + \sqrt{\pi}(\erf(u) - \erf(v))\left((2v^2 + 1) e^{-v^2} + 2uv(2u^2 - 1)e^{-u^2}\right).
\label{diff-f}
\end{align} 
With an abuse of notation denote by $S(v)$ the r.h.s. of (\ref{diff-f}), that is
\begin{align}
S(v):= & 2u(3-2u^2)e^{-u^2 - v^2} +  4 (u^2-1) v e^{-2u^2} - 2 v  e^{-2v^2}\nonumber\\
& + \sqrt{\pi}(\erf(u) - \erf(v))\left((2v^2 + 1) e^{-v^2} + 2uv(2u^2 - 1)e^{-u^2}\right).
\label{def-S}
\end{align}
Proving that $S(v)\geq 0$ for $0\leq v\leq u$ will prove that  the mapping $u\to F(u,v)$ is increasing in $[v,\infty)$ for all $v\geq 0$. But since $F(v,v)=0$ for $v\geq 0$, 
this will prove that $F(u,v)\geq 0$ for  $0\leq v\leq u$.

We have
\begin{align}
S^\prime(v):=\frac{d}{dv}S(v)=&-8 u v e^{-u^2 - v^2} + 4(u^2 - 1)e^{-2u^2} + 4(v^2 - 1)e^{-2v^2} \nonumber\\
&+ 2\sqrt{\pi} (\erf(u) - \erf(v))(v(1-2v^2) e^{-v^2} + u(2u^2 - 1)e^{-u^2}).
\label{diff-S}
\end{align}
If $S^\prime(v)\leq 0$ for $0\leq v\leq u$,  then $v\to S(v)$ is decreasing in $[0,u]$ for all $u\geq 0$. But $S(u)=0$, and therefore $S(v)\geq 0$ for $0\leq v\leq u$, which would prove that $F(u,v)\geq 0$ for $0\leq v\leq u$, as noted earlier.

The objective is now to show that $S^\prime(v)\leq 0$ for $0\leq v\leq u$. Denote $S^{\prime\prime}(v)=\frac{d}{dv}S^\prime(v)$. We have 
\begin{align}
S^{\prime\prime}(v)=-8e^{-2v^2} H(u),
\label{diff-2-S}
\end{align}
with (again with an abuse of notation)
\begin{equation}
H(u):=u\left(u^2 - 2v^2 + \frac{1}{2}\right)e^{-u^2 + v^2} + v^3 - \frac{5v}{2}- e^{v^2} \sqrt{\pi}(\erf(u) - \erf(v)) \left(v^4 - 2v^2+ \frac{1}{4}\right).
\label{def1-H}
\end{equation}
If  $H(u)\leq 0$ for $0\leq v\leq u$ then $S^{\prime\prime}(v) \geq 0$ for $0\leq v\leq u$, and $v\to S^\prime(v)$ is increasing in $[0,u]$ for all $u\geq 0$.
But $S^\prime(u)=-8 e^{-2u^2}<0$, which would give that $S^\prime(v)\leq 0$ for $0\leq v\leq u$.

Our objective is now to show that $H(u)\leq 0$ for $0\leq v\leq u$. We now differentiate $H(u)$ with respect to the variable $u$ to get rid of the error functions. We have
\begin{align}
\frac{d}{du}H(u)&= -2e^{-u^2 + v^2} \left(u^4 - (2v^2 + 1)u^2 + v^2(v^2 -1)\right),\nonumber\\
&= -2e^{-u^2 + v^2} (u^2-u _1)(u^2-u_2),
\label{eq:10}
\end{align}
with $u_1:=\frac{2v^2+1-\sqrt{8v^2+1}}{2}$ and  $u_2:=\frac{2v^2+1+\sqrt{8v^2+1}}{2}>0$. 
Notice that $\sqrt{u_1}\leq v < \sqrt{u_2}$ for $v\geq 0$.

When  $0\leq v\leq 1$ then  $u_1\leq 0$, and we conclude from (\ref{eq:10}) that $\frac{d}{du}H(u)\geq 0$ when $v\leq u\leq \sqrt{u_2}$ and
$\frac{d}{du}H(u)\leq 0$ when $u\geq \sqrt{u_2}$. Hence, $u\to H(u)$ increases in $[v, \sqrt{u_2})$ and decreases in $(\sqrt{u_2},\infty)$.
Its maximum is reached at $u=\sqrt{u_2}$.

When $v\geq 1$ then  $u_1\geq 0$, and $\frac{d}{du}H(u)\geq 0$ when $v\leq u\leq \sqrt{u_2}$ (Hint: use that that $v\geq \sqrt{u_1}$ for $v\geq 0$) and  $\frac{d}{du}H(u)\leq 0$ when $u\geq \sqrt{u_2}$.
Hence, $u\to H(u)$ increases in $[v, \sqrt{u_2})$ and decreases in $(\sqrt{u_2},\infty)$. Its maximum is reached at $u=\sqrt{u_2}$.

\begin{figure}[htbp]
\centering
\includegraphics[width=0.4\textwidth]{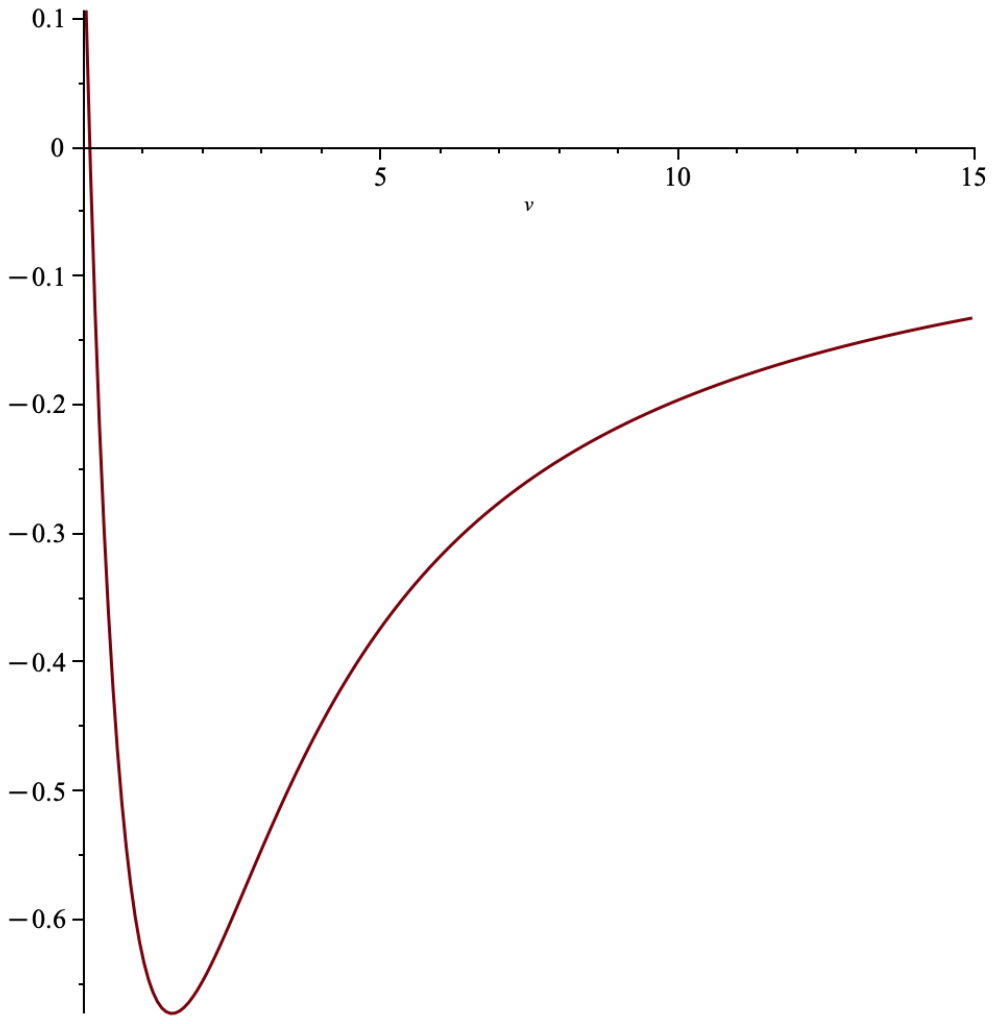}
\caption{$v\to H(\sqrt{u_2})$ for $v\in [0,15]$}
\label{fig-Hmax}
\end{figure}

This proves that 
\begin{equation}
\label{max-H}
\max_{v\leq u}H(u)= H(\sqrt{u_2}),\quad \forall v\geq 0.
\end{equation}
Figure \ref{fig-Hmax} displays the mapping $v\to H(\sqrt{u_2})$ for $v\in [0,15]$.  One can see  that this mapping is not negative for all $v \geq 0$  as
$H(0)=0.178$. However, this plot suggests that it is negative in (for instance) $[1/4,\infty)$ as $H(1/4)\approx -0.202$.

Given this situation, we will first establish a negative upper bound for $S^\prime(v)$ for all $v\in [0,1/4]$ in Section \ref{ssec:negative-ub}, and will then prove in 
Section \ref{ssec:bound-H} that $H(\sqrt{u_2})<0$ for $v\geq \frac{1}{4}$. As discussed earlier, these two results will prove Proposition \ref{prop:F-positive}.

\subsection{A negative upper bound for $S^\prime(v)$ when $v\in [0,1/4]$}
\label{ssec:negative-ub}

Let $a_0\in (0, \sqrt{1/2})$. We claim that
\begin{equation}
\label{bound-Gprime}
S^\prime(v)\leq \left\{\begin{array}{ll}
4(u^2 - 1)e^{-2u^2} +4(a_0^2-1 )e^{-2a_0^2} + 2\sqrt{\pi} \erf(u) a_0&\\
+2\sqrt{\pi}\erf(u)u(2u^2-1 )e^{-u^2} -2\sqrt{\pi}\erf(a_0)u(2u^2-1)e^{-u^2} , &\mbox{if $v\leq u\leq \sqrt{\tfrac{1}{2}}$,}\\
4(u^2 - 1)e^{-2u^2} + 4(a_0^2-1)e^{-2a_0^2}  +2\sqrt{\pi} \erf(u) a_0+2\sqrt{\pi}\erf(u) u(2u^2 - 1)e^{-u^2},&\mbox{if $u\geq \sqrt{\tfrac{1}{2}}$},
\end{array}
\right.
\end{equation}
when $0\leq v \leq a_0$

{\bf Proof of (\ref{bound-Gprime}):}
We have  for $0\leq v \leq a_0$ with  $a_0\in(0,1/\sqrt{2})$ and $u\geq 0$,
\begin{align*}
S^\prime(v)=&-8 u v e^{-u^2 - v^2} +4(u^2 -1)e^{-2u^2} + 4(v^2 - 1)e^{-2v^2}+ 2\sqrt{\pi}\erf(u)v(1-2v^2) e^{-v^2} \\
& + 2\sqrt{\pi} \erf(u)u(2u^2 - 1)e^{-u^2} - 2\sqrt{\pi}  \erf(v)(v(1-2v^2 ) e^{-v^2} - 2\sqrt{\pi}  \erf(v) u(2u^2 - 1)e^{-u^2},\\
&\leq 4(u^2-1 )e^{-2u^2} + 4(a_0^2 - 1)e^{-2a_0^2}+ 2\sqrt{\pi}\erf(u)a_0 + 2\sqrt{\pi} \erf(u)u(2u^2 - 1)e^{-u^2}\\
&  - 2\sqrt{\pi}  \erf(v) u(2u^2 - 1)e^{-u^2},
\end{align*}
where we have used that (i) $u v e^{-u^2 - v^2}\geq 0$, (ii) $(v^2 - 1)e^{-2v^2}\leq (a_0^2 - 1)e^{-2a_0^2}$ as $v\to (v^2 - 1)e^{-2v^2}$ is increasing in $[0,\sqrt{1/2}]$,  (iii)
$\erf(u)v(1-2v^2) e^{-v^2}\leq  \erf(u)a_0$ since $\erf(u)\geq 0$,  and (iv) $\erf(v)v(1-2v^2 ) e^{-v^2} \geq 0$.

Inequalities in  (\ref{bound-Gprime})  follow by noting that $- 2\sqrt{\pi}  \erf(v) u(2u^2 - 1)e^{-u^2}\leq - 2\sqrt{\pi}  \erf(a_0) u(2u^2 - 1)e^{-u^2}$ if $0\leq u\leq \sqrt{1/2}$ and 
 $- 2\sqrt{\pi}  \erf(v) u(2u^2 - 1)e^{-u^2}\leq 0$ if $u\geq \sqrt{1/2}$. \hfill\done
 
 Rewrite (\ref{bound-Gprime}) as 
 \[
 S^\prime(v)\leq \left\{\begin{array}{ll}
 f_1(u)+f_2(u),&\mbox{if $0\leq u\leq \sqrt{\tfrac{1}{2}}$,}\\
 f_1(u),&\mbox{if $u\geq \sqrt{\tfrac{1}{2}}$,}
 \end{array}
 \right.
 \]
 where 
 \begin{align}
 f_1(u)&=4(u^2 - 1)e^{-2u^2} + 4(a_0^2-1)e^{-2a_0^2}  +2\sqrt{\pi} \erf(u) a_0+2\sqrt{\pi}\erf(u) u(2u^2 - 1)e^{-u^2},\\
 f_2(u)&=-2\sqrt{\pi}\erf(a_0)u(2u^2-1)e^{-u^2}.
 \end{align}
 
 {\bf Proof that $f_1(u)\leq 0$ when $u\geq \sqrt{\tfrac{1}{2}}$ and $a_0=\tfrac{1}{4}$:}
 \[
 f_1(u)=g_1(u)+C_1+2\sqrt{\pi}\erf(u)(a_0+h_1(u)),
 \]
with $g_1(u)=4(u^2 - 1)e^{-2u^2}$, $C_1=4(a_0^2-1)e^{-2a_0^2}\approx  -3.309$, and $h_1(u)=u(2u^2-1)e^{-u^2}$.  Since $\erf(u)\leq 1$ and $a_0+h_1(u)\geq 0$,
we have
\[
 f_1(u)\leq g_1(u)+C_1+2\sqrt{\pi}(a_0+h_1(u)).
\]
On the other hand, $g_1^\prime(u)= 8e^{-2u^2}u(3-2u^2)$, which vanishes when $u=0$ and $u=\sqrt{\tfrac{3}{2}}$. Therefore, 
\[
\max_{u\geq \sqrt{\tfrac{1}{2}}} g_1(u)=g_1(\sqrt{\tfrac{3}{2}})=2e^{-3}\approx 0.0995.
\]
Also, $h_1^\prime(u)=-e^{-u^2}(4u^4-8u^2+1)= -4e^{-u^2}\left(u^2 - (1 -\tfrac{\sqrt{3}}{2})\right)\left(u^2 - (1 +\tfrac{\sqrt{3}}{2})\right)$. The smaller root $t_1=1-\tfrac{\sqrt{3}}{2}\approx 0.366$ does not lie in the domain
$\{u \geq \sqrt{1/2}\}$ whereas the larger root $t_2=1+\tfrac{\sqrt{3}}{2}$ does. Hence,
\[
\max_{u\geq \sqrt{\tfrac{1}{2}}} h_1(u)=h_1(\sqrt{t_2} )= \sqrt{t_2}(2t_2-1)e^{-t_2}\approx 0.577.
\]
Putting these bounds together yields
\[
f_1(u) \leq g_1\left(\sqrt{\tfrac{3}{2}}\right)+C_1+2\sqrt{\pi} (a_0 + h_1(\sqrt{t_2} )) \approx  0.0995 -3.881 +2.933=-0.276<0,
\]
for $u\geq \sqrt{\tfrac{1}{2}}$. \hfill\done

 {\bf Proof that $f_1(u)+f_2(u)\leq 0$ when $0\leq u\leq \sqrt{\tfrac{1}{2}}$ and $a_0=\tfrac{1}{4}$:}
 
 We have
 \begin{align*}
f_1(u)+f_2(u) =&4(u^2 - 1)e^{-2u^2} + C_1 +2\sqrt{\pi} \erf(u) a_0+ k_1(u),
 \end{align*}
where $k_1(u):=2\sqrt{\pi}e^{-u^2} u(2u^2 - 1) (\erf(u)-\erf(a_0))$ and where $C_1\approx -3.309$ was defined earlier 
 
 For $0\leq u\leq \sqrt{\tfrac{1}{2}}$, we have $u^2 -1 \leq-\tfrac{1}{2}$. Thus 
 \[
 4(u^2 - 1)e^{-2u^2} \leq -2 e^{-2u^2}\leq -2 e^{-1}\approx -0.735.
 \]
 Also, $2\sqrt{\pi} \erf(u) a_0\leq  2\sqrt{\pi} \erf(\sqrt{1/2}) a_0\approx 0.605$.
 
 We now bound $k_1(u)=2\sqrt{\pi}e^{-u^2} u(2u^2 - 1) (\erf(u)-\erf(a_0))$ from above on the interval $[0,\sqrt{1/2}]$. On $[a_0,\sqrt{1/2}]$, $k_1(u)$ is negative since $\erf(u)\geq \erf(a_0)$. On $u\in [0,a_0]$ we have
 \[
 k_1(u)=2\sqrt{\pi}e^{-u^2} u(1-2u^2) (\erf(a_0)-\erf(u))\leq 2\sqrt{\pi} a_0 \erf(a_0)\approx 0.244.
  \]
 Collecting these bounds, we obtain
 \[
 f_1(u)+f_2(u)\leq  -2 e^{-1} + C_1 +\frac{1}{2}\sqrt{\pi} \erf(\sqrt{1/2}) + \frac{1}{2}\sqrt{\pi} \erf(1/4) \approx -0.735 -3.909 +0.605 +0.244= -3.192<0,
 \]
 for $0\leq u\leq \sqrt{\frac{1}{2}}$.


\subsection{Proving that $H(\sqrt{u_2})$ for $v\geq 1/4$}
\label{ssec:bound-H}

Define $\alpha_0=\sqrt{3}/2$, $\alpha_1=\frac{\sqrt{8+2\sqrt{13}}}{2}\approx 1.95$, and $\alpha_2=\sqrt{5/2}\approx1.58$.   
We will see later on that $\alpha_0$, $\alpha_1$ and $\alpha_2$  each plays a particular role.

We will successively proved that  $H(\sqrt{u_2})<0$ for $v\in [1/4,\sqrt{1-\alpha_0}]$, $v\in [\sqrt{1-\alpha_0}, \sqrt{1+\alpha_0}]$,  $v\in [\sqrt{1+\alpha_0},\alpha_2]$, and
$v\in [\alpha_2,\infty)$ ($\sqrt{1-\alpha_0}\approx 0.336$ and $\sqrt{1+\alpha_0}\approx 1.336$).

We first list useful properties in Section \ref{sssec:properties} and derive upper bounds on $H(\sqrt{u_2})$ in Section \ref{sssec:bounds-H}.

\subsubsection{Useful properties}
\label{sssec:properties}

(a) Let $f_3(v)=\sqrt{u_2}-v$. Notice, from the definition of $u_2$, that $f_3(v)\geq 0$ for $v\geq 0$. With $f_4(v)=\sqrt{8 v^2+1}$, we have
\[
f_3^\prime(v)=\frac{2v\left(1+\frac{2}{f_4(v)}\right)}{\sqrt{4v^2+2 +2 f_4(v)}}-1,\quad
f_3^{\prime\prime}(v)=\frac{6(v^2 + f_4(v)+1)}{(1+2v^2+f_4(v)) f_2(v)^3\sqrt{4v^2+2 +2 f_4(v)}}.
\]
Since $f_3^{\prime\prime}(v)>0$ for $v\geq 0$, we see that the mapping $v\to f_3^\prime(v)$ is increasing in $[0,\infty)$. The latter property and  $\lim_{v\to\infty}f_3^\prime(v)=0$ imply
that $f_3^\prime(v)\leq 0$ for $v\geq 0$, which in turn implies that 
\begin{equation}
\label{f1-decreasing}
v\to f_3(v) \,\, \hbox{ is decreasing on }[0,\infty).
\end{equation}
Last, $f_3(0)=1$ and $\lim_{v\to\infty} f_3(v)=\frac{1}{\sqrt{2}}$.

(b)  The only zero of $u_2-2 v^2+\frac{1}{2}$ in $[0,\infty)$ is located at $\alpha_1=\frac{\sqrt{8+2\sqrt{13}}}{2}$.  In particular,
\begin{equation}
\label{sign-1}
u_2-2 v^2+\frac{1}{2}=\left\{\begin{array}{ll}
< 0&\mbox{for $v> \frac{\sqrt{8+2\sqrt{13}}}{2}$,}\\
>0&\mbox{for $0\leq v<\frac{\sqrt{8+2\sqrt{13}}}{2}$.}
\end{array}
\right. 
\end{equation}
The derivative of $u_2-2 v^2+\frac{1}{2}$ is $2v\left(\frac{2}{\sqrt{8v^2+1}}-1\right)$; in the region $\{v>0\}$ this derivative vanishes at $v=\frac{\sqrt{6}}{4}\approx 0.612$, is
positive in $(0,\frac{\sqrt{6}}{4})$ and is negative in $(\frac{\sqrt{6}}{4}, \infty)$. Hence, the mapping $v\to u_2-2 v^2+\frac{1}{2}$ increases in $(0,\frac{\sqrt{6}}{4})$
and decreases in $(\frac{\sqrt{6}}{4},\infty)$.

(c) The only zero of $v^3  - \frac{5v}{2}$ in $(0,\infty)$ is located at $\alpha_2=\sqrt{\frac{5}{2}}\approx 1.581$. In particular,
\begin{equation}
\label{sign-2}
v^3  - \frac{5v}{2} =\left\{\begin{array}{ll}
<0&\mbox{for $0<v< \sqrt{\frac{5}{2}}$,}\\
> 0&\mbox{for $v>  \sqrt{\frac{5}{2}}$}.
\end{array}
\right.
\end{equation}
Also, $v\to v^3  - \frac{5v}{2}$ is decreasing in $[0,\sqrt{5/6})$ and increasing in $(\sqrt{5/6},\infty)$ ($\sqrt{5/6}\approx 0.912$).

(d) The polynomial $-v^4 +2v^2 -\frac{1}{4}$ has two zeros in $[0,\infty)$ located at $\sqrt{1-\frac{\sqrt{3}}{2}}\approx 0.366$ and  $\sqrt{1+\frac{\sqrt{3}}{2}}\approx 1.366$.
In particular,
\begin{equation}
\label{zeros-term-erf}
-v^4 +2v^2 -\frac{1}{4}=\left\{\begin{array}{ll}
>0&\mbox{for $\sqrt{1-\frac{\sqrt{3}}{2}} < v< \sqrt{1+\frac{\sqrt{3}}{2}}$,}\\
<0&\mbox{for $0< v\leq  \sqrt{1-\frac{\sqrt{3}}{2}}$ and $v> \sqrt{1+\frac{\sqrt{3}}{2}}$.}
\end{array}
\right.
\end{equation}
Also, $v\to -v^4 +2v^2 -\frac{1}{4}$ is increasing in $[0,1)$ and decreasing in $(1,\infty)$.

\subsubsection{Upper bounds for $H(\sqrt{u_2})$}
\label{sssec:bounds-H}

By using the two-sided inequalities in (\ref{two-sided-inq-Erf}) along with (\ref{zeros-term-erf}), we get from (\ref{def1-H})
\begin{equation}
\label{bound1-H}
H(\sqrt{u_2)}\leq K_1(v):=\sqrt{u_2}\left(u_2 - 2v^2 + \frac{1}{2}\right)e^{v^2-u_2} + v^3  - \frac{5v}{2} -2 e^{v^2-u_2}(\sqrt{u_2}-v)\left(v^4 - 2v^2+ \frac{1}{4}\right),
\end{equation}
for $0\leq v\leq  \sqrt{1-\frac{\sqrt{3}}{2}}$ and $v\geq \sqrt{1+\frac{\sqrt{3}}{2}}$,
and 
\begin{equation}
\label{bound2-H}
H(\sqrt{u_2)}\leq K_2(v):=\sqrt{u_2}\left(u_2 - 2v^2 + \frac{1}{2}\right)e^{v^2-u_2} + v^3  - \frac{5v}{2} -2(\sqrt{u_2}-v)\left(v^4 - 2v^2+ \frac{1}{4}\right),
\end{equation}
for  $\sqrt{1-\frac{\sqrt{3}}{2}}\leq v\leq  \sqrt{1+\frac{\sqrt{3}}{2}}$.


\subsubsection{Proof that $H(\sqrt{u_2})<0$ for $v\in [1/4, \sqrt{1-\alpha_0}]$}
\label{sssec:1}
Thanks to the bound in (\ref{bound1-H}) it is sufficient to prove that $K_1(v)<0$ for $\frac{1}{4}\leq v\leq  \sqrt{1-\frac{\sqrt{3}}{2}}$.

Define $T_1(v)=\sqrt{u_2}\left(u_2 - 2v^2 + \frac{1}{2}\right)e^{v^2-u_2}$, $T_2(v)=v^3  - \frac{5v}{2}$, and $T_3(v)= 2 e^{v^2-u_2}(\sqrt{u_2}-v)\left(-v^4 + 2v^2- \frac{1}{4}\right)$ so that
\begin{equation}
\label{K1-sum}
K_1(v)=T_1(v)+T_2(v)+T_3(v). 
\end{equation}
We have
\begin{equation}
\label{diff-T1}
T_1^\prime(v)=\frac{v(8v^4 - 6v^2(\sqrt{8v^2 + 1} +4) + 1)e^{-\frac{1}{2}-\frac{\sqrt{8v^2 + 1}}{2}}}{\sqrt{8v^2+1}\sqrt{4v^2 + 2 + 2\sqrt{8v^2 + 1}}}.
\end{equation}
Since
\[
8v^4 - 6v^2(\sqrt{8v^2 + 1} +4) + 1\leq 8\left(\sqrt{1-\frac{\sqrt{3}}{2}}\right)^4 - 6\left(\left(\frac{1}{4}\right)^2 \sqrt{8\left(\frac{1}{4}\right)^2 + 1} +4\right) + 1\approx -0.815,
\]
for $\frac{1}{4}\leq v \leq \sqrt{1-\frac{\sqrt{3}}{2}}$, we conclude that $T^\prime_1(v)\leq 0$ for $ \frac{1}{4}\leq v \leq \sqrt{1-\frac{\sqrt{3}}{2}}$. Hence,
$\max_{\frac{1}{4}\leq v\leq \sqrt{1-\frac{\sqrt{3}}{2}}}T_1(v)\leq T_1\left(\frac{1}{4}\right)\approx 0.552$.

Last, the property that $v\to T_2(v)$ is decreasing in $[0, \sqrt{\frac{5}{6}})$ (see Section \ref{sssec:properties}-(c)) coupled with $\sqrt{1-\frac{\sqrt{3}}{2}}<\sqrt{\frac{5}{6}}$, ensures
that  $\max_{\frac{1}{4}\leq v\leq \sqrt{1-\frac{\sqrt{3}}{2}}}T_2(v)=T_2\left(\frac{1}{4}\right)\approx -0.609$.
Finally, since $T_3(v)\leq 0$ for  $v\in [\frac{1}{4},\sqrt{1-\frac{\sqrt{3}}{2}}]$ by (\ref{zeros-term-erf}), we conclude that $K_1(v)\leq  T_1\left(\frac{1}{4}\right)+ T_2\left(\frac{1}{4}\right)\approx -0.057$ for $v\in [\frac{1}{4},\sqrt{1-\frac{\sqrt{3}}{2}}]$.\hfill\done


\subsubsection{Proof that $H(\sqrt{u_2})<0$ for $v\in [\sqrt{1-\alpha_0}, \sqrt{1+\alpha_0}]$}
Thanks to the bound in (\ref{bound2-H}) it is sufficient to prove that $K_2(v)<0$ for $v\in I:=[\sqrt{1-\alpha_0}, \sqrt{1+\alpha_0}]$.

Define  $T_4(v)= -2 (\sqrt{u_2}-v)\left(v^4 - 2v^2+ \frac{1}{4}\right)$ so that $K_2(v)=T_1(v)+T_2(v)+T_4(v)$, where 
we recall that $T_1=\sqrt{u_2}\left(u_2 - 2v^2 + \frac{1}{2}\right)e^{v^2-u_2}$ and 
$T_2(v)=v^3  - \frac{5v}{2}$. $T_1(v)$ and $T_4(v)$  are both positive on $I$, while $T_2(v)$ is negative on $I=[\sqrt{1-\alpha_0}, \sqrt{1+\alpha_0}]$.

Let $[i_0,i_1]\subset I$, $i_0<i_1$.

$\bullet$  Differentiating $A(v):=8v^4 - 6v^2\left(\sqrt{8v^2 + 1} +4\right)+1$, gives 
\[
A^\prime(v)=\frac{4v(8v^2 \sqrt{8v^2 + 1} -36v^2 -12 \sqrt{8v^2 + 1} -3)}{\sqrt{8v^2 + 1} }.
\]
On $I$, 
\begin{align*}
&8v^2 \sqrt{8v^2 + 1} -36v^2 -12 \sqrt{8v^2 + 1} -3\\
&\leq 8(\sqrt{1+\alpha_0})^2 \sqrt{8(\sqrt{1+\alpha_0})^2 + 1}-36(\sqrt{1-\alpha_0})^2  -12  \sqrt{8(\sqrt{1-\alpha_0})^2 + 1} -3 \approx -31.713, 
\end{align*}
so that $A^\prime(v)<0$ on $I$ and $A(v)$ is decreasing on $I$. Since $A(\sqrt{1-\alpha_0})\approx -3.228$, we conclude from (\ref{diff-T1}) that $v\to T_1(v)$ is decreasing on $I$.
In particular, on $[i_0,i_1]$,
\[
T_1(v)\leq T_1(i_0);
\]
$\bullet$ $v\to T_2(v)$ is decreasing on $[0,\sqrt{5/6})$ and increasing on $(\sqrt{5/6},\infty)$. Therefore,  on $[i_0,i_1]$,
\[
T_2(v)\leq \left\{\begin{array}{ll}
T_2(i_0),&\mbox{if $i_1\leq \sqrt{5/6}$},\\
\max\{T_2(i_0),T_2(i_1)\},&\mbox{if $i_0\leq \sqrt{5/6}< i_1$},\\
T_2(i_1),&\mbox{if $i_0> \sqrt{5/6}$}.
 \end{array}
 \right.
 \]
 
$\bullet$  $B(v):= 2(\sqrt{u_2}-v) \geq 0$ for all $v\geq 0$, and $B(v)\leq B(i_0)$ for $v\in [i_0,i_1]$ since $v\to B(v)$ is decreasing on $[0,\infty)$ as shown in Section \ref{sssec:properties}-(a).
Define  $C(v)=-v^4 + 2v^2- \frac{1}{4}$. $C(v)$ is positive  on $I$, increasing in $[0,1)$ and decreasing in $(1,\infty)$. Therefore, for  $v\in [i_0,i_1]$,
\[
C(v)\leq \left\{\begin{array}{ll}
C(i_1),&\mbox{if $i_1\leq 1$},\\
C(1),&\mbox{if $i_0\leq 1< i_1$},\\
C(i_0),&\mbox{if $i_0> 1$}.
 \end{array}
 \right.
 \]
Since $T_4(v)=B(v)C(v)$ the above implies that
\[
T_4(v)\leq \left\{\begin{array}{ll}
B(i_0)C(i_1),&\mbox{if $i_1\leq 1$},\\
B(i_0)C(1),&\mbox{if $i_0\leq 1< i_1$},\\
B(i_0)C(i_0),&\mbox{if $i_0> 1$}.
 \end{array}
 \right.
 \]
In summary,  on $[i_0,i_1]\subset [\sqrt{1-\sqrt{3}/2}, \sqrt{1+\sqrt{3}/2}]$,
\begin{equation}
\label{bounds-K2}
K_2(v)\leq M(i_0,i_1):=\left\{\begin{array}{ll}
T_1(i_0)+T_2(i_0)+ B(i_0)C(i_1),&\mbox{if $i_1\leq \sqrt{5/6}$},\\
T_1(i_0)+\max\{T_2(i_0),T_2(i_1)\}+ B(i_0)C(i_1),&\mbox{if $i_0\leq \sqrt{5/6}<i_1\leq 1$},\\
T_1(i_0)+\max\{T_2(i_0),T_2(i_1)\})+ B(i_0)C(1),&\mbox{if $i_0\leq \sqrt{5/6}$ and $ i_1>1$},\\
T_1(i_0)+T_2(i_1)+ B(i_0)C(1),&\mbox{if $\sqrt{5/6}<i_0\leq 1<i_1$},\\
T_1(i_0)+T_2(i_1)+ B(i_0)C(i_0),&\mbox{if $i_0>1$}.
 \end{array}
 \right.
 \end{equation}

Partition the interval $I$ as follows: $I=\cup_{i=1}^{14} I_i$, with $I_1=[\sqrt{1-\alpha_0},0.5]$, $I_2=[0.5,0.6]$, $I_3=[0.6,0.7]$, $I_4=[0.7,0.8]$, $I_5=[0.8,0.85]$, $I_6=[0.85,0.9]$, $I_7=[0.95,1]$, $I_8=[1,1.05]$,
$I_9=[1.05,1.1]$,  $I_{10}=[1.1,1.15]$, $I_{11}=[1.15,1.2]$, $I_{12}=[1.2,1.25]$, $I_{13}=[1.25,1.3]$, and $I_{14}=[1.3,\sqrt{1+\alpha_0}]$.

We find $M(\sqrt{1-\alpha_0},0.5)\approx-0.022$, $M(0.5,0.6)\approx -0.075$, $M(0.6,0.7)\approx -0.041$, 
$M(0.7,0.8)\approx -0.008$, $M(0.8,0.85)\approx -0.056$, $M(0.85,0.95)\approx -0.002$, $M(0.95,1)\approx -0.026$,
$M(1,1.05)\approx -0.017$, $M(1.05,1.1)\approx -0.008$, $M(1.1,1.15)\approx -0.017$, $M(1.15,1.2)\approx -0.046$,
$M(1.2,1.25)\approx -0.099$, $M(1.25,1.3)\approx -0181$, and $M(1.3,\sqrt{1+\alpha})\approx -0.247$.

This shows that $K_2(v)<0$ on $I$.

\begin{remark}
Selecting less intervals for partioning interval $I$ may lead to an upper bound for $K_2(v)$ that is positive on some of these intervals. For instance, if instead of $[0.5.0.6]$, which yields
 $M(0.5,0.6)\approx -0.075$, we select
$[0.5,0.7]$ then $M(0.5,0.7)\approx 0.154$. Even the interval $[0.5,0.65]$ is not good as $M(0.5,0.65)\approx 0.041$. This is the price to pay for upper-bounding $H(\sqrt{u_2})$ by
$K_2(v)$ on $I$ (see (\ref{bound2-H})).
\end{remark}


\subsubsection{Proof that $H(\sqrt{u_2})<0$ for $v\in [\sqrt{1+\alpha_0},\alpha_2]$}

For convenience, we now denote $u_2=v^2 +\frac{1+\sqrt{8v^2+1}}{2}$ by $u_2(v)$.
We will use the bound in  (\ref{bound1-H}), also written as $K_1(v)=T_1(v)+T_2(v)+T_3(v)$, where $T_1$, $T_2$ and $T_3$ are defined in Section \ref{sssec:1}.

Since 
\[
8v^4 - 6v^2(\sqrt{8v^2 + 1} +4) + 1\leq 8\alpha_2^4 - 6(1+\alpha_0)( \sqrt{8(1+\alpha_0)} + 4) +1\approx -45.4<0,
\]
for $v\in [\sqrt{1+\alpha_0},\alpha_2]$, we conclude from (\ref{diff-T1}) that $T_1(v)$ is decreasing in $[\sqrt{1+\alpha_0},\alpha_2]$.

On the other hand, $T_2(v)=v^3 -\frac{5}{2}v$ is increasing in $[\sqrt{5/6}, \infty)$ (see Section \ref{sssec:properties}-(c)), thereby implying that it is 
increasing in $\subset  [\sqrt{1+\alpha_0},\alpha_2]$ as $\sqrt{5/6}<\sqrt{1+\alpha_0}$.

Let $[i_0,i_1]\subset  [\sqrt{1+\alpha_0},\alpha_2]$ with $i_0<i_1$. The above implies that $T_1(v)\leq T_1(i_0)$ and $T_2(v)\leq T_2(i_1)$ for $v\in [i_0,i_1]$.

It remains to find an upper bound for $T_3(v)=-2 e^{v^2-u_2(v)}(\sqrt{u_2(v)}-v)\left(v^4 - 2v^2+\frac{1}{4}\right)$ on $[i_0,i_1]$.  
Since $v^4 - 2v^2+\frac{1}{4}\geq 0$ for $v\geq \sqrt{1+\alpha_0}$ and is decreasing in $[1,\infty)$ (see Section \ref{sssec:properties}-(d)), 
that $v\to \sqrt{u_2}-v$ is positive and decreasing in $[0,\infty)$
(see Section \ref{sssec:properties}-(a)), and that $v\to e^{v^2-u_2}=e^{-\frac{1+\sqrt{8v^2+1}}{2}}$ is obviously decreasing in $[0,\infty)$, we have that 
 $-T_3(v)=2 e^{v^2-u_2(v)}(\sqrt{u_2}-v)\left(v^4 - 2v^2+\frac{1}{4}\right)\geq 2 e^{i_0^2-u_2(i_0)} (\sqrt{u_2(i_0)}-i_0)\left(i_0^4 - 2i_0^2+\frac{1}{4}\right)$ for $v\in [i_0,i_1]$.
 
Collecting the upper bounds obtained for $T_1$, $T_2$, and $T_3$, we obtain
\begin{equation}
\label{max-K1}
\max_{i_0\leq v\leq i_1}K_1(v)\leq N(i_0,i_1)=T_1(i_0)+T_2(i_1)+T_3(i_0).
\end{equation}

Partition the interval $[\sqrt{1+\alpha_0},\alpha_2]$ as follows: $[\sqrt{1+\alpha_0},\alpha_2]=\cup_{j=1}^3 I_j$ with $I_1=[\sqrt{1+\alpha_0},\sqrt{1+\alpha_0}+0.1]$,
$I_2=[\sqrt{1+\alpha_0}+0.1,\sqrt{1+\alpha_0}+0.195]$, and $I_3=[\sqrt{1+\alpha_0}+0.195, \alpha_2]$. We find that $N(\sqrt{1+\alpha_0} ,\sqrt{1+\alpha_0}+0.1)\approx -0.319$,
$N(\sqrt{1+\alpha_0}+0.1,\sqrt{1+\alpha_0}+0.195)\approx -0.003$, and $N(\sqrt{1+\alpha_0}+0.195,\alpha_2)\approx -0.0003$.

This shows from (\ref{max-K1}) that $K_1(v)<0$ for $v\in [\sqrt{1+\alpha_0},\alpha_2]$, and from (\ref{bound1-H}) that $H(\sqrt{u_2})<0$ for $v\in [\sqrt{1+\alpha_0},\alpha_2]$.


\subsubsection{Proof that $H(\sqrt{u_2})<0$ for $v\in [\alpha_2,\infty)$}
\label{sssec:2}

When $v>\alpha_2$ one can no longer use the bound $K_1(v)$ (given in (\ref{bound1-H})) as  $K_1(v)$ becomes rapidly positive. For instance, $K_1(v)>0$ 
for $v>\alpha_2+0.01$ (as can be seen by plotting $v\to K_1(v)$).  We will instead proceed as follows.

First, we know (cf. Section \ref{sssec:properties}-(b)) that $T_1(v)=\sqrt{u_2}\left(u_2 - 2v^2 + \frac{1}{2}\right)e^{v^2-u_2}\leq 0$ for $v\geq \alpha_1$. On the other hand,
since 
\[
8v^4 - 6v^2(\sqrt{8v^2 + 1} +4) + 1\leq 8\alpha_1^4 - 6\alpha_2^2( \sqrt{8\alpha_2^2 + 1} +4)+1\approx -175<0,
\]
for $v\in [\alpha_2,\alpha_1]$, we conclude from (\ref{diff-T1}) that $T_1(v)$ is decreasing in $[\alpha_2,\alpha_1]$, yielding $T_1(v)\leq T_1(\alpha_2)$ for $v\in [\alpha_2,\alpha_1]$. Therefore,
\begin{equation}
\label{bound3-H}
H(\sqrt{u_2})\leq \left\{\begin{array}{ll}
T_1(\alpha_2)+  v^3  - \frac{5v}{2} -\sqrt{\pi} e^{v^2}(\erf(\sqrt{u_2}) - \erf(v)) \left(v^4 - 2v^2+ \frac{1}{4}\right),&\mbox{for $\alpha_2\leq v< \alpha_1$,}\\
 v^3  - \frac{5v}{2} -\sqrt{\pi} e^{v^2}(\erf(\sqrt{u_2}) - \erf(v)) \left(v^4 - 2v^2+ \frac{1}{4}\right),&\mbox{for $v\geq \alpha_1$.}
\end{array}
\right.
\end{equation}
We find that $T_1(\alpha_2)\approx 0.111$.

To find a negative upper-bound for $H(\sqrt{u_2})$ for $v\in  [\alpha_2,\infty)$ we will use the bounds \cite[Inq. 7.1.13, p. 298]{AS65},
\begin{equation}
\label{bounds-erf}
1-\frac{2e^{-x^2}}{\sqrt{\pi}(x+\sqrt{x^2+4/\pi)}}
\leq \erf(x)\leq 1-\frac{2e^{-x^2}}{\sqrt{\pi}(x+\sqrt{x^2+2})}, \quad x\geq 0.
\end{equation}
Applying these bounds to the r.h.s. of (\ref{bound3-H}) gives we obtain (recall that $v^4 - 2v^2+ \frac{1}{4}>0$ for $v\geq \alpha_2$ -- see Section \ref{sssec:properties}-(d))
\begin{equation}
\label{bound4-H}
H(\sqrt{u_2})\leq \left\{\begin{array}{ll}
T_1(\alpha_2)+  v^3  - \frac{5v}{2}-\left(\frac{2}{v+\sqrt{v^2 + 2}}- \frac{2e^{v^2-u_2}}{\sqrt{u_2}+\sqrt{u_2 +4/\pi}}\right) \left(v^4 - 2v^2+ \frac{1}{4}\right) ,&\mbox{for $\alpha_2\leq v< \alpha_1$,}\\
 v^3  - \frac{5v}{2} -\left(\frac{2}{v+\sqrt{v^2 + 2}}- \frac{2e^{v^2-u_2}}{\sqrt{u_2}+\sqrt{u_2 +4/\pi}}\right) \left(v^4 - 2v^2+ \frac{1}{4}\right),&\mbox{for $v\geq \alpha_1$.}
\end{array}
\right.
\end{equation}

\begin{lemma}
\label{lem:bound}
For $v>0$,
\begin{equation}
\label{lem:bound10}
\frac{2}{v+\sqrt{v^2 + 2}}- \frac{2e^{v^2-u_2}}{\sqrt{u_2}+\sqrt{u_2 +4/\pi}}\geq \frac{1}{v}-\frac{1}{2v^3}-\frac{e^{-\sqrt{2}v}}{v},
\end{equation}
\end{lemma}
{\it Proof:} Let $g_1(x)=\sqrt{1+x}-\left(1+\frac{x}{2}-\frac{x^2}{8}\right)$. We find
$g_1^\prime(x)=\frac{1}{2\sqrt{1+x}}-\frac{1}{2}+\frac{x}{4}$ and $g_1^{\prime\prime}(x)=\frac{1}{4}\left(1-\frac{1}{(1+x)^{3/2}}\right)\geq 0$ for $x\geq 0$.
The latter implies that $v\to g_1^\prime(x)$ is increasing in $[0,\infty)$, and since $g_1^\prime(0)=0$, we conclude that $v\to g_1(x)$ is also increasing in $[0,\infty)$.
The latter property and $g_1(0)=0$ imply that 
\begin{equation}
\label{bound-g1}
g_1(x)\geq 0, \quad \forall x\geq 0,
\end{equation}
By noting that $\frac{2}{v+\sqrt{v^2 + 2}}=\sqrt{v^2 +2} -v$, we get from (\ref{bound-g1}) (Hint: set $x=\frac{2}{v^2}$ for $v>0$)
\begin{equation}
\label{lem:bound11}
\frac{2}{v+\sqrt{v^2 + 2}} =v\left(\sqrt{1+\frac{2}{v^2}}-1\right)\geq \frac{1}{v}-\frac{1}{2v^3}, \quad \forall v>0.
\end{equation}
On the other hand, $v^2-u_2=-\frac{1+\sqrt{8v^2 +1}}{2} \leq -\sqrt{2}v$ and $\sqrt{u_2}+\sqrt{u_2+4/\pi}\geq 2\sqrt{u_2}\geq 2v$ since $u_2\geq v$ for $v\geq 0$.
Therefore,
\begin{equation}
\label{lem:bound12}
\sqrt{u_2}+\sqrt{u_2 +4/\pi} \leq\frac{e^{-\sqrt{2}v}}{v},\quad \forall v>0.
\end{equation}
Combining (\ref{lem:bound11}) and (\ref{lem:bound12}) gives (\ref{lem:bound10}). \hfill\done

Applying Lemma \ref{lem:bound} to (\ref{bound4-H}) gives (recall that $v^4 -2v^2+1/4\geq 0$ for $v\geq \sqrt{1+\alpha_0}$, and that $\alpha_1> \sqrt{1+\alpha_0}$)
\begin{equation}
\label{bound5-H}
H(\sqrt{u_2})\leq \left\{\begin{array}{ll}
T_1(\alpha_2)-\frac{10v^2 -1}{8v^3} +\frac{e^{-\sqrt{2}v}}{v}\left(v^4 -2 v^2 +\frac{1}{4}\right) ,&\mbox{for $\alpha_2\leq v< \alpha_1$,}\\
-\frac{10v^2 -1}{8v^3} +\frac{e^{-\sqrt{2}v}}{v}\left(v^4 -2 v^2 +\frac{1}{4}\right),&\mbox{for $v\geq \alpha_1$.}
\end{array}
\right.
\end{equation}
For $v\geq \frac{1}{2\sqrt{2}}$, $v^4-2v^2 +\frac{1}{4}\leq v^4$. Since $\frac{1}{2\sqrt{2}} <\alpha_2$, we get from (\ref{bound5-H})
\begin{equation}
\label{bound6-H}
H(\sqrt{u_2})\leq \left\{\begin{array}{ll}
T_1(\alpha_2)  -\frac{10v^2 -1}{8v^3} +v^3e^{-\sqrt{2}v} ,&\mbox{for $\alpha_2\leq v< \alpha_1$,}\\
  -\frac{10v^2 -1}{8v^3} +v^3e^{-\sqrt{2}v},&\mbox{for $v\geq \alpha_1$.}
\end{array}
\right.
\end{equation}
We first focus on $ -\frac{10v^2 -1}{8v^3} +v^3e^{-\sqrt{2}v}$.
Let us show $ -\frac{10v^2 -1}{8v^3} +v^3e^{-\sqrt{2}v}<0$, or equivalently, that $(10v^2-1)e^{\sqrt{2}v}>8v^6$ for $v\geq \alpha_2$.

$(10v^2-1)e^{\sqrt{2}v}>8v^6$  is equivalent to $g_1(v):=\log(10x^2-1)+x\sqrt{2}-6\log x -\log 8>0$. Differentiating $g_1(x)$ gives
\[
g_1^\prime(x)=\frac{10\sqrt{2} x^3 -40 x^2 -\sqrt{2} x +6}{10x^2-1}.
\]
This derivative vanishes whenever the polynomial $p_1(x)=10\sqrt{2} x^3 -40 x^2 -\sqrt{2} x +6$ vanishes. Since $p_1(0)=6$, $p_1(1/\sqrt{2})=-10$,  $\lim_{x\to-\infty} p_1(x)=-\infty$,
and $\lim_{x\to\infty} p_1(x)=\infty$
we conclude that this polynomial of degree $3$ has three real roots, one located in $(-\infty,0)$, one located in $(0,1/\sqrt{2})$, and one located in $(1/\sqrt{2},\infty)$.
Call $v^*$ the zero of $p_1(x)$ located in $(1/\sqrt{2},\infty)$. Hence, the mapping $v\to g_1(v)$ is decreasing in $(1/\sqrt{2},v^*)$ and increasing in $(v^*,\infty)$, which
implies that it takes its smallest value at $v=v^*$ in the interval $(1/\sqrt{2},\infty)$. Solving for $p_1(x)=0$ we find that $v^*\approx 2.81$.

Consequently, (a) $g_1(v)\geq g_1(\alpha_2)$ if $v\in [\alpha_2,\alpha_1]$ since  $1/\sqrt{2}<\alpha_2<\alpha_1<v^*$, and (b)  $g_1(v)\geq g_1(v^*)$ for $v\geq \alpha_1$.
Since $g(v^*)\approx 0.051$, property (b) implies that $H(\sqrt{u_2})<0$ for $v\geq \alpha_1$.

It remains to show that $H(\sqrt{u_2})<0$ for $\alpha_2\leq v<\alpha_1$ or, equivalently from  (\ref{bound6-H}), that 
$T_1(\alpha_2) -\frac{10v^2 -1}{8v^3}+v^3e^{-\sqrt{2}v}\leq 0$ for $\alpha_2\leq v<\alpha_1$. 

Easy algebra gives
\begin{equation}
\label{ident-1}
 -\frac{10v^2 -1}{8v^3} +v^3e^{-\sqrt{2}v}=v^3 e^{-\sqrt{2}v}\left(1-e^{g_1(v)}\right).
 \end{equation}
Let $v\in [\alpha_2,\alpha_1)$.  We have from (\ref{ident-1})
\begin{align}
\label{inq10}
T_1(\alpha_2) -\frac{10v^2 -1}{8v^3} +v^3e^{-\sqrt{2}v}&=T_1(\alpha_2)+v^3 e^{-\sqrt{2}v}\left(1-e^{g_1(v)}\right),\nonumber\\
&\leq T_1(\alpha_2)+v^3 e^{-\sqrt{2}v}\left(1-e^{g_1(\alpha_2)}\right),
\end{align}
since we have shown in (a) above  that $g_1(v)\geq g_1(\alpha_2)$ when $v\in [\alpha_2,\alpha_1)$.

Since $g_1(\alpha_2)\approx 26$, we have  $v^3e^{-\sqrt{2}v} \left(1-e^{g_1(\alpha_2)}\right)<\alpha_2^3e^{-\sqrt{2}\alpha_1}\left(1-e^{g_1(\alpha_2)}\right)$ and, from
(\ref{inq10}),
\begin{align*}
T_1(\alpha_2) -\frac{10v^2 -1}{8v^3}+v^3e^{-\sqrt{2}v}
&\leq T_1(\alpha_2)+\alpha_2^3 e^{-\sqrt{2}\alpha_1}\left(1-e^{g_1(\alpha_2)}\right)\approx -0.088<0.
\end{align*}
This concludes the proof that $H(\sqrt{u_2})<0$ for $v\geq \alpha_2$.



\section{Proof of Proposition \ref{prop:asymptotic-expo}}
\label{sec:proof-prop:asymptotic-expo}

Let $q(x)=\lambda e^{-\lambda x}$, $q_\gamma (x)=\lambda_\gamma e^{-\lambda_\gamma x}$ for $x\geq 0$, and
$Y=\log \frac{q_\gamma(X)}{q(X)}= a_\gamma +(\lambda-\lambda_\gamma)X$, where $a_\gamma:=\log (\lambda_\gamma/\lambda)$.
We assume that $\lim_\gamma \lambda_\gamma=\lambda$.
Further introduce $r(x)=\mu e^{-\mu x}$ for $x\geq 0$. For $x\in\R$, we have
\[
F_r(x):=\Prob_r(Y<x)=\int_0^\infty {\bf 1}_{\{a_\gamma +(\lambda-\lambda_\gamma)z< x\}} r(z)dz
=\left\{\begin{array}{ll}
\int_{\frac{[a_\gamma -x]^+}{\lambda_\gamma-\lambda}}^\infty 
\mu  e^{-\mu z}dz =e^{-\mu\frac{[a_\gamma -x]^+}{\lambda_\gamma-\lambda}}, &\mbox{if $\lambda_\gamma>\lambda$},\\
\int_0^{\frac{[x-a_\gamma ]^+}{\lambda-\lambda_\gamma}}
\mu  e^{-\mu z}dz =1-e^{-\mu \frac{[x-a_\gamma]^+}{\lambda-\lambda_\gamma}}, &\mbox{if $\lambda_\gamma<\lambda$},
\end{array}
\right.
\]
so that 
\begin{equation}
f_r(x):=\frac{d}{dx}F_r(x)=\left\{\begin{array}{ll}
\frac{\mu}{\lambda_\gamma-\lambda}  e^{-\mu\frac{a_\gamma -x}{\lambda_\gamma-\lambda}} {\bf 1}_{\{x<  a_\gamma\}}, &\mbox{if $\lambda_\gamma>\lambda$},\\
\frac{\mu}{\lambda-\lambda_\gamma} e^{-\mu\frac{x-a_\gamma}{\lambda-\lambda_\gamma}}{\bf 1}_{\{x>  a_\gamma\}},&\mbox{if $\lambda_\gamma<\lambda$}.
\end{array}
\right.
\label{pdf-f-r}
\end{equation}
Using (\ref{lifetime-sup}) we obtain (Hint: $a_\gamma<0$ when $\lambda_\gamma<\lambda$)
\begin{align}
\E_{r}[(Y-y){\bf 1}{\{Y\geq y\}}\,|\, Y\geq y]&=\frac{\int_y^\infty {\bf 1}_{\{x\geq y\}} (x-y)f_r(x)dx}{\int_y^\infty  {\bf 1}_{\{x\geq y\}}  f_r(x)dx},\nonumber\\
&=
\left\{\begin{array}{ll}
\frac{\int_y^{a_\gamma} (x-y)e^{-\mu\frac{a_\gamma -x}{\lambda_\gamma-\lambda}}dx}{\int_y^{a_\gamma}e^{-\mu\frac{a_\gamma -x}{\lambda_\gamma-\lambda}}dx}
=\frac{\lambda-\lambda_\gamma}{\mu}+\frac{a_\gamma-y}{1-e^{-\mu\frac{a_\gamma -y}{\lambda_\gamma-\lambda}}},
&\mbox{for $0\leq y\leq a_\gamma$ and $\lambda_\gamma>\lambda$,}\\
\frac{\int_y^\infty (x-y)e^{-\mu\frac{x-a_\gamma}{\lambda-\lambda_\gamma}}dx}
{\int_y^\infty  e^{-\mu\frac{x-a_\gamma}{\lambda-\lambda_\gamma}}dx}
=\frac{\lambda-\lambda_\gamma}{\mu},&\mbox{for $y\geq 0$ and $\lambda_\gamma<\lambda$.}
\end{array}
\right.
\label{sup-expo}
\end{align}
We already conclude from the above that, when  $\lambda_\gamma<\lambda$,
\begin{equation}
\sup_{y\geq 0}\E_{r}[(Y-y){\bf 1}_{\{Y\geq y\}}\,|\, Y\geq y] =\left\{\begin{array}{ll}
\frac{\lambda-\lambda_\gamma}{\lambda}, &\mbox{when $r=q$,}\\
\frac{\lambda-\lambda_\gamma}{\lambda_\gamma}, &\mbox{when $r=q_\gamma$}.
\end{array}
\right.
\label{case1}
\end{equation}
Assume now that  $\lambda_\gamma>\lambda$. 
Define $f(y)=\frac{1}{d}+\frac{a_\gamma-y}{1-e^{-d(a_\gamma -y)}}$ with $d:=\frac{\mu}{\lambda_\gamma-\lambda}$, so that (cf. (\ref{sup-expo}))
\[
\sup_{y\geq 0}\E_{r}[(Y-y) {\bf 1}_{\{Y\geq y\}}\,|\, Y\geq y] =\sup_{0\leq y\leq a_\gamma} f(y).
\]
We find
\[
f^\prime(y)=\frac{-1+e^{-d(a_\gamma-y)}+d(a_\gamma-y)e^{-d(a-y)}}{\left(1-e^{-d(a_\gamma-y)}\right)^2}.
\]
The derivative of the numerator of $f^\prime(y)$ is equal to $d^2(a-y)e^{-d(a_\gamma -y})$, which is non-negative for $y\in [0,a_\gamma]$.
Consequently the mapping $y\to f(y)$ is non-decreasing in $[0,a_\gamma]$.  By L'H\^opital's rule we get that 
$\lim_{y\to a_\gamma}f(y)= \frac{2}{d}=\frac{2(\lambda_\gamma-\lambda)}{\mu}$, so that from (\ref{sup-expo})
\begin{equation}
\sup_{y\geq 0}\E_{r}[(Y-y){\bf 1}_{\{Y\geq y\}}\,|\, Y\geq y] =\left\{\begin{array}{ll}
\frac{2(\lambda_\gamma-\lambda)}{\lambda}, &\mbox{when $r=q$,}\\
\frac{2(-\lambda_\gamma-\lambda)}{\lambda_\gamma}, &\mbox{when $r=q_\gamma$},
\end{array}
\right.
\label{case2}
\end{equation}
when $\lambda_\gamma>\lambda$. We conclude from (\ref{case1})-(\ref{case2}) that $\lim_\gamma \sup_{y\geq 0}\E_{r}[Y-y\,|\, Y\geq y]=0$
for $r\in\{q,q_\gamma\}$.

Consider now the infimum. We have  from (\ref{lifetime-inf}) and (\ref{pdf-f-r})
\begin{align}
\E_{r}[(Y-y){\bf 1}{\{Y\leq y\}}\,|\, Y\leq y]&=\frac{\int_{-\infty}^y {\bf 1}_{\{x\leq y\}} (x-y)f_r(x)dx}{\int_{-\infty}^y  {\bf 1}_{\{x\leq y\}}  f_r(x)dx},\nonumber\\
&=
\left\{\begin{array}{ll}
\frac{\int_{-\infty}^y (x-y)e^{-\mu\frac{a_\gamma -x}{\lambda_\gamma-\lambda}}dx}{\int_{-\infty}^y e^{-\mu\frac{a_\gamma -x}{\lambda_\gamma-\lambda}}dx}
=\frac{\lambda-\lambda_\gamma}{\mu}, &\mbox{for $y\leq 0$ and $\lambda_\gamma>\lambda$,}\\
\frac{\int_{a_\gamma}^y (x-y)e^{-\mu\frac{x-a_\gamma}{\lambda-\lambda_\gamma}}dx}
{\int_{a_\gamma}^y  e^{-\mu\frac{x-a_\gamma}{\lambda-\lambda_\gamma}}dx}
=\frac{\lambda-\lambda_\gamma}{\mu}-\frac{y-a_\gamma}{1-e^{-\frac{\mu(y-a_\gamma)}{\lambda-\lambda_\gamma}}}
,&\mbox{for $a_\gamma\leq y\leq 0$ and $\lambda_\gamma<\lambda$.}
\end{array}
\right.
\label{inf-expo}
\end{align}
Mimicking the analysis for the supremum, we easily get from (\ref{inf-expo}) that $\lim_\gamma \inf_{y\leq 0} \E_{r}[(Y-y){\bf 1}{\{Y\leq y\}}\,|\, Y\leq y]=0$ for 
$r\in \{q,q_\gamma\}$. This concludes the proof. \hfill\done


\end{document}